\documentclass[graybox, envcountchap]{svmult}

\usepackage{mathptmx}        
\usepackage{amsmath}
\usepackage{amssymb}
\usepackage{color}
\usepackage{helvet}          
\usepackage{courier}         
\usepackage{dirtree}

\usepackage{makeidx}        
\usepackage{graphicx}        
\usepackage{subfig}

\usepackage{multicol}        
\usepackage[bottom]{footmisc}

\usepackage[misc]{ifsym}

\newcommand{\be}{\begin{eqnarray}}
\newcommand{\ee}{\end{eqnarray}}
\newcommand{\rar}{\rightarrow}

\begin{document}

\title*{Accreting Black Holes}

\author{Sourabh Nampalliwar and Cosimo Bambi}

\institute{Sourabh Nampalliwar (\Letter) \at Theoretical Astrophysics, Eberhard-Karls Universit\"at T\"ubingen, Auf der Morgenstelle 10, 72076 T\"ubingen, Germany, \email{sourabh.nampalliwar@uni-tuebingen.de} \and Cosimo Bambi \at Department of Physics, Fudan University, 2005 Songhu Road, Shanghai 200438, China, \email{bambi@fudan.edu.cn}}

\maketitle

\abstract{This chapter provides a general overview of the theory and observations of black holes in the Universe and on their interpretation. We briefly review the black hole classes, accretion disk models, spectral state classification, the AGN classification, and the leading techniques for measuring black hole spins. We also introduce quasi-periodic oscillations, the shadow of black holes, and the observations and the theoretical models of jets.}


\section{Theory of black holes: formation and masses}\label{s2-classes}

Black holes happen to be surprisingly simple objects. Only two parameters, the \emph{mass} $M$ and the \emph{spin} $J$, are thought to be sufficient to characterize a black hole in our Universe~\cite{r-bh-book}. The spin parameter cannot be arbitrary and must satisfy the constraint $J/M^2 \le 1$, which is the condition for the existence of the event horizon. There are no theoretical constraints on the value of the mass of a black hole, which may thus be arbitrarily small as well as arbitrarily large.

From astronomical observations, we have strong evidence of two classes of astrophysical black holes:
\begin{enumerate}
\item {\it Stellar-mass black holes}~\cite{r-bh-re-mc}, with masses $\sim 3-100 M_\odot$.
\item {\it Supermassive black holes}~\cite{r-bh-k-r}, with masses $> 10^5 M_\odot$.
\end{enumerate}
One would expect, and there is some evidence, that black holes with masses in the intermediate range should exist~\cite{r-bh-inter}. These are termed {\it intermediate-mass black holes}. Each of these classes is theorized to have a different past, present and future. We will discuss them separately.

\subsection{Stellar-mass black holes}

The most common formation channel for stellar-mass black holes is gravitational collapse. In lay terms, when a star runs out of fuel, the pressure inside is insufficient to hold the star against gravitational pull and the star collapses. For massive enough stars, the star collapses all the way to a singularity and a black hole is born.  

The initial mass of a stellar-mass black hole depends on the properties of the progenitor: its mass, its evolution, and the supernova explosion mechanism~\cite{r-bh-mass}. Depending on these details, the supernova remnant could be a neutron star, where the quantum neutron pressure can hold against the gravitational collapse, or a black hole. In fact, the lower bound on the black hole initial mass may come from the maximum mass for a neutron star: the exact value is currently unknown, since it depends on the equation of state of matter at super-nuclear densities, but it should be in the range of $2-3~M_\odot$. It is possible though, that a mass gap exists between the most massive neutron stars and the less massive black holes~\cite{r-bh-mass-gap}. An upper bound on stellar-mass black holes may be derived from the progenitor's metallicity. The final mass of the remnant is determined by the mass loss rate by stellar winds, which increases with the metallicity because heavier elements have a larger cross section than lighter ones, and therefore they evaporate faster. For a low-metallicity progenitor~\cite{r-bh-gg1,r-bh-gg2,r-bh-gg3}, the mass of the black hole remnant may be $M \lesssim 50$~$M_\odot$ or $M \gtrsim 150$~$M_\odot$. As the metallicity increases, black holes with $M \gtrsim 150$~$M_\odot$ disappear, because of the increased mass loss rate. Note, however, that some models do not find remnants with a mass above the gap, because stars with $M \gtrsim 150$~$M_\odot$ may undergo a runaway thermonuclear explosion that completely destroys the system, without leaving any black hole remnant~\cite{r-bh-gg1,r-bh-gg2}. Stellar-mass black holes may thus have a mass in the range of $3-100~M_\odot$. Until now, all the known stellar-mass black holes in X-ray binaries have a mass $M \approx 3-20~M_\odot$~\cite{r-bh-mass2}. Gravitational waves, on the other hand, have shown the existence of heavier stellar-mass black holes. In particular, the event called GW150914 was associated with the coalescence of two black holes with masses $M \approx 30$~$M_\odot$ that merged to form a black hole with $M \approx 60$~$M_\odot$~\cite{r-bh-gw150914}.

From stellar evolution studies, we expect that in our Galaxy there is a population of $10^8$-$10^9$~black holes formed at the end of the evolution of heavy stars~\cite{r-bh-bhnum0,r-bh-bhnum}, and the same number can be expected in similar galaxies. But with observations, we only know about 20~black holes with a dynamical measurement of the mass and about 50 without (it is thus possible that some of them are not black holes but neutron stars). This is because their detection is very challenging. The simplest scenario is when the black hole is in a binary system and has a companion star. The presence of a compact object can be discovered from the observation of a short timescale variability, the non-detection of a stellar spectrum, etc. The study of the orbital motion of the companion star can permit the measurement of the mass function~\cite{r-bh-mass2}
\be
f (M) = \frac{K_{\rm c}^3 P_{\rm orb}}{2 \pi G_{\rm N}} 
= \frac{M \sin^3 i}{\left( 1 + q \right)^2} \, ,
\ee
where $K_{\rm c} = v_{\rm c} \sin i$, $v_{\rm c}$ is the velocity of the companion star, $i$ is the angle between the normal of the orbital plane and our line of sight, $P_{\rm orb}$ is the orbital period of the system, $q = M_{\rm c}/M$, $M_{\rm c}$ is the mass of the companion, and $M$ is the mass of the dark object. If we can somehow estimate $i$ and $M_{\rm c}$, we can infer $M$, and in this case we talk about dynamical measurement of the mass. The dark object is a black hole if $M > 3$~$M_\odot$~\cite{r-bh-bh1,r-bh-bh2,r-bh-bh3}.

Note that, among astronomers, it is common to call ``black hole'' a compact object for which there is a dynamical measurement of its mass proving that $M > 3$~$M_\odot$. The latter indeed guarantees that the object is too heavy for being a neutron star. ``Black hole candidates'' are instead compact objects that are supposed to be black holes, for instance because of the detection of spectral features typical of black holes, but for which there is no dynamical measurement of their mass.

\begin{figure}[t]
\vspace{0.3cm}
\begin{center}
\includegraphics[width=10.0cm]{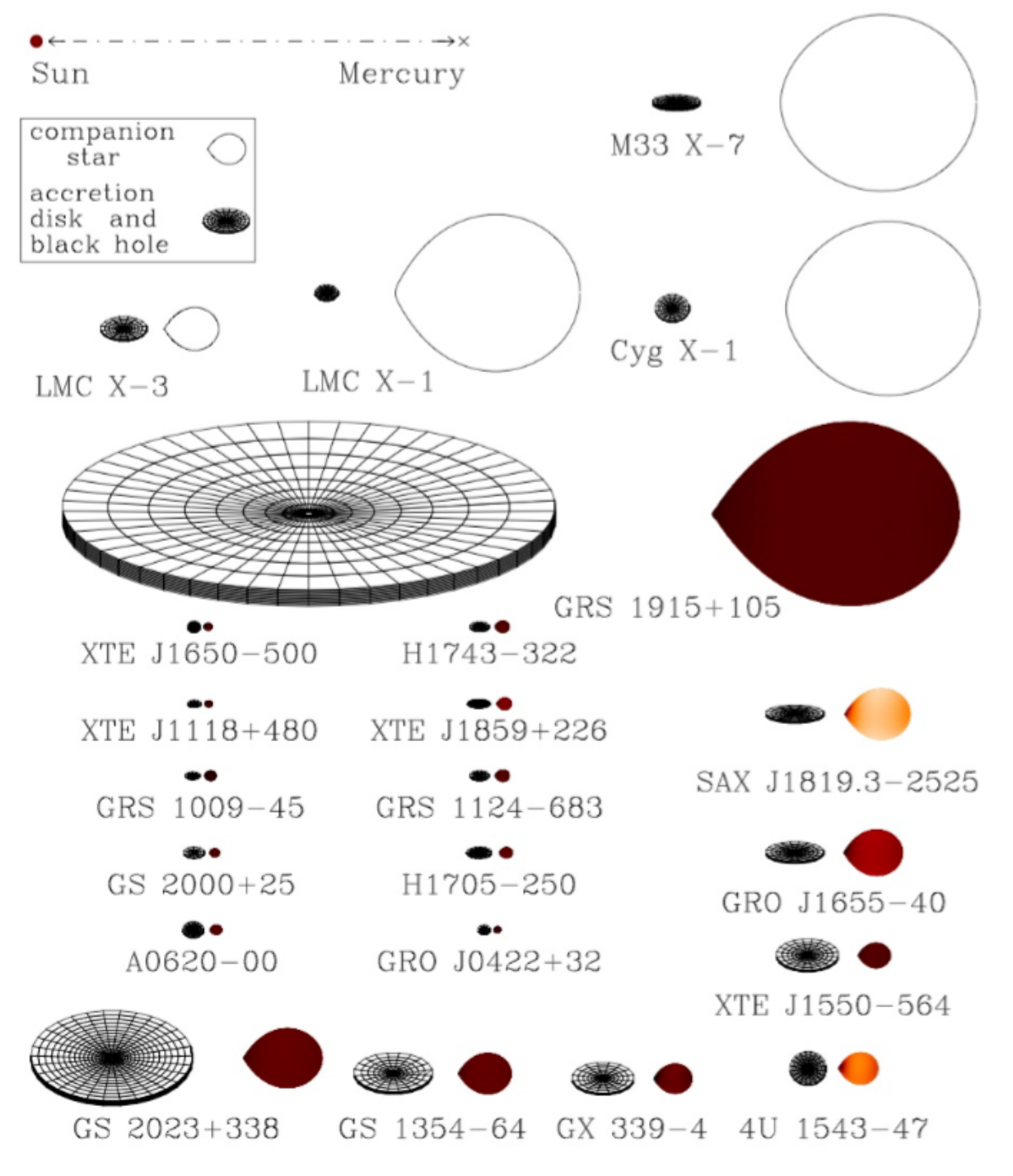}
\end{center}
\vspace{0.0cm}
\caption{Sketch of 22~X-ray binaries with a stellar-mass black hole confirmed by dynamical measurements. For every system, the black hole accretion disk is on the left and the companion star is on the right. The color of the companion star roughly indicates its surface temperature (from brown to white as the temperature increases). The orientation of the disks indicates the inclination angles of the binaries. For comparison, in the top left corner of the figure we see the system Sun-Mercury: the distance between the two bodies is about 50~million km and the radius of the Sun is about 0.7~million km. Figure courtesy of Jerome Orosz. \label{f-bhb}}
\end{figure}

Black holes in X-ray binaries (black hole binaries\footnote{Generally speaking, a {\it black hole binary} is a binary system in which at least one of the two bodies is a black hole, and a {\it binary black hole} is a binary system of two black holes.}) are grouped into two classes: {\it low-mass X-ray binaries} (LMXBs) and {\it high-mass X-ray binaries} (HMXBs). Here, ``low'' and ``high'' refers to the stellar companion, not to the black hole: in the case of LMXBs, the companion star has normally a mass $M_{\rm c} < 3$~$M_\odot$, while for HMXBs the companion star has $M_{\rm c} > 10$~$M_\odot$. Observationally, we can classify black hole binaries either as {\it transient X-ray sources} or {\it persistent X-ray sources}. LMXBs are usually transient sources, because the mass transfer is not continuos (for instance, at some point the surface of the companion star may expand and the black hole strips some gas): the system may be bright for a period ranging from some days to a few months and then be in a quiescent state for months or even decades. Every year we discover 1-2~new objects, when they pass from their quiescent state to an outburst (see Section~\ref{s2-spectra}). Overall, we expect $10^3$-$10^4$~LMXBs in the Galaxy~\cite{r-bh-lm1,r-bh-lm2}. HMXBs are persistent sources: the mass transfer from the companion star to the black hole is a relatively regular process (typically it is due to the stellar wind of the companion) and the binary is a bright source at any time without quiescent periods.
Fig.~\ref{f-bhb} shows 22~X-ray binaries with a stellar-mass black hole confirmed by dynamical measurements. To have an idea of the size of these systems, the figure also shows the Sun (whose radius is 0.7~million km) and the distance Sun-Mercury (about 50~million km). The black holes have a radius $< 100$~km and cannot be seen, but we can clearly see their accretion disks formed from the transfer of material from the companion star. The latter may have a quite deformed shape (in particular, we can see some cusps) due to the the tidal force produced by the gravitational field of the black hole. Among the sources listed in the figure, Cygnus~X-1 (Cyg~X-1 in Fig.~\ref{f-bhb}), LMC~X-1, LMC~X-3, and M33~X-7 are HMXBs, while all other systems are LMXBs. Among these HMXBs, only Cygnus~X-1 is in our Galaxy. Among the LMXBs, there is GRS~1915+105, which is quite a peculiar source: since 1992, it is a bright X-ray source in the sky, so it can be considered a persistent source. This is probably because of its large accretion disk, which can provide enough material at any time.

\begin{figure}[t]
\vspace{0.3cm}
\begin{center}
\includegraphics[width=8.7cm]{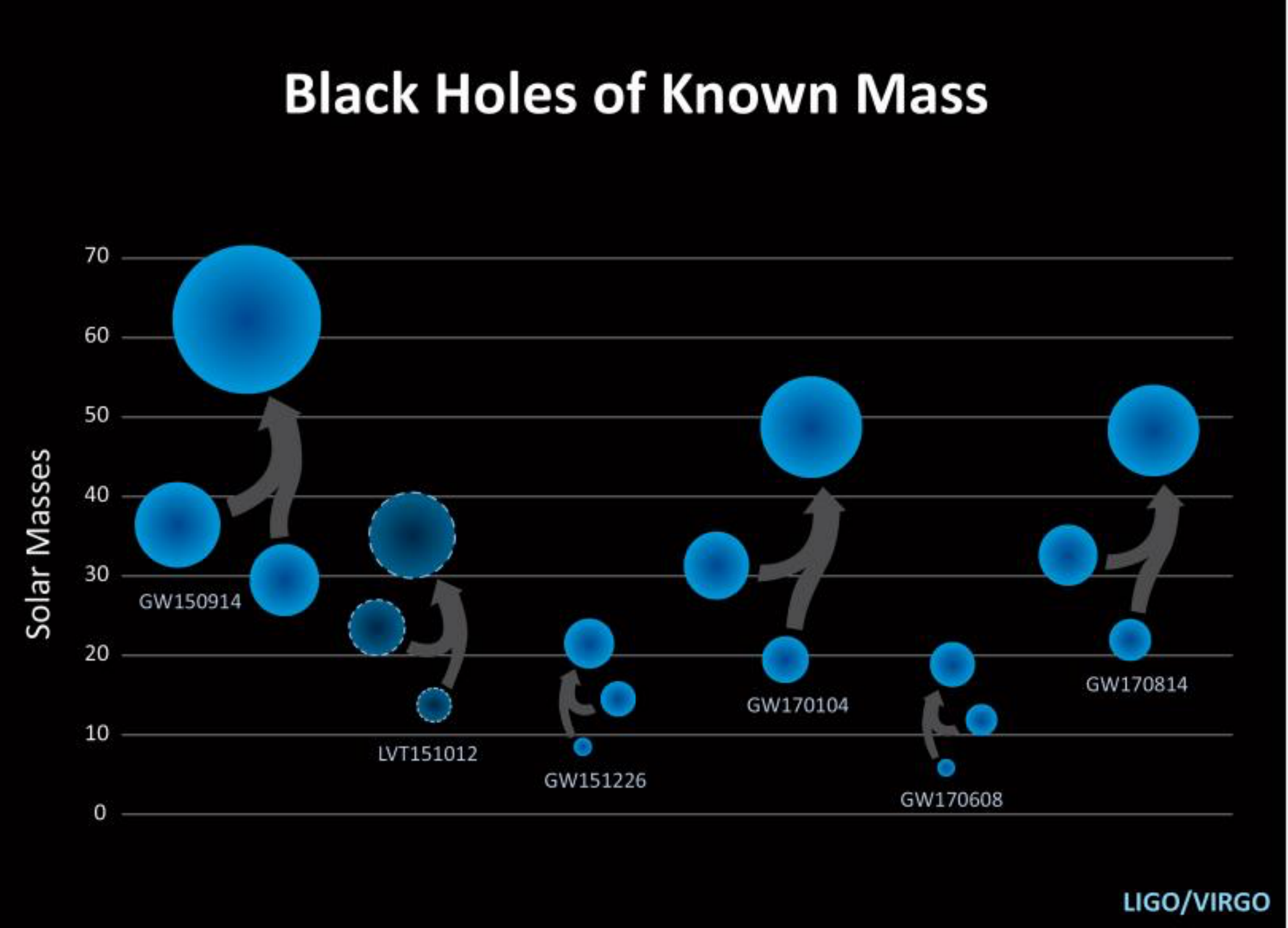}
\end{center}
\vspace{0.0cm}
\caption{Masses of the first black holes observed with gravitational waves, with the two initial objects merging into a larger one, as shown by the arrows. Image Credit: LIGO/NSF/Caltech/SSU Aurore Simmonet.  \label{f-bhb2}}
\end{figure}

Black holes in compact binary systems (black hole-black hole or black hole-neutron star) can be detected with gravitational waves when the signal is sufficiently strong. Fig.~\ref{f-bhb2} shows the first detections by the LIGO/Virgo collaboration. The name of the event is classified as GW (gravitational wave event) and then the date of detection: for example, GW150914 was detected on 14~September~2015. LVT151012 is not classified as a gravitational wave event because the signal to noise ratio was not large enough to qualify as a detection\footnote{LVT stands for LIGO/Virgo transient.}. For every event, the figure shows the two original black holes as well as the final one after merger.

Isolated black holes are much more elusive. In principle, they can be detected by observing the modulation of the light of background stars due to the gravitational lensing caused by the passage of a black hole along the line of sign of the observer~\cite{r-bh-lensing}.

\subsection{Supermassive black holes}
The formation channels of supermassive black holes are not well established. The gigantic masses of supermassive black holes are not thought to be natal, but acquired. Accretion has been shown to be an effective mechanism for growing the masses of black holes. In fact, some models suggest the possibility of super-Eddington accretion, and this may indeed be a possible path to the rapid growth of supermassive black holes~\cite{r-bh-super-e1}. Another possibility is merger of several black holes. But the question of the progenitor, or \emph{seed}, remains open. See~\cite{r-bh-mv} for a review of the possible formation channels.

\begin{figure}[t]
\vspace{0.2cm}
\begin{center}
\includegraphics[width=8.0cm]{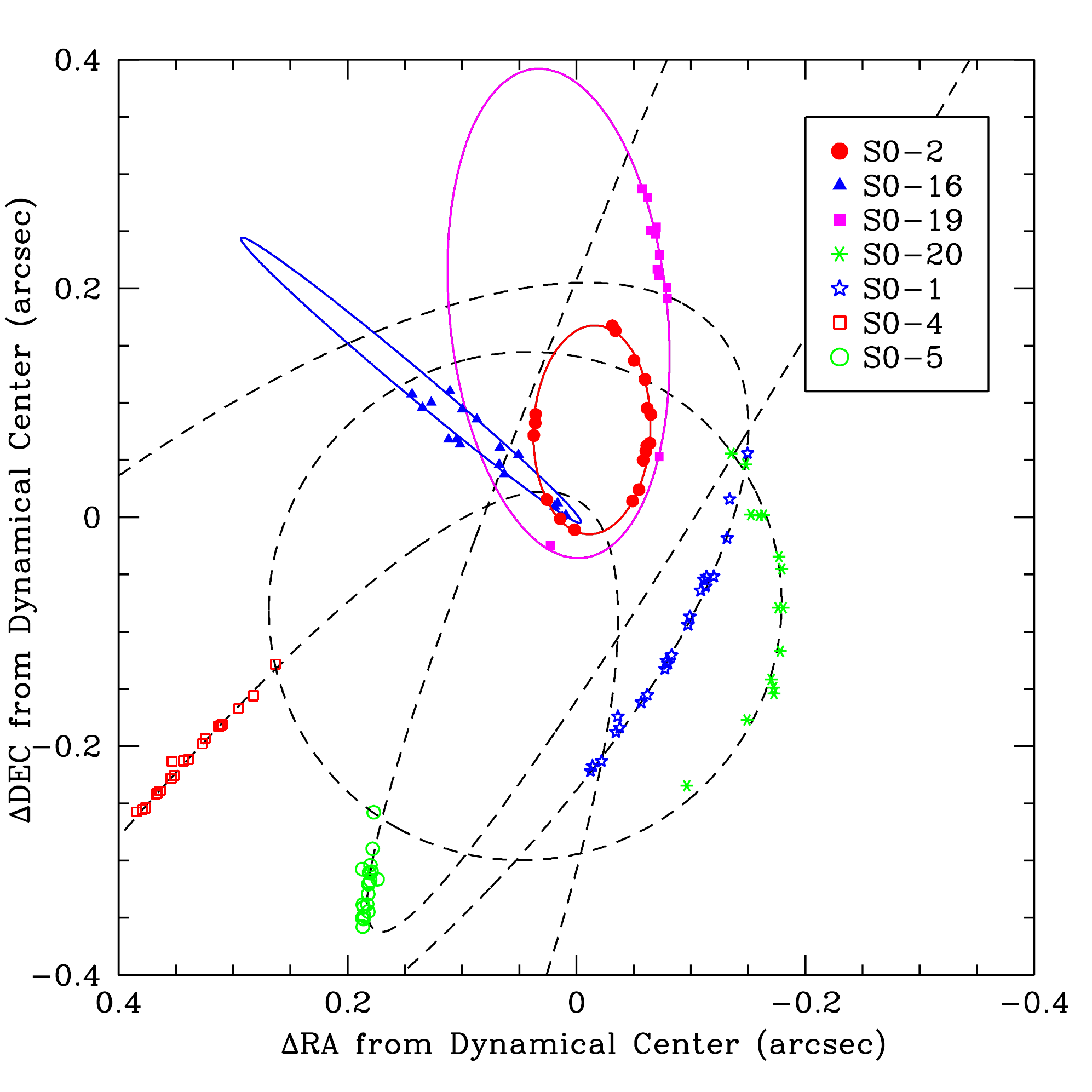}
\end{center}
\vspace{-0.2cm}
\caption{Astrometric positions and orbital fits for seven stars orbiting the supermassive black hole at the center of the Galaxy. From~\cite{r-bh-c-ghez}. \copyright AAS. Reproduced with permission. \label{f-ghez05}}
\end{figure}

Astronomical observations show that at the center of many galaxies there is a large amount of mass in a relatively small volume. The standard interpretation is that these objects are supermassive black holes with $M \sim 10^5$--$10^{10}$~$M_\odot$. Strong constraints come from the center of our Galaxy and NGC~4258~\cite{r-bh-maoz}. For our Galaxy, we can study the Newtonian motion of individual stars and infer that at the center there is an object with a mass of $4\cdot 10^6$~$M_\odot$ (see Fig.~\ref{f-ghez05}). An upper bound on the size of this body can be obtained from the minimum distance approached by one of these stars, which is less than 45~AU and corresponds to $\sim 1,200$~$r_{\rm g}$ for a $4\cdot 10^6$~$M_\odot$ object. In the end, we can exclude the existence of a cluster of compact non-luminous bodies like neutron stars and therefore we can conclude that the most natural interpretation is that it is a supermassive black hole. In the case of NGC~4258, we can study the orbital motion of gas in the nucleus, and again we can conclude that the central object is too massive, compact, and old to be a cluster of neutron stars. In the case of other galaxies, it is not possible to put such constraints with the available data, but it is thought that every mid-size (like the Milky Way) or large galaxy has a supermassive black hole at its center\footnote{Exceptions may be possible: the galaxy A2261-BCG has a very large mass but it might not have any supermassive black hole at its center~\cite{r-bh-postman}.}. For smaller galaxies, the situation is more uncertain. Most models predict supermassive black holes at the center of lighter galaxies as well~\cite{r-bh-mv}, but there exist predictions of faint low-mass galaxies with no supermassive black hole at their centers~\cite{r-bh-mv07b,r-bh-mv07a}. Observations suggest that some small galaxies have a supermassive black hole and other small galaxies do not~\cite{r-bh-ferrarese,r-bh-amuse}.

\subsection{Intermediate-mass black holes}

Intermediate-mass black holes are, by definition, black holes with a mass between the stellar-mass and the supermassive ones, say $M \sim 10^2$--$10^5$~$M_\odot$. At the moment, there is no dynamical measurement of the mass of these objects, and their actual nature is still controversial. Among the possible formation channels, intermediate-mass black holes are expected to form at the center of dense stellar clusters, by mergers. 

Observational evidence for intermediate-mass black holes is inconclusive. The presence of an intermediate-mass black hole at the center of stellar clusters should increase the velocity dispersion in the cluster. Some studies suggest that there are indeed intermediate-mass black holes at the center of certain globular clusters~\cite{r-bh-clu1,r-bh-clu2}.
Some intermediate-mass black hole candidates are associated with ultra luminous X-ray sources~\cite{r-bh-ulxs}. These objects have an X-ray luminosity $L_X > 10^{39}$~erg/s, which exceeds the Eddington luminosity of a stellar-mass object, and they may thus have a mass in the range $10^2$--$10^5$~$M_\odot$. However, we cannot exclude the possibility that they are actually stellar-mass black holes (or neutron stars~\cite{r-bh-super-e2}) with non-isotropic emission and a moderate super-Eddington mass accretion rate~\cite{r-bh-mine}.
The existence of intermediate-mass black holes is also suggested by the detection of some quasi-periodic oscillations (QPOs, see Section~\ref{ss-qpos}) in some ultra-luminous X-ray sources. QPOs are currently not well understood, but they are thought to be associated to the fundamental frequencies of the oscillation of a particle around a black hole. Since the size of the system scales as the black hole mass, QPOs should scale as $1/M$, and some observations indicate the existence of compact objects with masses in the range $10^2$-$10^5$~$M_\odot$~\cite{r-bh-i-qpo}.


\section{Theory of black holes: evolution and spins}

Apart from mass, a typical black hole is expected to have some spin. Generally speaking, the value of the spin parameter of a black hole can be expected to be determined by the competition of three physical processes: the event creating the object, mergers, and gas accretion.

\subsection{Stellar-mass black holes}

In the case of black hole binaries, it is usually thought that the spin of a black hole is mainly natal and that the effect of the accretion process is negligible~\cite{r-bh-thin-acc-king}. The argument is that a stellar-mass black hole has a mass around 10~$M_\odot$. If the stellar companion is a few Solar masses, the black hole cannot significantly change its mass and spin angular momentum even after swallowing the whole star. If the stellar companion is heavy, its lifetime is too short: even if the black hole accretes at the Eddington rate, there is not enough time to transfer the necessary amount of matter to significantly change the black hole spin parameter. One may expect that a black hole cannot swallow more than a few $M_\odot$ from the companion star, and for a $10$~$M_\odot$ object this is not enough to significantly changes $a_*$~\cite{r-bh-thin-acc-king}.
If the black hole spin were mainly natal, its value should be explained by studying the gravitational collapse of massive stars. While there are still uncertainties in the angular momentum transport mechanisms of the progenitors of stellar-mass black holes, it is widely accepted that the gravitational collapse of a massive star with Solar metallicity cannot create fast-rotating remnants~\cite{r-bh-thin-acc-rem1,r-bh-thin-acc-rem2}. The birth spin of these black holes is expected to be low (see e.g.~\cite{r-bh-thin-acc-fragos} and references therein). 

Observations of spins of stellar-mass black holes contradict the above hypothesis. For instance, in the case of LMXBs, the black hole in GRS~1915+105 has $a_* > 0.98$~\cite{r-bh-cfm-1915} and $M = 12.4 \pm 2.0$~$M_\odot$~\cite{r-bh-thin-acc-reid}, while the stellar companion's mass is $M = 0.52 \pm 0.41$~$M_\odot$. In the case of HMXBs, the black hole in Cygnus~X-1 has $a_* > 0.98$~\cite{r-bh-cfm-cyg1,r-bh-cfm-cyg2} and $M = 14.8 \pm 1.0$~$M_\odot$, while the stellar wind from the companion is not an efficient mechanism to transfer mass. Very high spin values are also measured for 4U~1630-472, GS~1354-645, MAXI~J1535-571, and Swift~J1658.2, see Tab.~\ref{t-spin}.
While black holes in LMXBs and HMXBs should form in different environments, in both cases the origin of so high spin values is puzzling. In~\cite{r-bh-thin-acc-fragos}, the authors show that at least in the case of LMXBs, the accretion process immediately after the formation of a black hole binary may be very important and be responsible for the observed high spins. For HMXBs, possible channels for producing high spins are discussed in~\cite{r-bh-Qin:2018sxk}.

\subsection{Supermassive black holes}

The case of supermassive black holes in galactic nuclei is different. The initial value of their spin parameter is likely completely irrelevant: their mass has increased by several orders of magnitude from its original value, and the spin parameter has evolved accordingly. 

There are two primary channels of mass acquisition for supermassive black holes, mergers and accretion. On average, the capture of small bodies (\emph{minor merger}) in randomly oriented orbits should spin the black hole down, since the magnitude of the orbital angular momentum for corotating orbits is always smaller than the one for counterrotating orbits~\cite{r-bh-thin-scott}. In the case of random merger of two black holes with comparable mass (\emph{major merger}), the most probable final product is a black hole with $a_* \approx 0.70$, while fast-rotating objects with $a_* > 0.9$ should be rare~\cite{r-bh-thin-berti}.
On the other hand, accretion from a disk can potentially be a very efficient way to spin a compact object up\footnote{Unless the accretion proceeds via short episodes (chaotic accretion)~\cite{r-bh-thin-chaotic}, in which case it is effectively like minor mergers.}~\cite{r-bh-thin-berti}. In this case, black holes in active galactic nuclei (AGNs) may have a spin parameter close to the Thorne limit (see next section). Such a possibility seems to be supported by some observations; see e.g.~\cite{r-bh-thin-agn-soltan} and also the spin measurements from X-ray reflection spectroscopy in Tab.~\ref{t-spin-agn}.

\subsection{Thorne limit \label{ss-evolution}}

An accreting black hole changes its mass $M$ and spin angular momentum $J$ as it swallows more and more material from its disk. In the case of a Novikov-Thorne disk (see next section), it is relatively easy to calculate the evolution of these parameters. If we assume that the gas in the disk emits radiation until it reaches the radius of the innermost stable circular orbit (ISCO) and then quickly plunges onto the black hole, the evolution of the spin parameter $a_*$ is governed by the following equation~\cite{r-bh-thorne}
\be\label{eq-evolution}
\frac{d a_*}{d \ln M} = \frac{c}{r_{\rm g}} 
\frac{L_{\rm ISCO}}{E_{\rm ISCO}} - 2 a_* \, ,
\ee
where $E_{\rm ISCO}$ and $L_{\rm ISCO}$ are, respectively, the energy and the angular momentum per unit rest-mass of the gas at the ISCO radius. Assuming an initially non-rotating black hole of mass $M_0$, the solution of Eq.~(\ref{eq-evolution}) is
\be
\hspace{-0.6cm}
a_* =  
\begin{cases}
\sqrt{\frac{2}{3}}
\frac{M_0}{M} \left[4 - \sqrt{18 \, \frac{M_0^2}{M^2} - 2}\right] 
 & \text{if } M \le \sqrt{6} \, M_0 \, , \\
1 & \text{if } M > \sqrt{6} \, M_0 \, .
\end{cases}
\ee
The black hole spin parameter $a_*$ monotonically increases from 0 to 1 and then remains constant. $a_* = 1$ is the equilibrium spin parameter and is reached after the black hole has increased its mass by the factor $\sqrt{6} \approx 2.4$.

If we take into account the fact that the gas in the accretion disk emits radiation and that a fraction of this radiation is captured by the black hole, Eq.~(\ref{eq-evolution}) becomes
\be
\frac{d a_*}{d \ln M} = \frac{c}{r_{\rm g}} 
\frac{L_{\rm ISCO} + \zeta_{\rm L}}{E_{\rm ISCO} + \zeta_{\rm E}} - 2 a_* \, ,
\ee
where $\zeta_{\rm L}$ and $\zeta_{\rm E}$ are related to the amount of photons captured by the black holes and must be computed numerically. Now the equilibrium value of the spin parameter is not 1 but the so-called {\it Thorne limit} $a_*^{\rm Th} \approx 0.998$ (its exact numerical value depends on the emission properties of the gas in the disk)~\cite{r-bh-thorne}. Intuitively, the Thorne limit is lower than 1 because retrograde photons (i.e., those photons with angular momentum antiparallel to the black hole spin) have larger capture cross section and therefore they contribute to reduce the black hole spin.


\section{Accreting black holes in nature: modeling}\label{s2-disk}

Accretion is the process of material spiraling onto a black hole as a consequence of the gravitational pull of the black hole. It is quite commonplace in the Universe and is crucial for various techniques for studying black holes. 
Here we will discuss only the concepts relevant within the context of this book. For a review on the theory of black hole accretion models, see~\cite{r-bh-rev-abram}.

The morphology of the accretion flow is mainly determined by two factors: $i)$ the angular momentum of the accreting gas, and $ii)$ the mass accretion rate. Depending on the value of these two quantities, we have different accretion models and different electromagnetic spectra for the accretion flow. We will now discuss some of the accretion models. 

\begin{figure}[t]
\begin{center}
\includegraphics[width=0.2\textwidth]{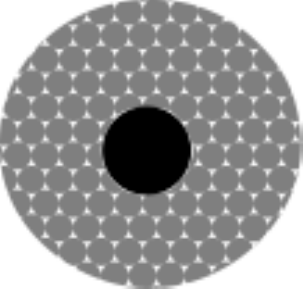}\vspace{0.5cm}\\
\includegraphics[width=0.6\textwidth]{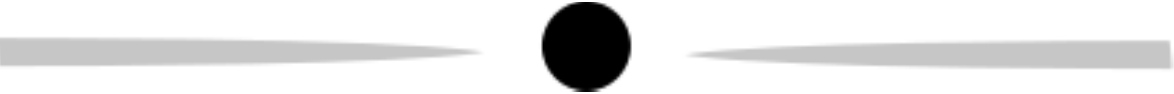} \vspace{0.5cm} \\
\includegraphics[width=0.6\textwidth]{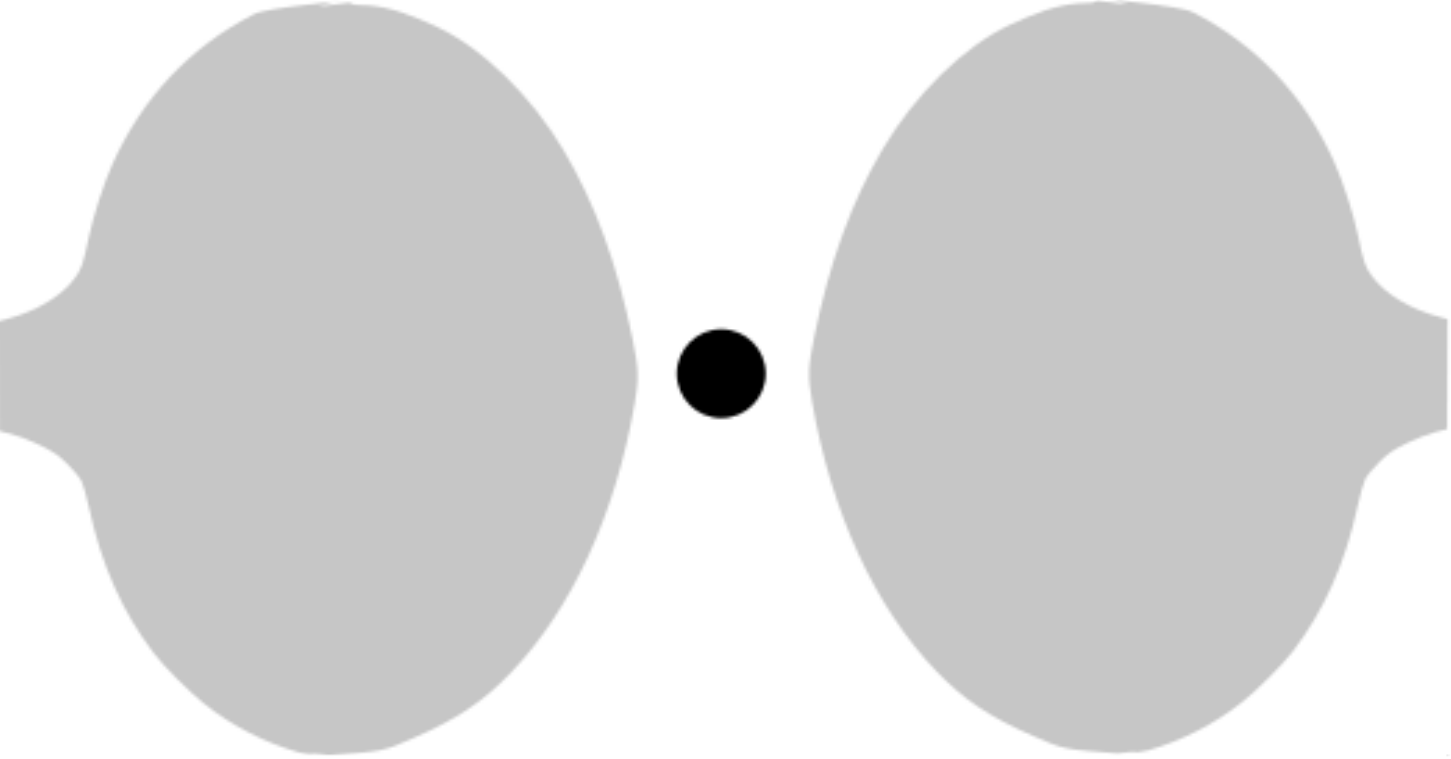} 
\end{center}
\caption{Sketch of Bondi accretion (top), thin disk (middle), and thick disk (bottom). The black hole is indicated by the black filled circle and the accretion flow is represented by the gray shape.
\label{f-disk}}
\end{figure}

In the case of spherically symmetric accretion (vanishing or negligible angular momentum of the accreting gas), we have the so-called {\it Bondi accretion} (Fig.~\ref{f-disk} top sketch), which essentially describes particles in radial free fall. In the Bondi accretion scenario, the radiative efficiency is very low~\cite{r-bh-bondi1,r-bh-bondi2}, i.e. the gravitational energy of the falling particles is mainly converted into their kinetic energy and lost into the black hole after crossing the event horizon.

In typical accretion flows around black holes the gas has a non-negligible angular momentum. In such cases, a disk is created. For stellar-mass black holes with a companion star, the disk is created by the mass transfer from the stellar companion to the black hole. In the case of supermassive black holes in galactic nuclei, the disk forms from the material in the interstellar medium~\cite{r-bh-fuel1} or as a result of galaxy merger~\cite{r-bh-fuel2,r-bh-fuel3}.
An accretion disk is {\it geometrically thin} ({\it thick}) if $h/r \ll 1$ ($h/r \sim 1$), where $h$ is the semi-thickness of the disk at the radial coordinate $r$. The disk is {\it optically thin} ({\it thick}) if $h \ll \lambda$ ($h \gg \lambda$), where $\lambda$ is the photon mean free path in the medium of the disk. If the disk is optically thick, we see the radiation emitted from the surface of the disk, like in the case of stars. The typical way to categorize disks is in terms of their Eddington ratio:
\begin{itemize}
\item {\it Thick disk}~\cite{r-bh-thick-disk1,r-bh-thick-disk2}: If the mass accretion rate is super-Eddington ($\dot{M}/\dot{M}_{\rm Edd} > 1$), the gas pressure makes the disk inflate. The disk is thus geometrically thick (Fig.~\ref{f-disk} bottom sketch). Since the particle density is high, the disk is optically thick and it cannot efficiently radiate away energy from its surface.
\item {\it Slim disk}~\cite{r-bh-slim-disk}: As the mass accretion rate decreases, the thickness of the disk decreases too. A slim disk describes the situation between a thick and a thin disk. For slim disks, roughly, $ 0.3 < \dot{M}/\dot{M}_{\rm Edd}  < 1$.
\item {\it Thin disk}~\cite{r-bh-thin-disk}: For moderate accretion rates ($ 0.05 < \dot{M}/\dot{M}_{\rm Edd}  < 0.3$), the gas pressure is negligible and we have a geometrically thin and optically thick disk (Fig.~\ref{f-disk} middle sketch). The radiative efficiency is high because the disk surface is large enough with respect to the mass accretion rate to radiate away the gravitational energy converted into heat. The standard model for thin disks around black holes is the Novikov-Thorne model~\cite{r-bh-ntm,r-bh-ntm2}, which is described in Section~\ref{ss-nt}.
\item {\it Advection-dominated accretion flow (ADAF)}~\cite{r-bh-adaf-disk-1,r-bh-adaf-disk-2}: If the mass accretion rate is very low ($ \dot{M}/\dot{M}_{\rm Edd}  <  0.05$), the disk evaporates and we have a low density accreting medium. Because of the low density, the medium is optically thin. The interaction rate between particles is low, so there is no efficient cooling mechanism: the gas temperature is very high, the gas is swallowed by the black hole without emitting much radiation, and the accretion luminosity is low. 
\end{itemize}

Tab.~\ref{t-flows} summarizes the accretion scenarios in terms of the specific angular momentum of the accreting gas $L$ (in units of $r_{\rm g} c$), the mass accretion rate $\dot{M}$ (in Eddington units), the geometrical and optical thickness of the disk and the radiative efficiency.

\begin{table}[h]
\centering
\vspace{0.5cm}
{\renewcommand{\arraystretch}{1.5}
\begin{tabular}{|cccccc|}
\hline
$\quad L c / G_{\rm N} M \quad$ & $\quad \dot{M}/\dot{M}_{\rm Edd} \quad$ & $\quad h/r \quad$ & $\quad h/\lambda \quad$ & $\quad \eta_{\rm r} \quad$ & Accretion Model\\
\hline
$\ll 1$ & Any & -- & Any & $\ll 0.1$ & Bondi Accretion \\
$> 1$ & $> 1$ & $\sim 1$ & $\gg 1$ & $\ll 0.1$ & Thick Disk \\
$> 1$ & $0.3 - 1$ & $< 1$ & $\gg 1$ & $< 0.1$ & Slim Disk \\
$> 1$ & $0.05 - 0.3$ & $\ll 1$ & $\gg 1$ & $\sim 0.1$ & Thin Disk \\
$\lesssim 1$ & $\ll 1$ & $\sim 1$ & $\ll 1$ & $\ll 0.1$ & ADAF \\
\hline
\end{tabular}}
\vspace{0.5cm}
\caption{Summary of the main scenarios of accretion processes around black holes and of their basic properties. The first column indicates the angular momentum of the accreting gas; the second column is for the mass accretion rate (in Eddington units); the third column indicates if the accretion disk is geometrically thick or thin; the fourth column shows if the accretion flow is optically thick or thin; the fifth column is for the radiative efficiency; the last column indicates the name of the accretion model. See the text for more details. \label{t-flows}}
\end{table}

\subsection{Novikov-Thorne disks \label{ss-nt}}

The Novikov-Thorne model is the standard framework for the description of geometrically thin and optically thick accretion disks around black holes. The main assumptions of the model are:
\begin{enumerate}
\item The accretion disk is geometrically thin ($h/r \ll 1$).
\item The accretion disk is perpendicular to the black hole spin.
\item The inner edge of the disk is at the ISCO.
\item The motion of the particle gas in the disk is determined by the gravitational field of the black hole, while the impact of the gas pressure is ignored. 
\end{enumerate}
For the full list of assumptions and a detailed discussion, see e.g.,~\cite{r-bh-book,r-bh-ntm2} and references therein. Here we point out a few important considerations.

Assumption~1 just points out that the Novikov-Thorne model is strictly applicable only to thin disks, even though it is often used for all black hole X-ray sources. This is in part because it is the only simple analytic model in the market and in part because it is often difficult to estimate the Eddington-scaled mass accretion rate (in particular for AGNs, where mass and distance are usually poorly constrained).

Assumption~2's validity depends on the origin and the evolution of the system. The cases of stellar-mass black holes and of supermassive black holes are somewhat different. Let us begin with the former. If the stellar-mass black hole is the final product of the supernova explosion of a heavy star in a binary, its natal spin should be along the same direction as the progenitor star's spin. Assuming a symmetric explosion without strong shocks and kicks, this would mean a spin axis orthogonal to the orbital plane of the binary~\cite{r-bh-thin-fragos2010}. A misalignment may be introduced by a non-symmetric supernova explosion and/or shocks and kicks, as well as in those systems formed through multi-body interactions (binary capture or replacement), where the orientation of the spin of the black hole and that of the orbital angular momentum of the binary are initially uncorrelated. 

Regardless of the natal spin and binary capture scenarios, at least the inner part of the disk -- which plays the most important role in spin measurements -- is expected to be equatorial, as a result of the Bardeen-Petterson effect~\cite{r-bh-thin-bp,r-bh-thin-bp2}\footnote{\emph{Bardeen-Petterson configuration} refers to a system in which the inner part of the disk is flat and perpendicular to the black hole spin, while the outer part is also flat but in the plane perpendicular to the angular momentum vector of the binary.}. The alignment timescale of thin disks has been estimated to be in the range $10^6$-$10^8$~yrs, and therefore the disk should be already adjusted in the black hole equatorial plane for not too young systems~\cite{r-bh-thin-6849} (but see~\cite{r-bh-thin-rm07m,r-bh-thin-rm08m} for more details). However, the actual timescale depends on parameters like the viscosity $\alpha$ which are usually not known~\cite{r-bh-thin-r_w1,r-bh-thin-r_w2}. Moreover, some numerical simulations do not observe the adjustment of the alignment of the disk~\cite{r-bh-thin-fragile1,r-bh-thin-fragile2}. Additionally, if the inner part of the disk is a hot, geometrically thick accretion flow, the inner disk precesses as a solid body about the black hole angular momentum axis~\cite{r-bh-thin-ingram}. Future X-ray spectropolarimetric measurements of the thermal spectrum of accretion disks will be able to check the validity of the assumption that the disk is in the equatorial plane (see, for instance, \cite{r-bh-book} and references therein).

In the case of supermassive black holes, the orientation of the accretion disk with respect to the black hole spin is expected to change during the evolution of the system, in particular because of galaxy mergers. However, in the absence of galactic mergers (or if enough time has passed since the previous merger), again the Bardeen-Petterson effect will make the inner part of the disk orthogonal to the black hole spin.

\begin{figure}[t]
\vspace{0.5cm}
\begin{center}
\includegraphics[width=9.0cm]{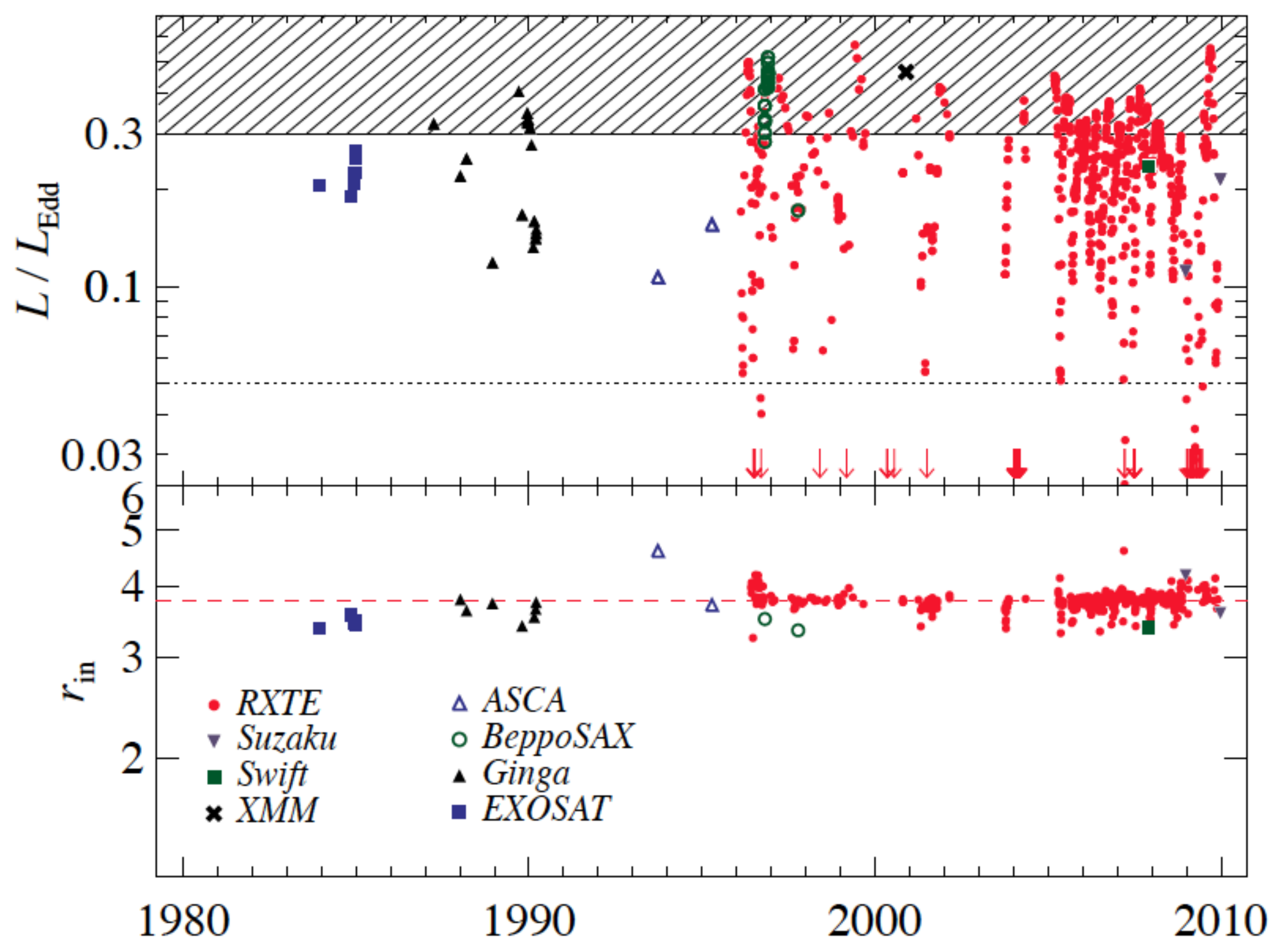}
\end{center}
\vspace{0.3cm}
\caption{Top panel: Accretion disk luminosity in Eddington units versus time for 766 spectra of LMC~X-3. The shaded region does not satisfy the thin disk selection criterion $L/L_{\rm Edd} < 0.3$, as well as the data below the dotted line, which marks $L/L_{\rm Edd} = 0.05$. Bottom panel: fitted value of the inner disk radius of the 411 spectra in the top panel that can meet the thin disk selection criterion. See the text for more details. From~\cite{r-bh-thin-constant}. \copyright AAS. Reproduced with permission.}
\label{f-steiner2010}
\end{figure}

Assumption 3 is crucial for doing spin measurements. This is because, assuming the Kerr metric, there is a one-to-one correspondence between the spin parameter $a_*$ and the ISCO radius. A measurement of the ISCO via the inner edge thus provides direct information about the spin of the black hole. 

Observations show that the inner edge of the disk does not change appreciably over several years when the source is in the soft state (see Section~\ref{s2-spectra} for the definition of soft state) with a luminosity between $\sim$5\% to $\sim$30\% of its Eddington limit. The most compelling evidence comes from LMC~X-3. The analysis of many spectra collected during eight X-ray missions and spanning 26~years shows that the radius of the inner edge of the disk is nearly constant~\cite{r-bh-thin-constant}, see Fig.~\ref{f-steiner2010}. The most natural interpretation is that the inner edge is associated to some intrinsic property of the geometry of the spacetime, namely the radius of the ISCO, and is not affected by variable phenomena like the accretion process.

Assumption 4 requires that the radial acceleration of the gas due to pressure gradients is negligible in comparison with the gravitational acceleration due to the black hole. This requires that, as the gas falls onto the black hole, its potential energy is transported away or radiated away, and only a negligible part is converted to internal energy of the gas~\cite{r-bh-ntm2}. This assumption holds in general for radiatively efficient accretion flows.

The accretion process in the Novikov-Thorne model can be summarized as follows. The particles of the accreting gas slowly fall onto the central black hole. When they reach the ISCO radius, they quickly plunge onto the black hole without emitting additional radiation. The total power of the accretion process is $L = \eta \dot{M} c^2$, where $\eta = \eta_{\rm r} + \eta_{\rm k}$ is the total efficiency, $\eta_{\rm r}$ is the radiative efficiency, and $\eta_{\rm k}$ is the fraction of gravitational energy converted to kinetic energy of jets/outflows. The Novikov-Thorne model assumes that $\eta_{\rm k}$ can be ignored, and therefore the radiative efficiency of a Novikov-Thorne accretion disk is
\be\label{eq-thin-ntradeff}
\eta_{\rm NT} = 1 - E_{\rm ISCO} \, ,
\ee 
where $E_{\rm ISCO}$ is the energy per unit rest-mass of the gas at the ISCO radius.

In spin measurements, it is clearly very important to select the observations and the sources in which the disk is geometrically thin and its inner edge is at the ISCO radius. In the case of the continuum-fitting method, one usually selects sources in the soft state with a strong contribution from the thermal disk emission. The luminosity of the source should be between $\sim$5\% and $\sim$30\% of the Eddington limit~\cite{r-bh-cfm-1915}. At lower luminosities, the disk may be truncated. In such a case, the inner edge of the disk would be at a radius larger than the ISCO and between the inner edge of the disk and the black hole there is probably an accretion flow that can be described by ADAF. For higher accretion rates, the gas pressure becomes more important, and the disk is no longer thin. In such a case, the inner edge of the disk might be at a radius slightly smaller than the ISCO.

\subsection{Disk-corona model}

\begin{figure}[b]
\begin{center}
\includegraphics[width=0.95\textwidth]{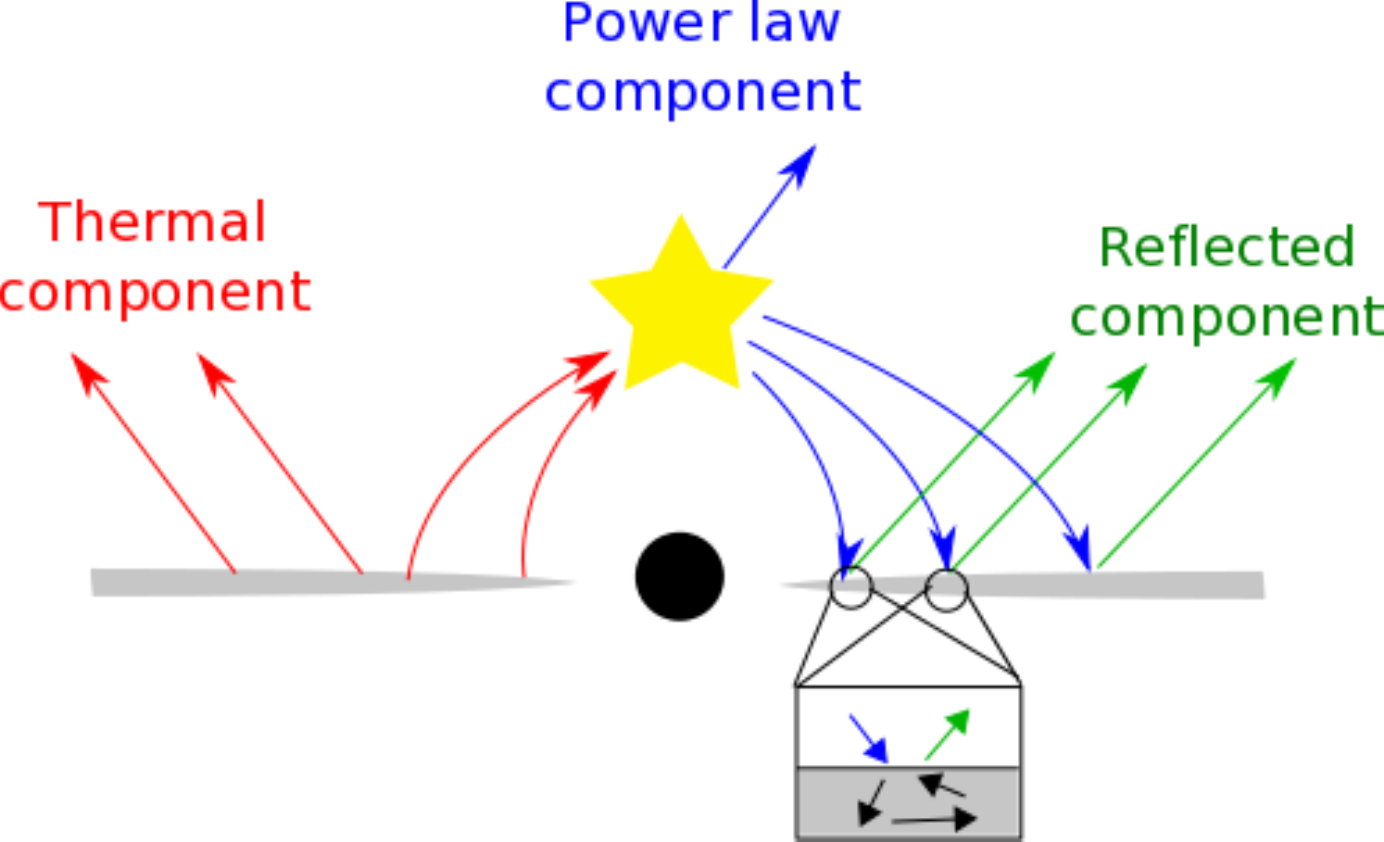}
\end{center}
\caption{Disk-corona model. The black hole is surrounded by a thin accretion disk with a multi-color blackbody spectrum (red arrows). Some thermal photons from the disk have inverse Compton scattering off free electrons in the corona, producing a power-law component (blue arrows). The latter also illuminates the disk, generating a reflection component (green arrows). \label{f-corona}}
\end{figure}

Observations have led us to theorize, in addition to a black hole with a disk, a corona around the black hole. An example of the disk-corona model is schematically illustrated in Fig.~\ref{f-corona}. The black hole accretes from a geometrically thin and optically thick disk. The disk emits as a blackbody locally and as a multi-color blackbody when integrated radially ({\it thermal component} indicated by the red arrows in Fig.~\ref{f-corona})~\cite{r-bh-multicolor}. The \emph{multi-color} feature comes from the fact that different parts of the disk have different temperatures. For a given radius of the disk, the temperature depends on the black hole mass and the mass accretion rate. Most of the radiation is emitted near the inner edge of the disk and is in the soft X-ray band (0.1-1~keV) for stellar-mass black holes and in the optical/UV band (1-10~eV) for supermassive black holes.

\begin{figure}[t]
\begin{center}
\subfloat[Lamppost corona]{\includegraphics[width=0.45\textwidth]{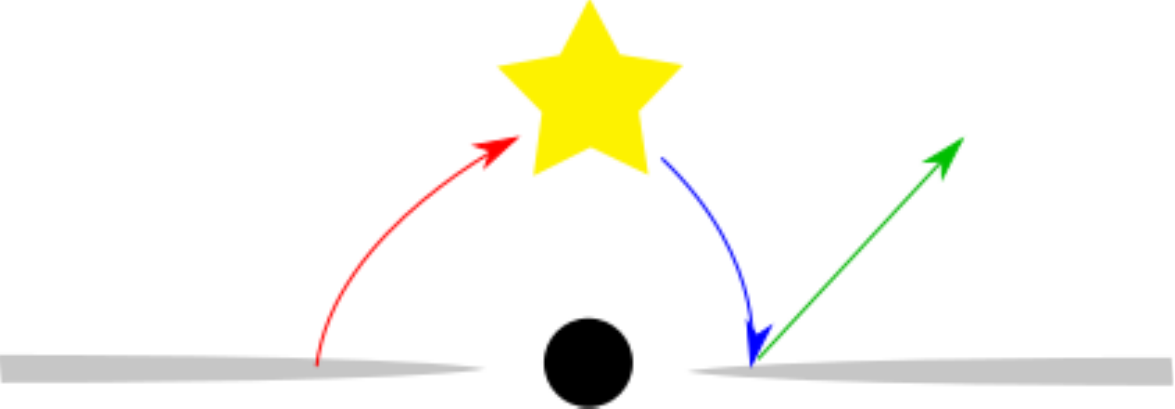}} \hfill
\subfloat[Sandwich corona]{\includegraphics[width=0.45\textwidth]{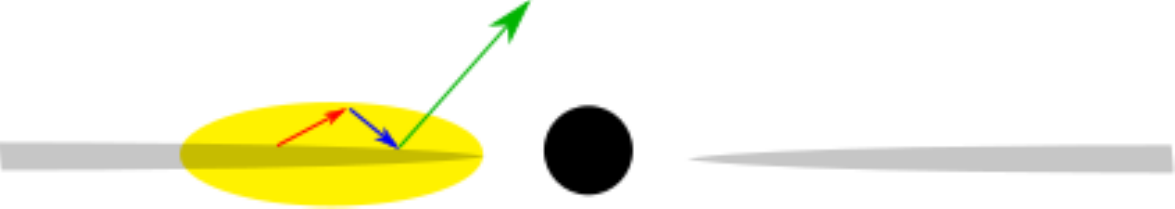}} \\
\subfloat[Spherical corona]{\includegraphics[width=0.45\textwidth]{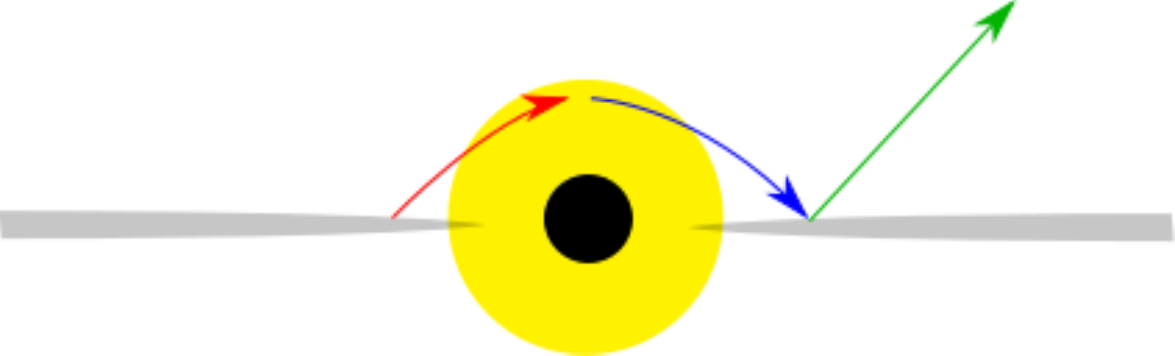}} \hfill
\subfloat[Toroidal corona]{\includegraphics[width=0.45\textwidth]{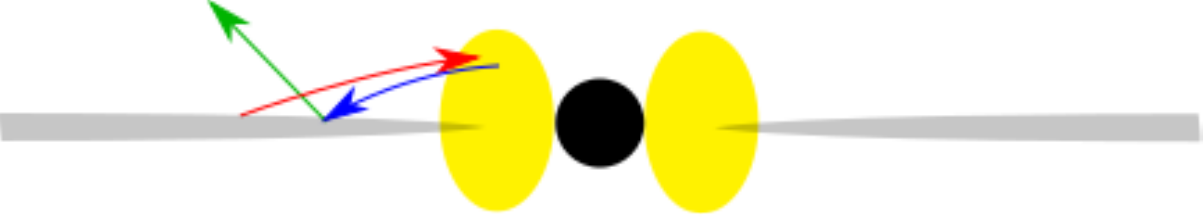}} 
\end{center}
\caption{Examples of possible corona geometries: lamppost geometry (top left panel), sandwich geometry (top right panel), spherical geometry (bottom left panel), and toroidal geometry (bottom right panel). \label{f-geocor} }
\end{figure}
\begin{figure}[h]
\begin{center}
\includegraphics[width=10.0cm]{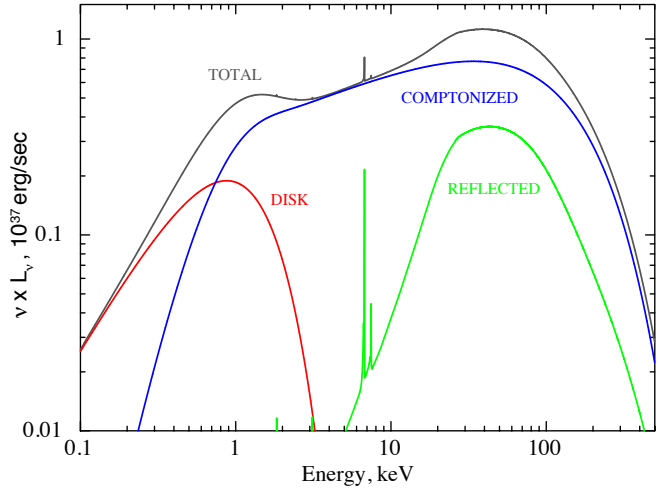}
\end{center}
\caption{Spectral components of an accreting black hole in the disk-corona model: disk's thermal spectrum (red), power-law component from the corona (blue), and reflection component from the illumination of the disk by the power-law component (green).}
\label{f-spectrum}
\end{figure}

The {\it corona} is a hotter ($\sim 100$~keV), usually compact and optically thin, cloud close to the black hole (the yellow region in Fig.~\ref{f-corona}), but its exact geometry is currently unknown~\cite{r-bh-corona--1,r-bh-corona--2,r-bh-corona--3}. Fig.~\ref{f-geocor} shows some coronal geometries proposed in literature. The lamppost corona is a point-like source along the spin axis of the black hole~\cite{r-bh-lamppost}. Such a possibility may be realized in the case the corona is the base of the jet. In the sandwich geometry, the corona would be the atmosphere above the accretion disk~\cite{r-bh-sandwich}. In the cases of spherical or toroidal geometries, the corona would be the accretion flow between the inner edge of the disk and the black hole. In all cases, inverse Compton scattering of the thermal photons from the accretion disk off free electrons in the corona produces a {\it power-law component} (or {\it Comptonized component}, blue arrows in Fig.~\ref{f-corona}) with a cut-off energy that depends on the temperature of the corona ($E_{\rm cut} \sim 100$-1000~keV)~\cite{r-bh-compton1,r-bh-compton2}.

The power-law component from the corona back-illuminates the accretion disk, producing a {\it reflection component} (green arrows in Fig.~\ref{f-corona}) with some fluorescent emission lines~\cite{r-bh-ref}. The strongest feature of the reflection component is usually the iron K$\alpha$ line, which is at 6.4~keV in the case of neutral or weakly ionized iron and shifts up to 6.97~keV for H-like iron ions, and the Compton hump at 10-30~keV. Fig.~\ref{f-spectrum} shows the resulting spectrum of the disk-corona geometry: we have the thermal component (in red) from the accretion disk, the power-law component (in blue) from the corona, and the reflection component (in green) from the illumination of the accretion disk by the power-law component. In the presence of jets, although the radiation from the jet is mostly in the radio/IR bands due to synchrotron radiation by accelerating particles, it may extend to the X-ray and $\gamma$-ray bands and then there is an additional jet component in the X-ray spectrum (see Sec.~\ref{app-2} for more details on jets).


\section{Accreting black holes in nature: classification}
An accreting black hole can be found in different ``spectral states'', which are characterized by the luminosity of the source and by the relative contribution of its spectral components (thermal, power-law, reflection)~\cite{r-bh-spst,r-bh-spst2}. Although the spectral state classification is purely phenomenological, i.e. based on the observed X-ray spectrum, we expect a correlation (not completely understood as of now) to exist between spectral states and accretion flow configurations. The spectral classification of accreting black holes is different for stellar-mass and supermassive black holes. 

\subsection{Stellar-mass black holes}\label{s2-spectra}
\begin{figure}[t]
\begin{center}
\includegraphics[width=8.7cm]{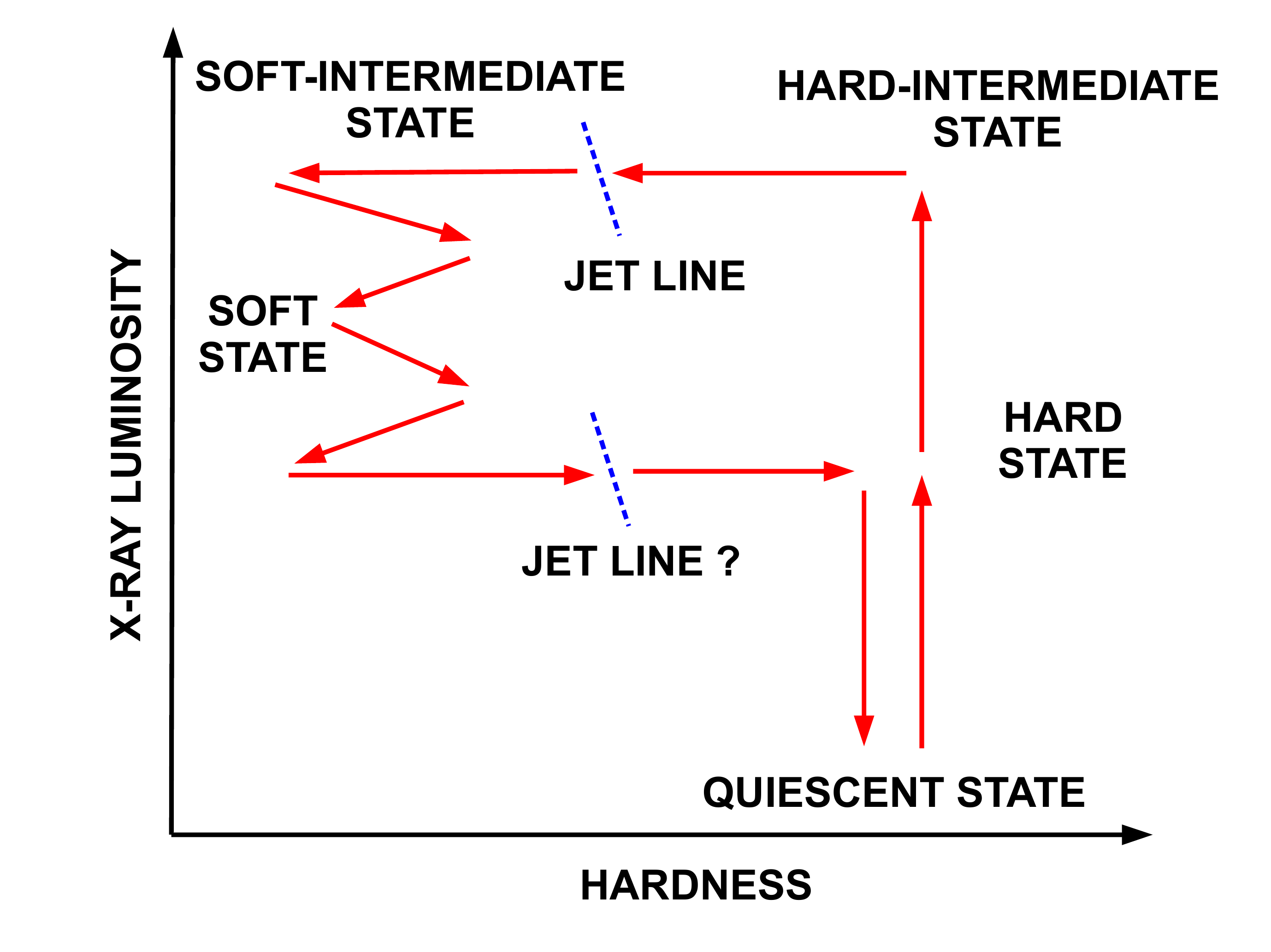}
\end{center}
\vspace{-0.5cm}
\caption{Evolution of the prototype of an outburst in the hardness-intensity diagram. The source is initially in a quiescent state. At the beginning of the outburst, the source enters the hard state, then moves to some intermediate states, to the soft state, and eventually returns to a quiescent state. See the text for more details. \label{f-hid}}
\end{figure}

Let us begin with the case of a stellar-mass black hole in an X-ray transient. The object typically spends most of the time in a {\it quiescent state} with a very low accretion luminosity ($L/L_{\rm Edd} < 10^{-6}$). At a certain point, the source has an {\it outburst} and becomes a bright X-ray source in the sky ($L/L_{\rm Edd} \sim 10^{-3}$--1). The quiescent state is determined by a very low mass accretion rate, namely a very low amount of material transfers from the companion star to the black hole. When there is a sudden increase of the mass accretion rate (for instance, the companion star inflates and the black hole strips material from the surface of the companion), the outburst happens. The object may be in a quiescent state for several months or even decades. An outburst typically lasts from some days to a few months (roughly the time that the black hole takes to swallow the material that produced the outburst). During an outburst, the spectrum of the source changes.
The {\it hardness-intensity diagram} (HID)~\cite{r-bh-spst,r-bh-spst2} is a useful tool for describing the outburst. Fig.~\ref{f-hid} illustrates the typical life-cycle of an outburst. The $x$-axis is the source hardness, which is the ratio between its luminosity in the hard and soft X-ray bands (e.g., between the luminosity in the 6-10 and 2-6~keV bands). The $y$-axis can be any measure of the intensity of the source, e.g., the X-ray luminosity or the number of counts on the instrument. The specific hardness-intensity diagram depends on the source (e.g. the interstellar absorption) and on the instrument (e.g. its effective area at different energies), but, qualitatively it turns out to be very useful for studying transient sources.

The relation between spectral states and accretion flow can be understood noting that the intensity of the thermal component is mainly determined by the mass accretion rate and the position of the inner edge of the accretion disk, while the contributions of the power-law and reflection components depend on the properties of the corona (its location, extension, geometry, etc.). In particular, the local flux of the disk's thermal component is approximately proportional to the mass accretion rate and the inverse of the cube of the disk's radius, $\mathcal{F}(r) \propto \dot{M}/r^3$~\cite{r-bh-lasota4}. When the mass accretion rate is low and the inner edge of the disk is at large radii, the thermal component is weak, whereas, when the mass accretion rate is high and the the disk inner edge is close, the thermal component is strong. Similarly, the power-law and the reflection components are strong when the corona is large and close to the disk, whereas they are weak when the corona is small and far away from the disk. The relative contribution of all these components depends on the material around the black hole, and determines the spectral state.

{\it Quiescent state ---} The source is initially in a quiescent state: the mass accretion rate and the luminosity are very low (the source may in fact be too faint to be detected) and the spectrum is hard. The inner edge of the accretion disk is truncated at a radius significantly larger than the ISCO and the accretion process around the black hole is described by ADAF. In this phase, the low-density accretion flow close to the black hole may act as the corona, which would thus be spherical and large.

{\it Hard state {\rm (or corona-dominated state)} ---} At the beginning of the outburst, the spectrum is hard and the source becomes brighter and brighter because the mass accretion rate increases ($L/L_{\rm Edd}$ starts from $\sim 10^{-3}$ and can reach values up to $\sim 0.7$ in some cases). The spectrum is dominated by the power-law  and reflection components. The thermal component is subdominant, and the temperature of the inner part of the disk may be low ($\sim$0.1~keV or even lower), but it increases as the luminosity of the source rises. The inner edge of the disk is initially farther away than the ISCO, but it moves to the ISCO as the luminosity increases, and it may reach the ISCO at the end of the hard state. During the hard state, compact mildly relativistic steady jets are common, but the exact mechanism producing these jets is currently unknown. Observations point out a compact corona~\cite{r-bh-corona--2,r-bh-corona--3}, which may be the base of the jet in a lamppost geometry.

{\it Intermediate states ---} 
The mass accretion rate rises, so the contribution of the thermal component increases. The power-law and the reflection components get weaker, probably because of a variation in the geometry/properties of the corona. As a consequence, the source moves to the left part of the HID. We first have the {\it hard-intermediate state} and then the {\it soft-intermediate state}. During the transition, transient highly relativistic jets are observed, which is denoted by the {\it jet line} in Fig.~\ref{f-hid}. If the hardness of the source oscillates near the jet line, we can observe several transient jets.

{\it Soft state {\rm (or disk-dominated state)} ---} The thermal spectrum of the disk is the dominant component in the spectrum and the inner part of the disk temperature is around 1~keV. If the luminosity of the source is between $\sim$5\% to $\sim$30\% of its Eddington luminosity, the disk inner edge is at the ISCO~\cite{r-bh-cfm-1915}, and the accretion disk should be well described by the Novikov-Thorne model~\cite{r-bh-edd1a,r-bh-edd1b}. In the soft state, we do not observe any kind of jet\footnote{For instance, in the corona lamppost geometry, the corona may be the base of the jet. This could explain why, in the soft state, we do not see jets and the power-law and reflection components are weak.}. However, strong winds and outflows are common (while they are absent in the hard state). The luminosity of the source may somewhat decrease and the hardness may change, while remaining on the left side of the HID.

At a certain point, the transfer of material decreases, leading to the end of the outburst. The contribution of the thermal spectrum of the disk decreases and, as a consequence, the hardness of the source increases. The source re-enters the soft-intermediate state, the hard-intermediate state, then the hard state, and eventually, when the hardness is high, the luminosity drops down and the source returns to the quiescent state till the next outburst. Between the soft-intermediate and the hard-intermediate states, we may observe transient jets, but the existence of a jet line is not clear here. Every source follows the path shown in Fig.~\ref{f-hid} counter-clockwise, but there are differences among different sources and even for the same source among different outbursts.

In the case of stellar-mass black holes in persistent X-ray sources, there is no outburst, but we can still use the HID. The most studied source is Cygnus~X-1 (the other persistent sources are in nearby galaxies, so they are fainter and more difficult to study). This object spends most of the time in the hard state, but it occasionally moves to a softer state, which is usually interpreted as a soft state. LMC X-1 is always in the soft state. LMC X-3 is usually observed in the soft state, rarely in the hard state, and there is no clear evidence that this source can be in an intermediate state.


\subsection{Supermassive black holes}
In the case of supermassive black holes, there are at least two important differences. First, because the size of the system scales as the mass (e.g., 1~day for a 10~$M_\odot$ black hole corresponds to 3,000~years for a $10^7$~$M_\odot$ black hole) the study of the evolution of a specific system is rendered impossible on human timescales. Second, the temperature of the disk is in the optical/UV range for a supermassive black hole, as compared to stellar-mass black holes where it is the X-ray band. It is possible though to employ the same spectral state classification as above for supermassive black holes (see, for instance, \cite{r-bh-spst} and references therein). Here, we will classify supermassive black holes according to their luminosity and spectral features.

Active galactic nuclei (AGNs) are very bright galactic nuclei, powered by the mass accretion onto their central supermassive black hole. The term AGN is usually used to indicate the same supermassive black hole as well. Fig.~\ref{f-galaxies} shows the AGN classification, groups and subgroups, and the corresponding fraction of members. While it is thought that most galaxies have a supermassive black hole at their center, only a small fraction of them host an AGN. In most galaxies, the central supermassive object is ``dormant'', like the supermassive black hole in our Galaxy, Sgr~A$^*$, which has a luminosity of the order of $10^{-7}$ in Eddington units. 

\begin{figure}[h]
\dirtree{%
.1 \textbf{Galaxies}.
.2 Non-active galaxies [93\%].
.2 Active galaxies [7\%].
.3 LINERs [15\%].
.3 Star-forming galaxies [85\%].
.3 AGNs [0.5\%].
.4 Radio-loud AGNs [10\%].
.5 Radio galaxies.
.5 Radio-loud quasars.
.5 Blazars.
.6 BL Lac objects.
.6 OOV quasars.
.4 Radio-quiet AGNs [90\%].
.5 Seyfert-ELFs.
.5 Radio-quiet quasars.
.5 NELGs.
.5 Seyfert galaxies [95\%].
.6 Seyfert 1s [30\%].
.6 Seyfert 1.5s, 1.8s, 1.9s [10\%].
.6 Seyfert 2s [60\%].
}
\caption{Sketch of the AGN family and of its subgroups. This classification has to be taken with caution, because different authors may use slightly different classifications. The diagram shows also the fraction of members in each subgroup. AGNs represent only 0.035\% of the galactic nuclei. Most of the AGNs are radio-quiet and belong to the class of Seyfert galaxies.\label{f-galaxies}}
\end{figure}

\begin{figure}[h]
\begin{center}
\includegraphics[width=0.99\textwidth]{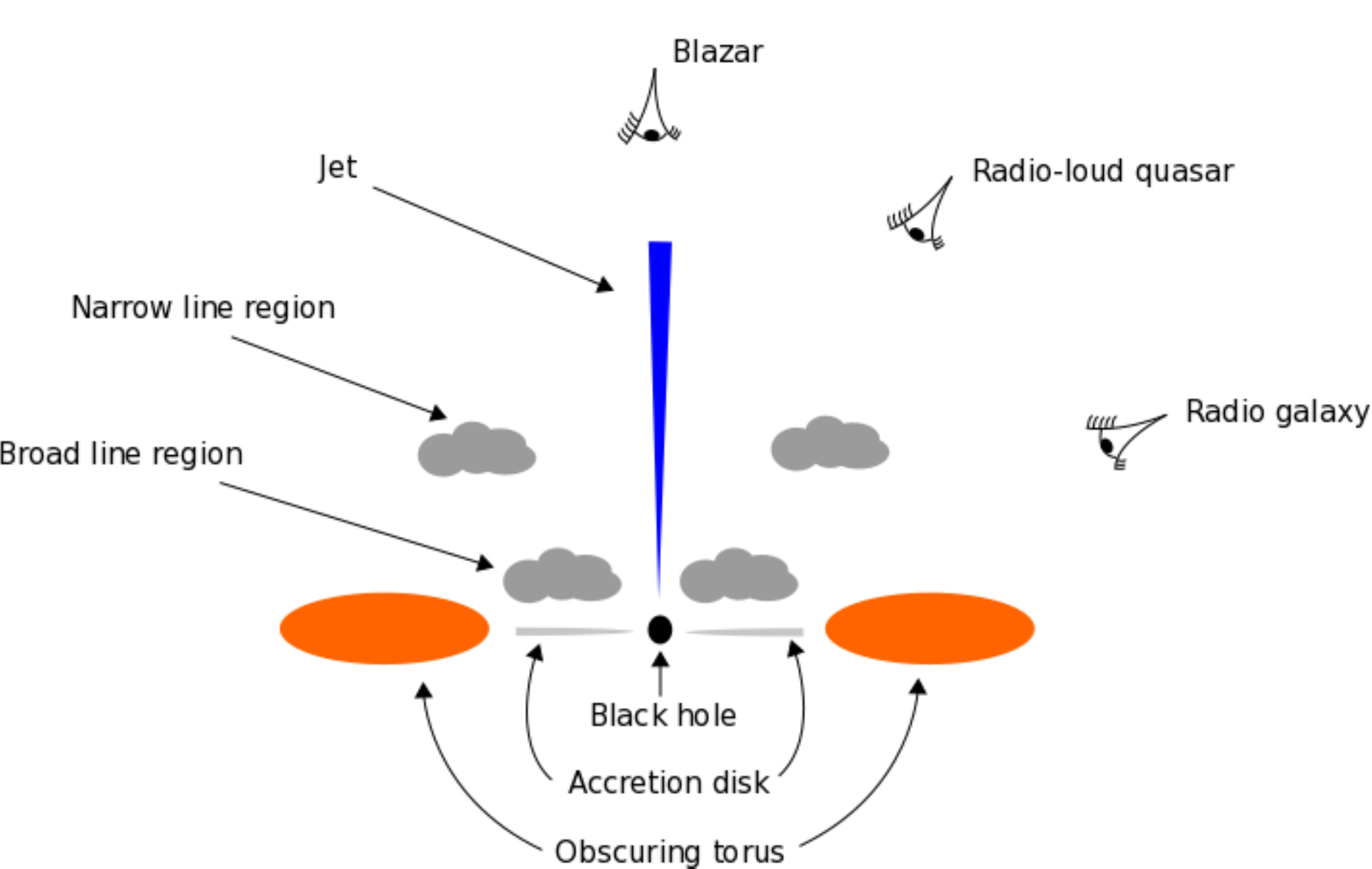}
\end{center}
\caption{Sketch of a radio-loud AGN according to the unified AGN model~\cite{r-bh-Urry:1995mg}. The black hole is surrounded by an accretion disk, which may be obscured by a dusty torus. The broad line region is close to the black hole and there are clouds orbiting with high velocity. The narrow line region is relatively far from the black hole and there are clouds moving at lower velocity. Depending on the angle between the jet and the line of sight of the observer, the AGN can appear as a blazar, as a radio-loud quasar, or as a radio galaxy, as shown in the picture.}
\label{f-agn1}
\end{figure}
\begin{figure}[h]
\begin{center}
\includegraphics[width=0.99\textwidth]{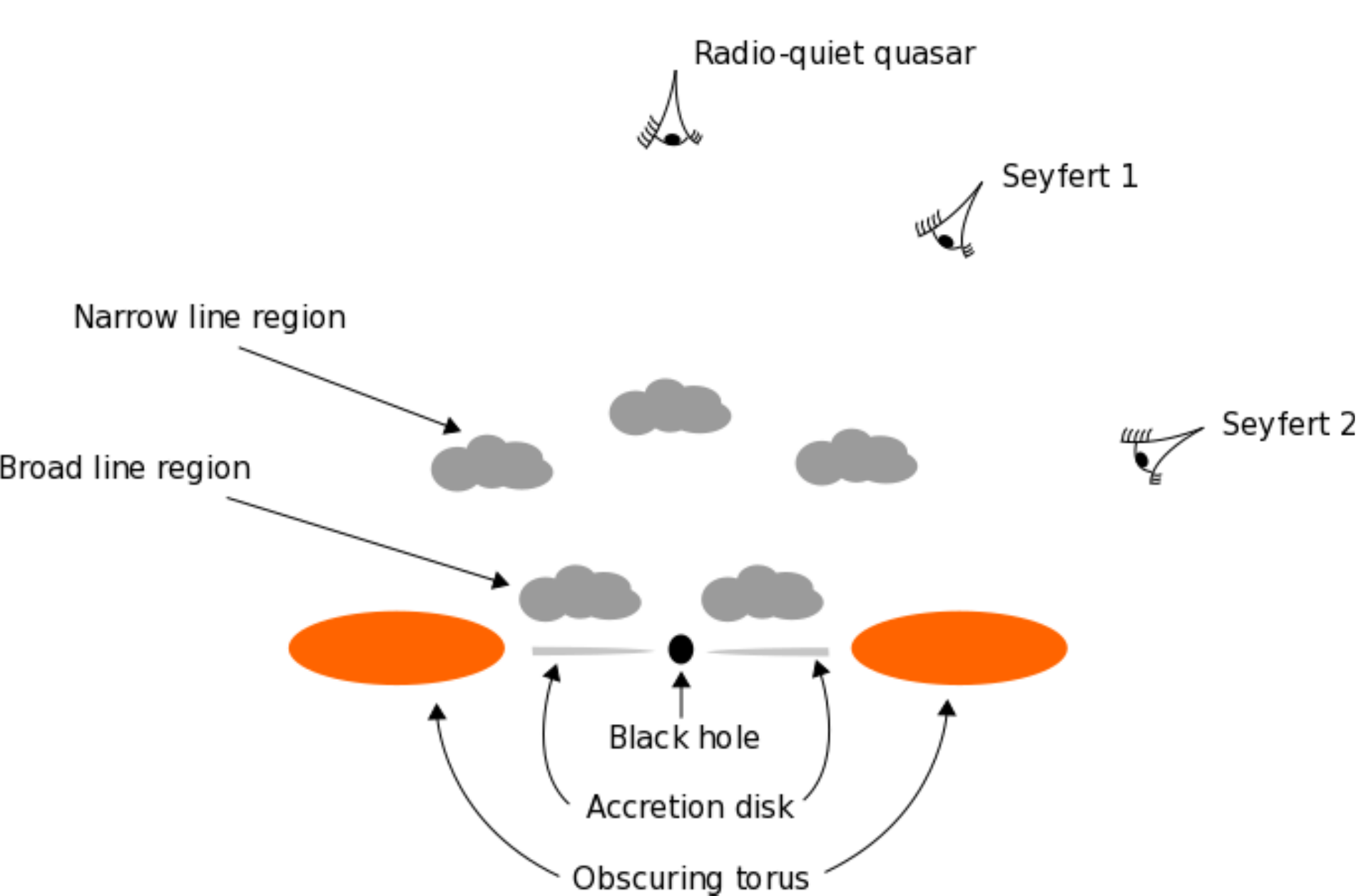}
\end{center}
\caption{As in Fig.~\ref{f-agn1}, in the case of a radio-quiet AGN. Depending on the viewing angle of the observer, the AGN can appear as a radio-quiet quasar, as a Seyfert~1 galaxy, or as a Seyfert~2 galaxy.}
\label{f-agn2}
\end{figure}

About 93\% of the galaxies are non-active. Among the 7\% of the active galaxies, most of them are star-forming galaxies or low-ionization nuclear emission-line regions (LINERs). The latter are sometimes considered AGNs. Proper AGNs are relatively rare: they are in 0.5\% of the active galaxies, which means only in 0.035\% of all galaxies.

AGNs are mainly classified according to their luminosity and spectral features. It is thus useful to briefly review their possible spectral components:
\begin{enumerate}
\item {\it Radio emission} from jets with the typical spectrum from synchrotron radiation.
\item {\it IR emission} from the thermal spectrum of the accretion disk, which is reprocessed by gas and dust around the nucleus. This occurs when the accretion disk is obscured by gas and dust. 
\item {\it Optical continuum} mainly from the thermal spectrum of the accretion disk, and in part from possible jets.
\item {\it Narrow optical lines} from cold material orbiting relatively far from the supermassive black hole. The orbital velocity of this material is 500-1,000~km/s.
\item {\it Broad optical lines} from cold material orbiting close to the supermassive black hole. The orbital velocity of this material is 1,000-5,000~km/s. The lines are broad due to Doppler boosting.
\item {\it X-ray continuum} from a hot corona and possible jets. 
\item {\it X-ray lines} from fluorescence emission of the gas in the accretion disk illuminated by the X-ray continuum. The iron K$\alpha$ line at 6.4~keV is usually one of the most prominent lines.
\end{enumerate}

The AGN classification is sometimes confusing, some objects may not be easily associated to a specific group, and different authors may use different classifications. With reference to Fig.~\ref{f-galaxies}, we see that AGNs can be grouped into two categories, radio-quiet and radio-loud AGNs. In the radio-quiet AGN category, the jet component is absent or negligible, so the radio luminosity is low. Radio-quiet AGNs may be grouped into four classes: Seyfert extremely luminous far infrared galaxies (Seyfert-ELFs), Seyfert galaxies, narrow emission line galaxies (NELGs), and radio-quiet quasars. The classification is based on a number of properties. For instance, Seyfert galaxies have an optical continuum and emission lines. Seyfert 1s have both narrow and broad emission lines, while Seyfert 2s have only narrow emission lines. Seyfert galaxies of type 1.5, 1.8, and 1.9 are grouped according to their spectral appearance. The radio-loud AGN category has powerful jets, which may be powered by the black hole spin. Radio-loud AGNs can be grouped into three classes: radio galaxies, radio-loud quasars, and blazars. Blazars are characterized by rapid variability and by polarized optical, radio and X-ray emission. They are divided into BL Lacertae objects (BL Lac objects) and optically violent variable quasars (OVV quasars). OVV quasars have stronger broad emission lines than BL Lac objects.

According to the unified AGN model~\cite{r-bh-Urry:1995mg}, all AGNs are essentially the same kind of objects. The difference appears because they are observed from different viewing angles. Figs.~\ref{f-agn1} and \ref{f-agn2} illustrate the idea of the unified AGN model. In the former figure, depending on the viewing angle, we observe a blazar, a radio-loud quasar and then a radio galaxy. In the latter figure, we have a similar situation. Depending on the viewing angle of the observer, the AGN can appear as a radio-quiet quasar, as a Seyfert~1 galaxy, or as a Seyfert~2 galaxy.


\section{Accreting black holes in nature: observational techniques}\label{s2-spin}

Any astrophysical black holes should be completely characterized by its mass $M$ and its spin parameter $a_*$. It is relatively easy to measure the mass of a black hole, by studying the orbital motion of gas or of individual stars around the compact object. Spin measurements are much more challenging. The spin has no ``gravitational effects'' in Newtonian gravity. This is not the case in general relativity, and the spin alters the gravitational field around a massive body. However, any spin effect is strongly suppressed at larger radii, so black hole spin measurements require to probe the strong gravity region close to the black hole event horizon. As of now, there are two leading techniques to measure black hole spins by studying the X-ray radiation emitted by the gas in the inner part of the accretion disk: the so-called {\it continuum-fitting method}, which consists in the analysis of the thermal spectrum of thin accretion disks and is usually applicable to stellar-mass black holes only, and {\it X-ray reflection spectroscopy} (or {\it iron line method}), which is based on the study of the disk's reflection spectrum, can be applied to both stellar-mass and supermassive black holes, and is currently the only available method to measure the spins of supermassive black holes. There are a few other proposed techniques for measuring black hole spins with electromagnetic radiation. Among these, the most promising one is probably the detection of {\it quasi-periodic oscillations} (QPOs).

\subsection{Continuum-fitting method}

\begin{figure}[t]
\begin{center}
\includegraphics[trim=10mm 0mm 10mm 0mm,width=5.3cm]{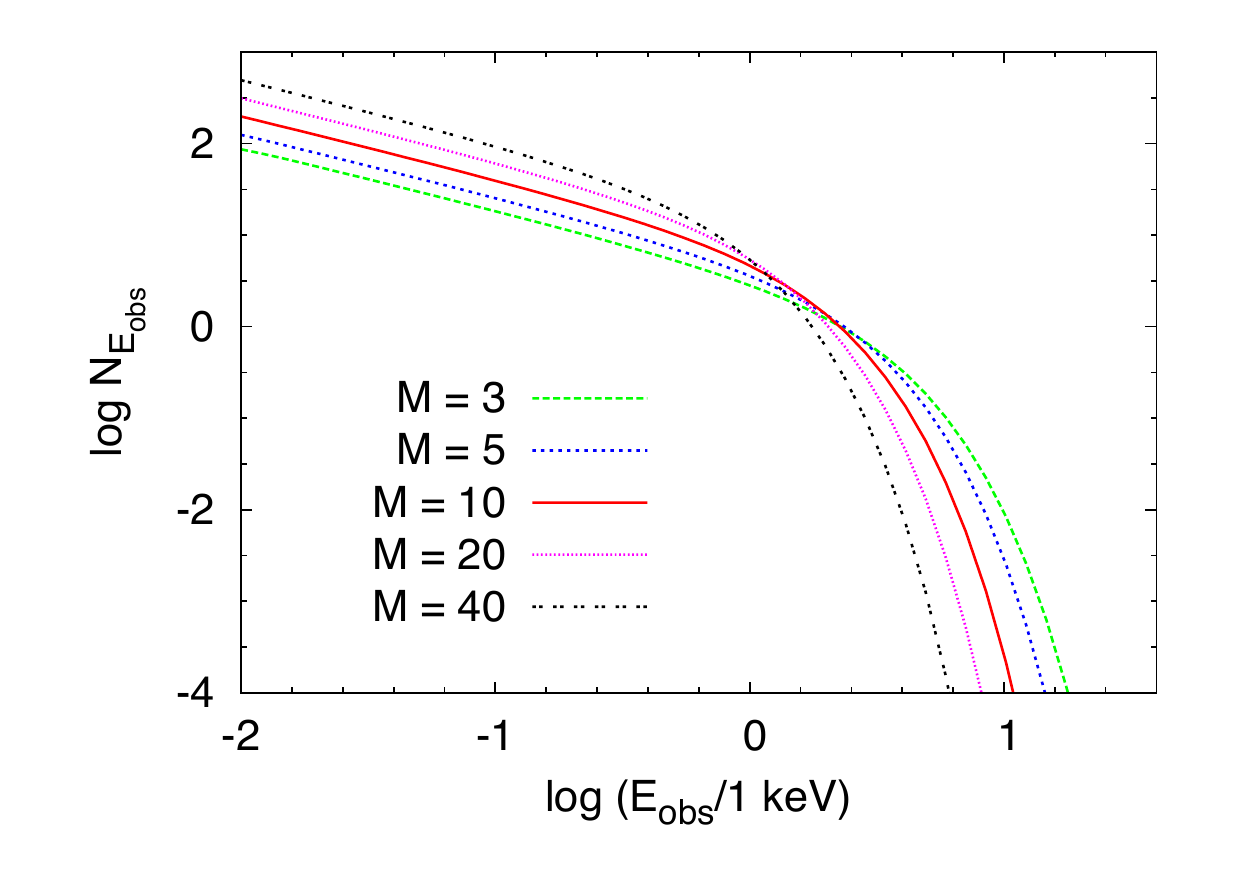}
\hspace{0.5cm}
\includegraphics[trim=10mm 0mm 10mm 0mm,width=5.3cm]{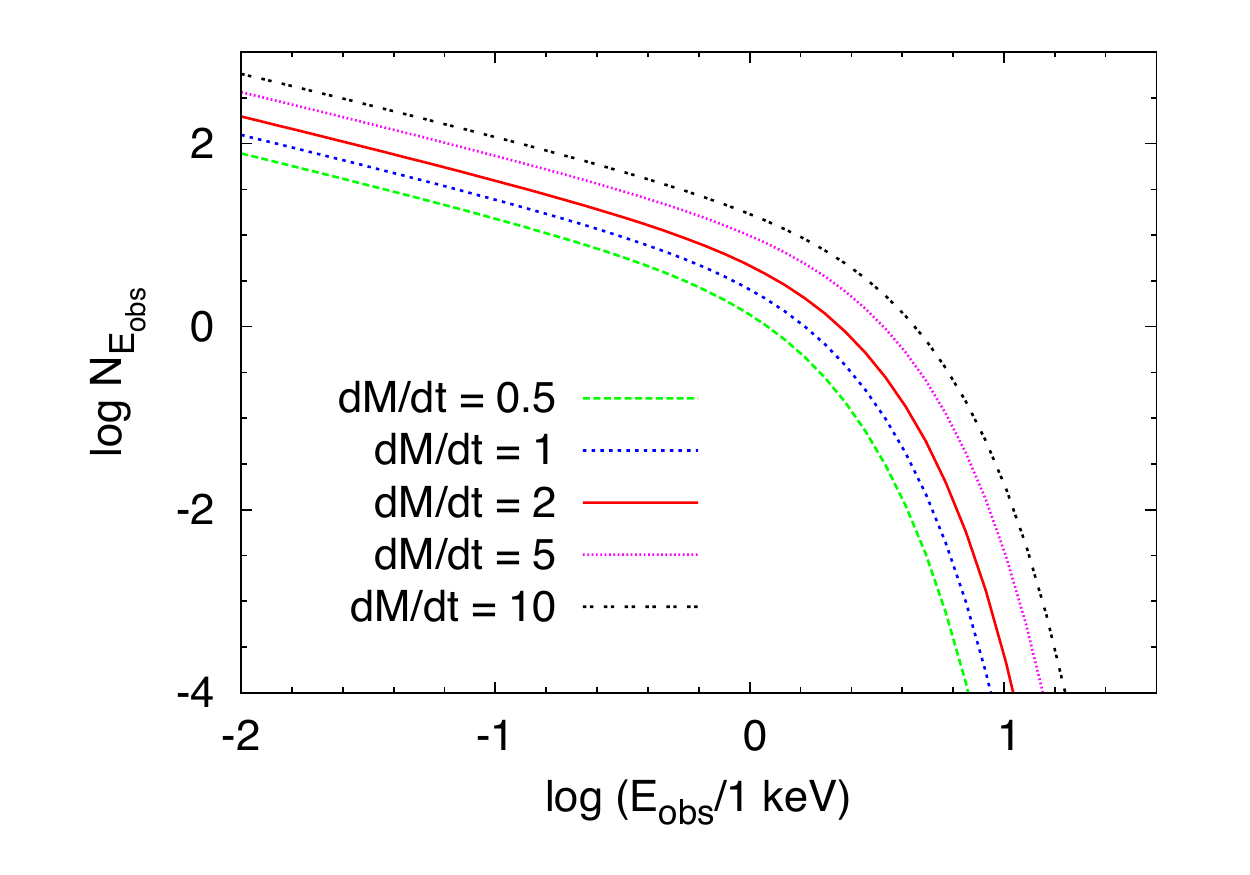} \\
\vspace{0.3cm}
\includegraphics[trim=10mm 0mm 10mm 0mm,width=5.3cm]{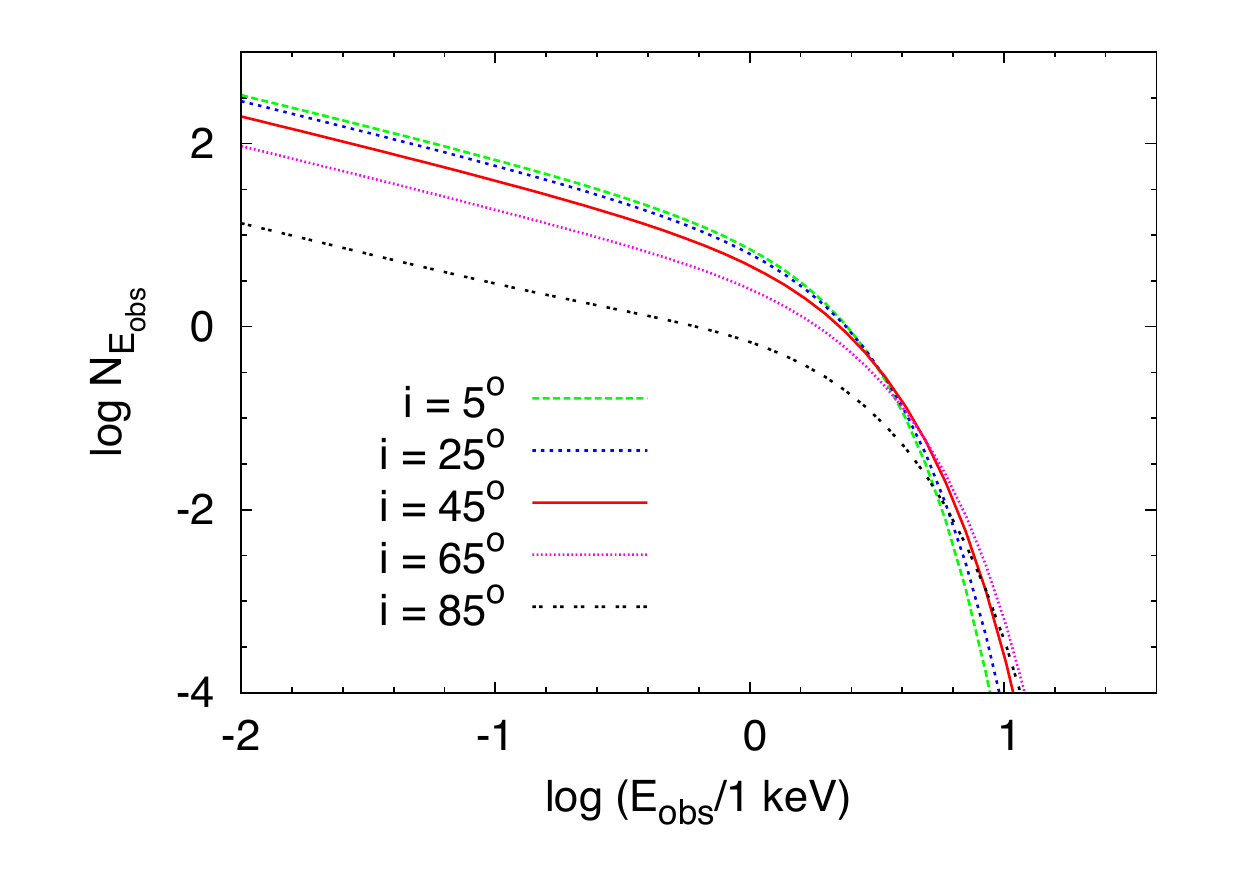}
\hspace{0.5cm}
\includegraphics[trim=10mm 0mm 10mm 0mm,width=5.3cm]{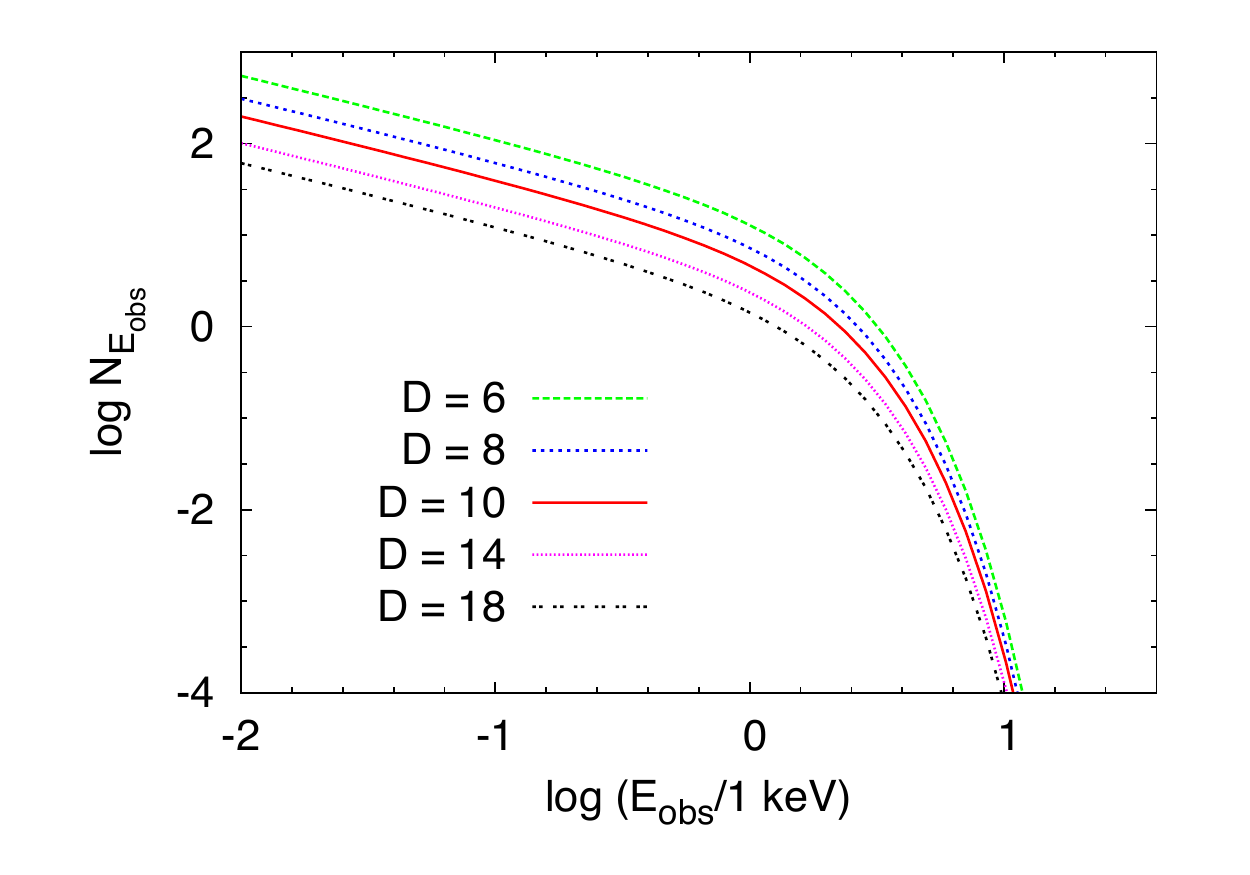} \\
\vspace{0.3cm}
\includegraphics[trim=10mm 0mm 10mm 0mm,width=5.3cm]{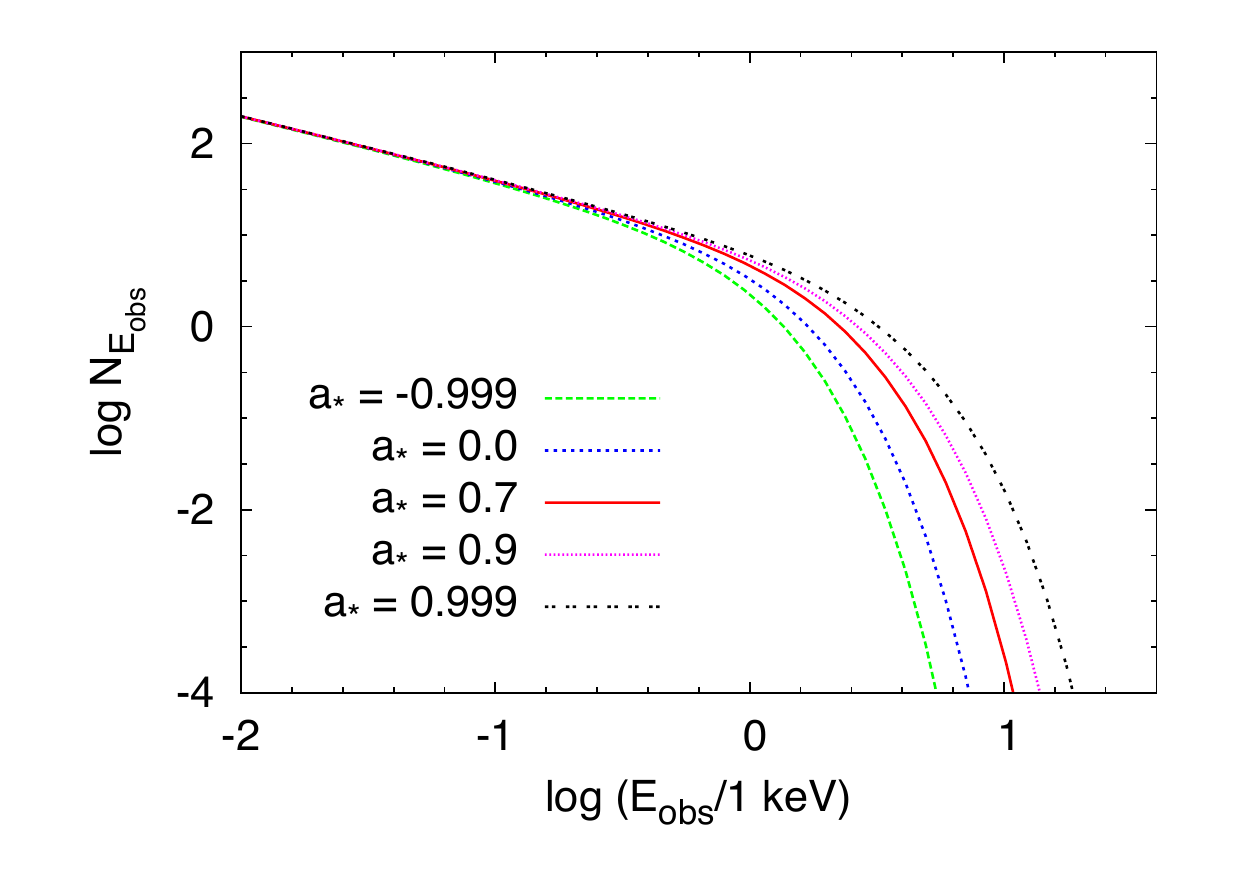}
\end{center}
\vspace{-0.3cm}
\caption{Impact of the model parameters on the thermal spectrum of a thin disk: mass $M$ (top left panel), mass accretion rate $\dot{M}$ (top right panel), viewing angle $i$ (central left panel), distance $D$ (central right panel), and spin parameter $a_*$ (bottom panel). When not shown, the values of the parameters are: $M = 10$~$M_\odot$, $\dot{M} = 2 \cdot 10^{18}$~g~s$^{-1}$, $D = 10$~kpc, $i = 45^\circ$, and $a_*=0.7$. $M$ in units of $M_\odot$, $\dot{M}$ in units of $10^{18}$~g~s$^{-1}$, $D$ in kpc, and flux density $N_{E_{\rm obs}}$ in photons~keV$^{-1}$~cm$^{-2}$~s$^{-1}$. \label{f-cfm1}}
\end{figure}

Within the Novikov-Thorne model, we can derive the time-averaged radial structure of the accretion disk from the fundamental laws of the conservation of rest-mass, energy, and angular momentum~\cite{r-bh-ntm2}. The time-averaged energy flux emitted from the surface of the disk is
\be
\mathcal{F} (r) = \frac{\dot{M} c^2}{4 \pi r_{\rm g}^2} F(r) \, ,
\ee
where $\dot{M}=dM/dt$ is the time-averaged mass accretion rate, which is independent of the radial coordinate, and $F(r)$ is a dimensionless function of the radial coordinate that becomes roughly of order 1 at the disk inner edge (see~\cite{r-bh-ntm2} for more details). Assuming that the disk is in local thermal equilibrium, its emission is blackbody-like and at any radius we can define an effective temperature $T_{\rm eff} (r)$ from the time-averaged energy flux as $\mathcal{F} = \sigma T^4_{\rm eff}$, where $\sigma$ is the Stefan-Boltzmann constant. 
Novikov-Thorne disks with the inner edge at the ISCO radius are realized when the accretion luminosity is between $\sim$5\% to $\sim$30\% of the Eddington limit of the object~\cite{r-bh-cfm-1915} and this is confirmed by theoretical~\cite{r-bh-edd1a,r-bh-edd1b} and observational studies~\cite{r-bh-thin-constant}. At lower luminosities, the disk is more likely truncated at a radius larger than the ISCO, and we have an ADAF between the inner edge of the disk and the black hole. At higher luminosities, the gas pressure becomes important, the inner part of the disk is not thin any longer, and the inner edge might be at a radius slightly smaller than the ISCO~\cite{r-bh-rin-thick}. 
Requiring $\dot{M} \sim 0.1 \, \dot{M}_{\rm Edd}$ as the condition for Novikov-Thorne disks, we can get a rough estimate of the effective temperature of the inner part of the accretion disk
\be
T_{\rm eff} \sim 
\left( \frac{0.1 \; \dot{M}_{\rm Edd} c^2}{4 \pi \sigma r_{\rm g}^2} \right)^{1/4} 
\sim \left( \frac{10 \; M_\odot}{M} \right)^{1/4} \text{keV} \, ,
\ee
and we can see that the disk's thermal spectrum is in the soft X-ray band for stellar-mass black holes and in the optical/UV band for the supermassive ones.

The continuum-fitting method is the analysis of the thermal spectrum of geometrically thin and optically thick accretion disks of black holes in order to measure the black hole spin parameter $a_*$~\cite{r-bh-cfm1,r-bh-cfm2,r-bh-cfm3,r-bh-cfm4}. The technique is normally used for stellar-mass black holes only, because the spectrum of supermassive black holes is in the optical/UV band where dust absorption limits the capability of accurate measurements.
The model describing the thermal spectrum of an accretion disk around a Kerr black hole depends on five parameters: the black hole mass $M$, the mass accretion rate $\dot{M}$, the inclination angle of the disk with respect to the line of sight of the observer $i$, the distance of the source from the observer $D$, and the black hole spin parameter $a_*$. The impact of these five parameters on the shape of the spectrum is illustrated in Fig.~\ref{f-cfm1}.
Note that it is impossible to infer all the five model parameters from the data of the spectrum of a thin disk, because the spectrum's shape is too simple and there is a degeneracy. However, if we can get independent measurements of $M$, $D$, and $i$, usually from optical observations, it is possible to fit the data and measure $a_*$ and $\dot{M}$. Presently, there are about ten stellar-mass black holes with a spin measurement obtained from the continuum-fitting method, see Tab.~\ref{t-spin}.

\begin{table*}[t]
\centering
{\renewcommand{\arraystretch}{1.2}
\begin{tabular}{|ccccccc|}
\hline 
BH Binary & \hspace{0.1cm} & $a_*$ (Continuum) & \hspace{0.1cm} & $a_*$ (Iron) & \hspace{0.1cm} & \hspace{0.3cm} Principal References \hspace{0.3cm} \\
\hline 
4U~1630-472 && --- && $0.985^{+0.005}_{-0.014}$ && \cite{r-bh-King:2014sja} \\
GRS~1915+105 && $> 0.98$ && $0.98 \pm 0.01$ && \cite{r-bh-cfm-1915,r-bh-cfm-1915b} \\
Cygnus~X-1 && $> 0.98$ && $> 0.95$ && \cite{r-bh-cfm-cyg1,r-bh-cfm-cyg2,r-bh-cfm-cyg3,r-bh-cfm-cyg4,r-bh-cfm-cyg5,r-bh-cfm-cyg6} \\
GS~1354-645 && -- && $> 0.98$ && \cite{r-bh-cfm-gs1354} \\
MAXI~J1535-571 && --- && $> 0.98$ && \cite{r-bh-iron-maxi15a,r-bh-iron-maxi15b} \\
Swift~J1658.2 && --- && $> 0.96$ && \cite{r-bh-iron-swift-xu} \\
LMC~X-1 && $0.92 \pm 0.06$ && $0.97^{+0.02}_{-0.25}$ && \cite{r-bh-cfm-lmcx1,r-bh-cfm-lmcx1b} \\
GX~339-4 && $< 0.9$ && $0.95\pm0.03$ && \cite{r-bh-cfm-gx339,r-bh-cfm-gx339b,r-bh-cfm-gx339c,r-bh-cfm-gx339d} \\
V404~Cyg && --- && $> 0.92$ && \cite{r-bh-Walton:2016fso} \\
GRS~1716-249 && --- && $> 0.92$ && \cite{r-bh-Tao:2019yhu} \\
XTE~J1752-223 && --- && $0.92 \pm 0.06$ && \cite{r-bh-cfm-1752,r-bh-i-1752} \\
Swift~J174540.2 && --- && $> 0.9$ && \cite{r-bh-Mori:2019iwz} \\
MAXI~J1836-194 && --- && $0.88 \pm 0.03$ && \cite{r-bh-cfm-maxi} \\
XTE~J1650-500 && --- && $0.84 \sim 0.98$ && \cite{r-bh-cfm-1650} \\
M33~X-7 && $0.84 \pm 0.05$ && --- && \cite{r-bh-cfm-liu08} \\
4U~1543-47 && $0.80 \pm 0.10^\star$ && --- && \cite{r-bh-cfm-sh06} \\
GRS~1739-278 && --- && $0.8 \pm 0.2$ && \cite{r-bh-Miller:2014sla} \\
IC10~X-1     &&  $\gtrsim0.7$  && --- && \cite{r-bh-cfm-st16} \\
Swift~J1753.5 && --- && $0.76^{+0.11}_{-0.15}$ && \cite{r-bh-cfm-swift} \\
GRO~J1655-40 && $0.70 \pm 0.10^\star$ && $> 0.9$ && \cite{r-bh-cfm-sh06,r-bh-cfm-swift} \\
GS~1124-683 && $0.63^{+0.16}_{-0.19}$ && --- && \cite{r-bh-cfm-gou_novamus} \\
XTE~J1652-453 && --- && $< 0.5$ && \cite{r-bh-cfm-1652} \\
XTE~J1550-564 && $0.34 \pm 0.28$ && $0.55^{+0.15}_{-0.22}$ && \cite{r-bh-cfm-xte} \\
LMC~X-3 && $0.25 \pm 0.15$ && --- && \cite{r-bh-cfm-lmcx3} \\
H1743-322 && $0.2 \pm 0.3$ && --- && \cite{r-bh-cfm-h1743} \\
A0620-00 &&  $0.12 \pm 0.19$ && --- && \cite{r-bh-cfm-62} \\
\hspace{0.3cm} XMMU~J004243.6 \hspace{0.3cm} && $< -0.2$ && --- && \cite{r-bh-cfm-m31} \\
\hline 
\end{tabular}}
\vspace{0.4cm}
\caption{Summary of the continuum-fitting and iron line measurements of the spin parameter of stellar-mass black holes. See the references in the last column for more details. Note: $^\star$These sources were studied in an early work of the continuum-fitting method, within a more simple model, and therefore the published 1-$\sigma$ error estimates are doubled following~\cite{r-bh-cfm4}. \label{t-spin}}
\end{table*}

\begin{table}[t]
\centering
{\renewcommand{\arraystretch}{1.2}
\begin{tabular}{|ccccc|}
\hline 
Object & \hspace{0.1cm} & $a_*$ (Iron) & \hspace{0.1cm} & \hspace{0.3cm} Principal References \hspace{0.3cm} \\
\hline 
\hspace{0.3cm} IRAS~13224-3809 \hspace{0.3cm} && $> 0.99$ && \cite{r-bh-suzaku} \\
Mrk~110 && $> 0.99$ && \cite{r-bh-suzaku} \\
NGC~4051 && $> 0.99$ && \cite{r-bh-ngc4051} \\
1H0707-495 && $> 0.98$ && \cite{r-bh-suzaku,r-bh-1h0707} \\
RBS~1124 && $> 0.98$ && \cite{r-bh-suzaku} \\
NGC~3783 && $> 0.98$ && \cite{r-bh-ngc3783} \\
Fairall~9 && $0.973^{+0.003}_{-0.003}$ && \cite{r-bh-fairall9} \\
NGC~1365 && $0.97^{+0.01}_{-0.04}$ && \cite{r-bh-ngc1365a,r-bh-ngc1365b} \\
Swift~J0501-3239 && $> 0.96$ && \cite{r-bh-suzaku} \\
PDS~456 && $> 0.96$ && \cite{r-bh-suzaku} \\
Ark~564 && $0.96^{+0.01}_{-0.06}$ && \cite{r-bh-suzaku} \\
3C120 && $> 0.95$ && \cite{r-bh-3c120} \\
Mrk~79 && $> 0.95$ && \cite{r-bh-mrk79} \\
NGC~5506 && $0.93^{+0.04}_{-0.04}$ && \cite{r-bh-shangyu} \\
MCG-6-30-15 && $0.91^{+0.06}_{-0.07}$ && \cite{r-bh-mcg63015a,r-bh-mcg63015b} \\
Ton~S180 && $0.91^{+0.02}_{-0.09}$ && \cite{r-bh-suzaku} \\
1H0419-577 && $> 0.88$ && \cite{r-bh-suzaku} \\
IRAS~00521-7054 && $> 0.84$ && \cite{r-bh-iras521} \\
Mrk~335 && $0.83^{+0.10}_{-0.13}$ && \cite{r-bh-suzaku,r-bh-mrk335} \\
Ark~120 && $0.81^{+0.10}_{-0.18}$ && \cite{r-bh-suzaku,r-bh-ark120} \\
Swift~J2127+5654 && $0.6^{+0.2}_{-0.2}$ && \cite{r-bh-swift2127} \\
Mrk~841 && $> 0.56$ && \cite{r-bh-suzaku} \\
\hline 
\end{tabular}}
\vspace{0.4cm}
\caption{Summary of spin measurements of supermassive black holes reported in the literature. See the references in the last column for more details. \label{t-spin-agn}}
\end{table}

\subsection{X-ray reflection spectroscopy}

X-ray reflection spectroscopy (or the iron line method) refers to the study of the reflection component. This technique can be applied to both stellar-mass and supermassive black holes and is currently the only available method to measure the spin of supermassive black holes~\cite{r-bh-i1,r-bh-i2}.

\begin{figure}[t]
\begin{center}
\includegraphics[width=10.0cm]{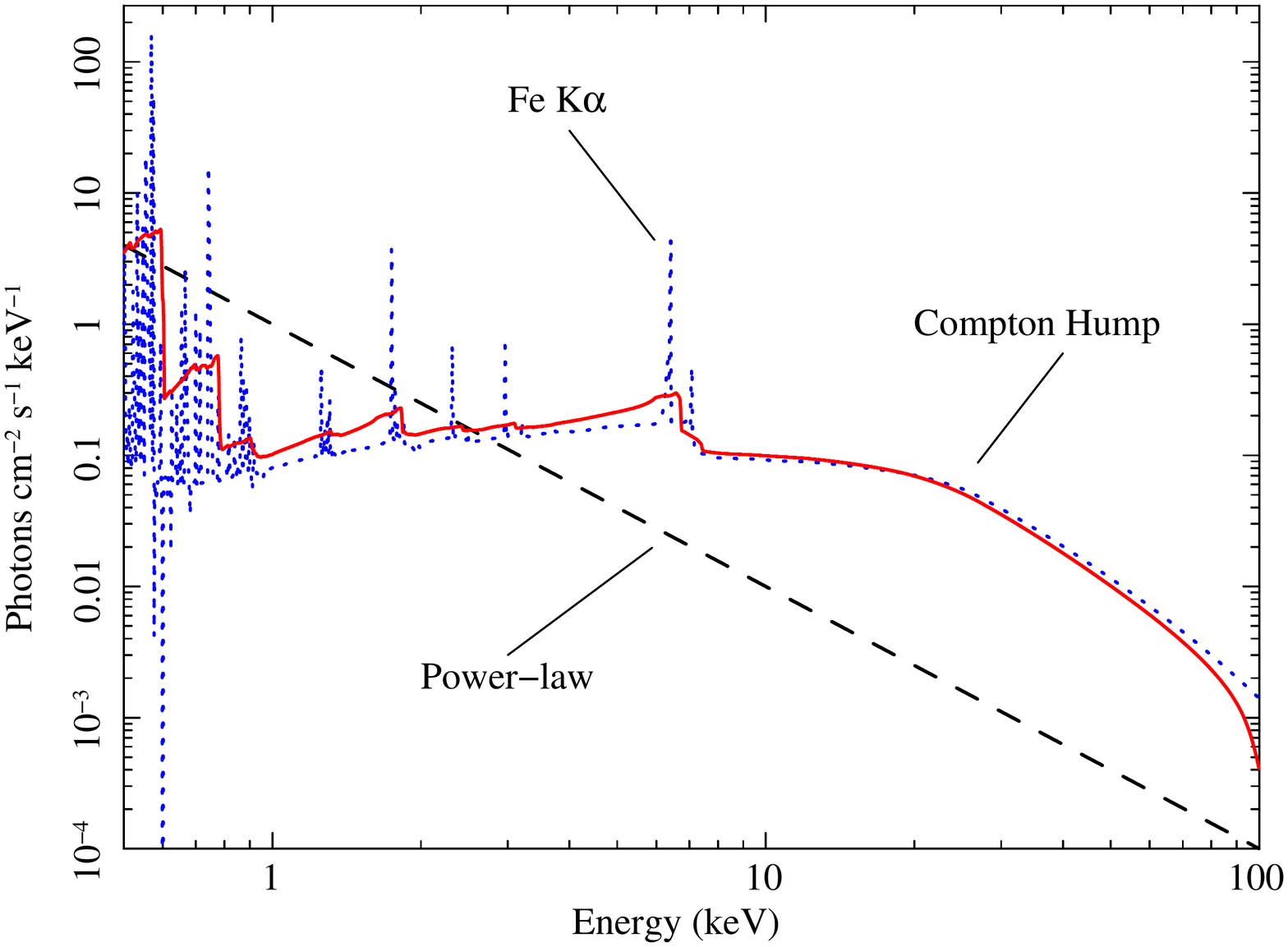}
\end{center}
\vspace{-0.5cm}
\caption{Reflection spectrum for a neutral accretion disk irradiated by a power-law continuum with photon index $\Gamma = 2$. The dashed black line indicates the power-law continuum from the corona; the dotted blue curve is for the reflection spectrum at the emission point in the rest-frame of the gas (only atomic physics is involved); the solid red curve is for the reflection spectrum of a non-rotating black hole at the detection point and is blurred by relativistic effects (gravitational redshift, Doppler boosting, light bending). From~\cite{r-bh-i2}, reproduced with permission. \label{f-laura}}
\end{figure}

The most prominent feature of the reflection spectrum is usually the iron K$\alpha$ line\footnote{A K$\alpha$ line results from the transition of an electron from a $p$ orbit of the L shell (quantum number 2) to the K shell (quantum number 1). The line is actually a doublet with slightly different energies, K$\alpha_1$ and K$\alpha_2$, respectively for the transitions $2 p_{1/2} \rar 1 s$ and $2 p_{3/2} \rar 1 s$ using the atomic notation. A K$\beta$ line results from the transition of an electron from a $p$ orbit of the M shell (quantum number 3) to the K shell. An L$\alpha$ line is emitted from transition of an electron from a $d$ orbit of the M shell to a $p$ orbit of the L shell.}. This is because iron is more abundant than other heavy elements (the iron-56 nucleus is more tightly bound than lighter and heavier elements, so it is the final product of nuclear reactions) and the probability of fluorescent line emission is also high (scaling as $Z^4$, where $Z$ is the atomic number). Moreover, X-ray detectors typically have high sensitivity around 6~keV and there are no other atomic lines around this energy. The iron K$\alpha$ line is a very narrow feature in the rest-frame of the emitter, while the one observed in the reflection spectrum of black holes can be very broad and skewed, as the result of relativistic effects occurring in the strong gravity region of the object (gravitational redshift, Doppler boosting, light bending)~\cite{r-bh-book,r-bh-i1,r-bh-i2,r-bh-i2003}. While the iron K$\alpha$ line is usually the strongest feature, accurate measurements of black hole spins require to fit the whole reflection spectrum, not just the iron line. Fig.~\ref{f-laura} shows the reflection spectrum for a neutral accretion disk: the unblurred reflection spectrum in the rest-frame of the gas is the dotted blue curve while the solid red curve is the blurred reflection spectrum of the accretion disk around a Schwarzschild black hole as detected by a distant observer.

\begin{figure}[t]
\begin{center}
\includegraphics[trim=10mm 0mm 10mm 0mm,width=5.3cm]{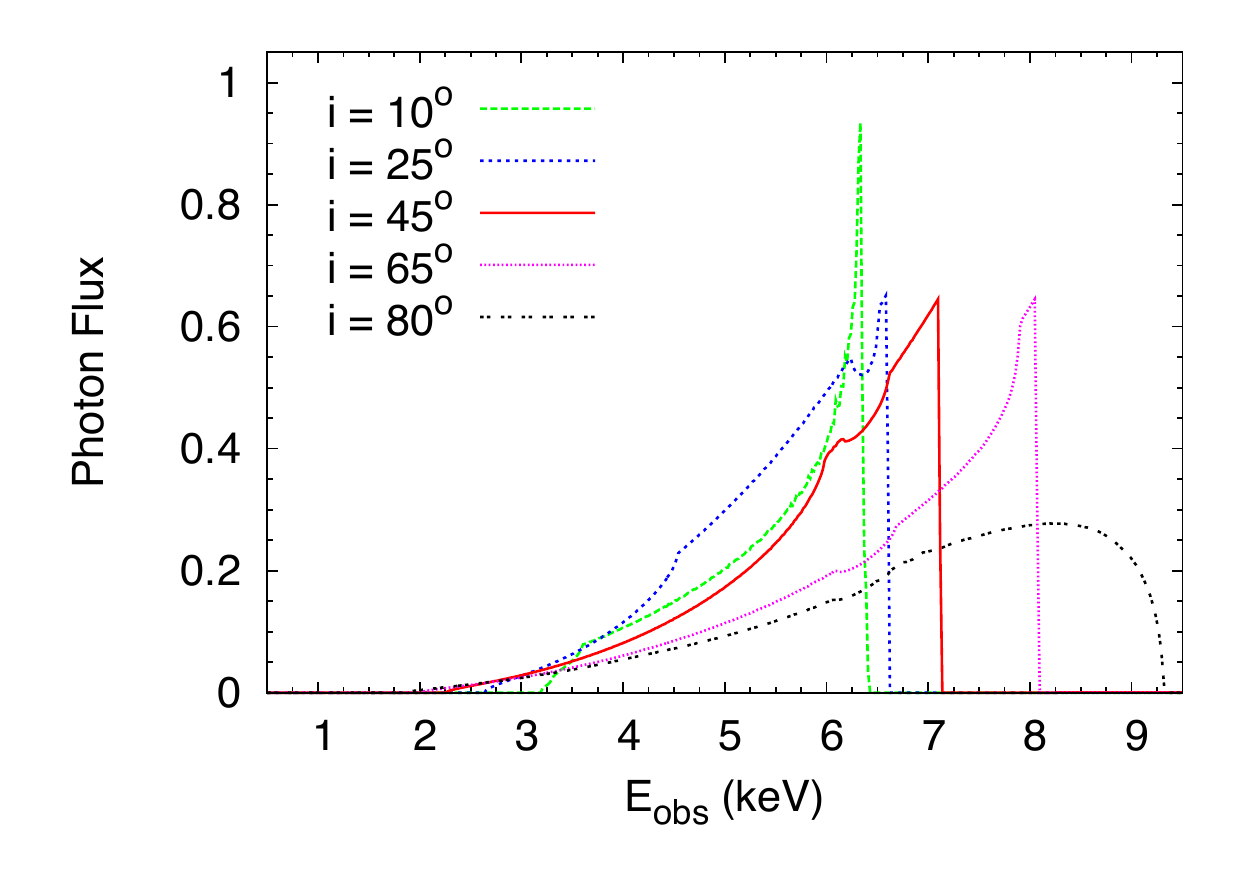}
\hspace{0.7cm}
\includegraphics[trim=10mm 0mm 10mm 0mm,width=5.3cm]{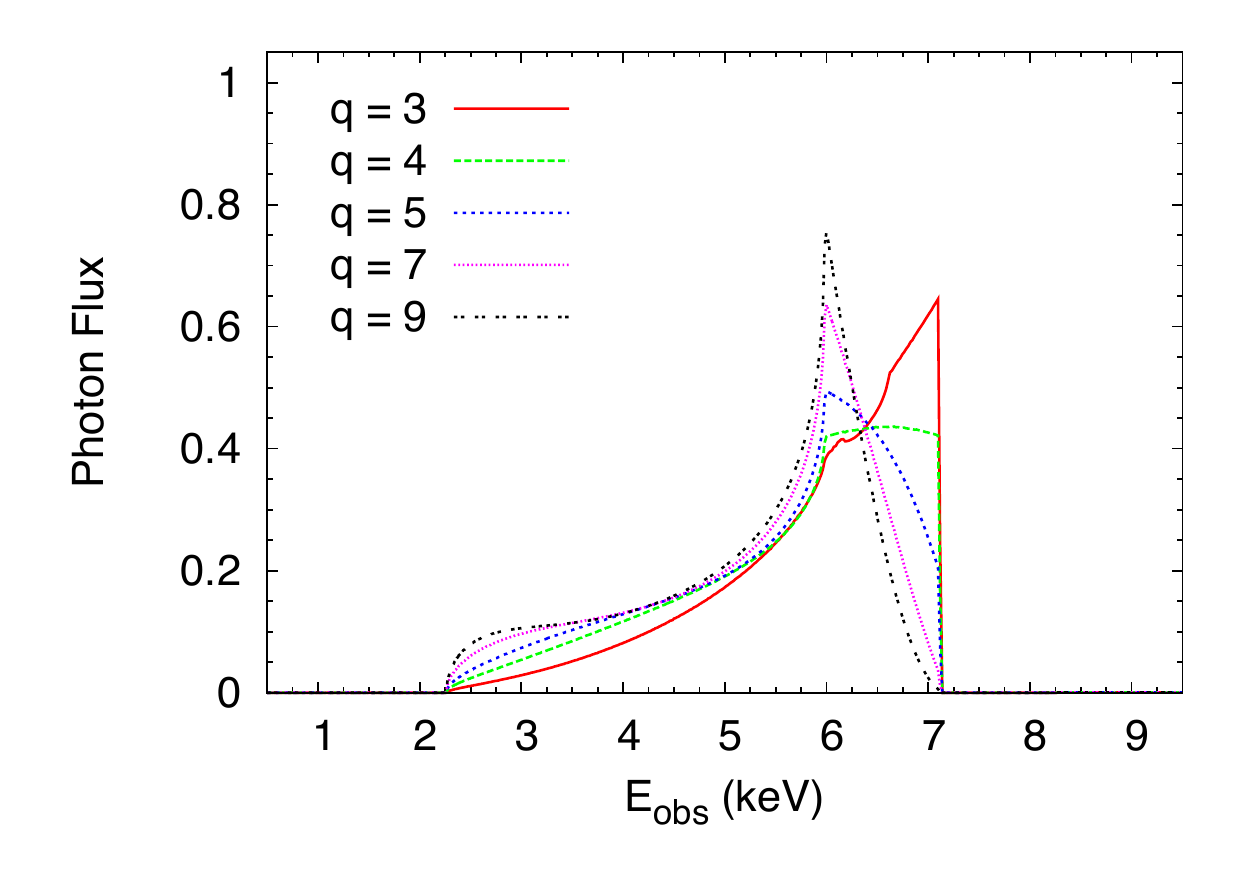} \\
\vspace{0.3cm}
\includegraphics[trim=10mm 0mm 10mm 0mm,width=5.3cm]{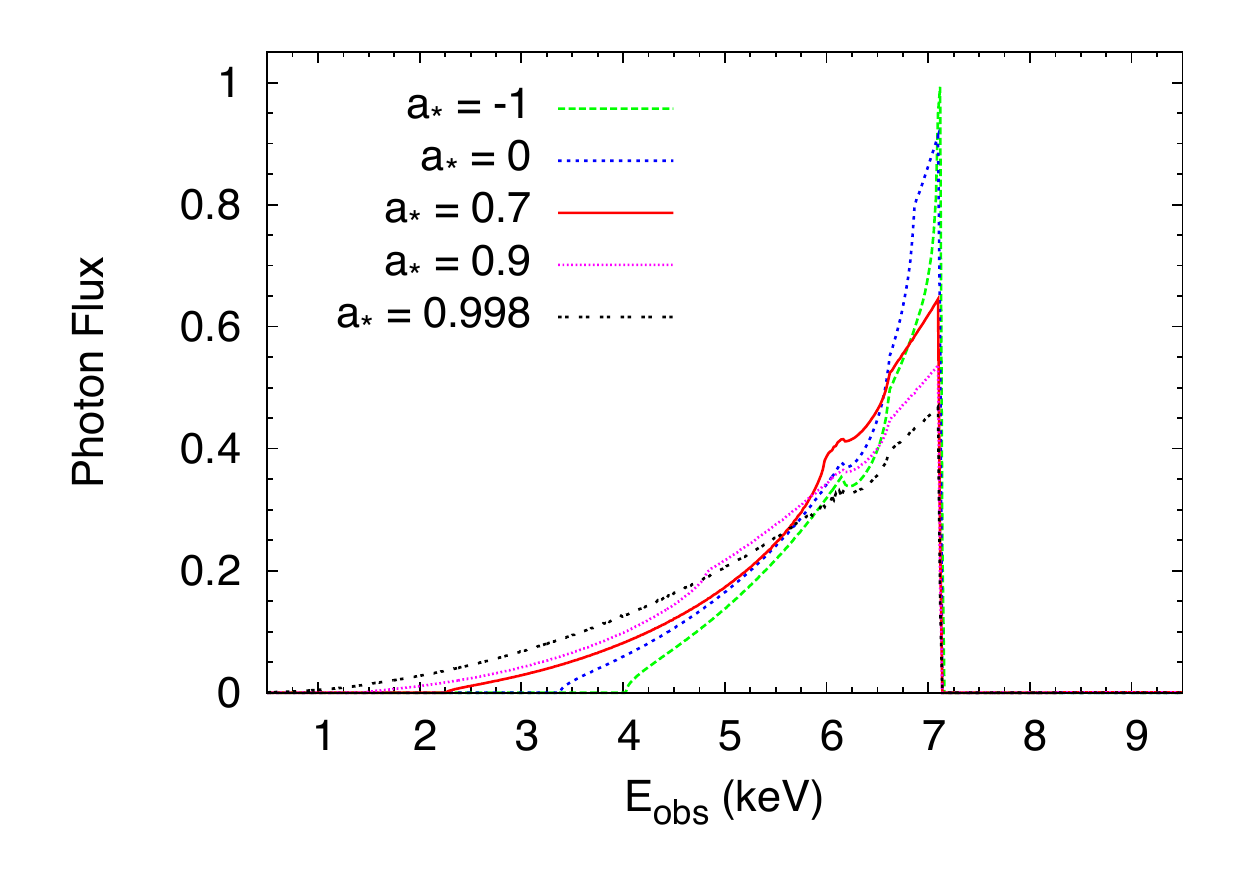}
\end{center}
\vspace{-0.3cm}
\caption{Impact of the the inclination angle of the disk $i$, the emissivity index $q$, and the spin parameter $a_*$ on the shape of an iron line at 6.4~keV emitted from a thin accretion disk. When not shown, the values of the parameters are: $i = 45^\circ$, $q = 3$, $a_*=0.7$, and $R_{\rm out} = 400$~$r_{\rm g}$. From~\cite{r-bh-book}, reproduced with permission. \label{f-iron}}
\end{figure}

Reflection models describing the reflection component of accretion disks around Kerr black holes depend on several parameters: the black holes spin $a_*$, the inner edge of the disk $R_{\rm in}$ (which may or may not be assumed at the ISCO radius), the outer edge of the disk $R_{\rm out}$, the inclination angle of the disk $i$, the metallicity (or the iron abundance), the ionization of the disk, and some parameters related to the emissivity profile of the disk. The latter is quite a crucial ingredient and depends on the geometry of the corona, which is currently unknown. A phenomenological approach is to model the emissivity profile with a power-law (the intensity on the disk is $I \propto 1/r^q$ where $q$ is the emissivity index) or with a broken power-law ($I \propto 1/r^{q_{\rm in}}$ for $r < R_{\rm br}$, $I \propto 1/r^{q_{\rm out}}$ for $r > R_{\rm br}$). In the case of supermassive black holes, it is often necessary to take the cosmological redshift $z$ into account. For stellar-mass black holes, their relative motion in the Galaxy is of order 100~km/s and the redshift can be ignored. Fig.~\ref{f-iron} shows the impact of the the inclination angle of the disk $i$, the emissivity index $q$ (assuming an emissivity profile described by a simple power-law $I \propto 1/r^q$), and the spin parameter $a_*$ on the shape of an iron line at 6.4~keV emitted from a thin accretion disk around a black hole.  A significant advantage of the iron line method is that it does not require independent measurements of the black hole mass $M$, the distance $D$, and the inclination angle of the disk $i$, three quantities that are required in the continuum-fitting method, are usually difficult to measure, and have large uncertainty. The reflection spectrum is independent of the former two, and can directly measure the inclination angle of the disk.

Current spin measurements of stellar-mass black holes with the iron line method are summarized in the third column in Tab.~\ref{t-spin} (see the corresponding references in the fourth column for more details). Note that some black holes have their spin measured with both the continuum-fitting and the iron line methods. In general, the two measurements agree (GRS~1915+105, Cygnus~X-1, LMC~X-1, XTE~J1550-564). For GX~339-4 and GRO~J1655-40, the two measurements are not consistent. There can be a few reasons for this discrepancy. The iron line method is usually applied when the source is in the hard state, when the reflection spectrum is stronger, but the disk may be truncated at a radius larger than the ISCO. This would lead to an underestimation of the black hole spin, but since the iron line method provides spin values higher than the continuum-fitting method in the case of GX~339-4 and GRO~J1655-40, this cannot be the reason for the discrepancy. Rather, since the continuum-fitting method crucially depends on independent measurements of the black hole mass $M$, the distance $D$, and the inclination angle of the disk $i$, large systematic uncertainties in these measurements may cause the continuum-fitting method to deviate. For instance, in the case of GRO~J1655-40 there are a few mass measurements reported in the literature, but they are not consistent amongst each other.

A summary of spin measurements of supermassive black holes with the iron line method is reported in Tab.~\ref{t-spin-agn} (see the references in the last column for more details and the lists of spin measurements in~\cite{r-bh-i1,r-bh-i2,r-bh-vasudevan} for a few more sources with a constrained spin). Note the very high spin of several objects. In part, this can be explained noting that fast-rotating black holes are brighter and thus the spin measurement is easier. If these measurements are correct, they would point out that these objects have been spun up by prolonged disk accretion and therefore would provide information about galaxy evolutions (see the discussion in Section~\ref{ss-evolution}). However, the very high spin measurements have to be taken with some caution, as they may be affected by large systematic uncertainties in the model employed to infer the black hole spin. For example, if the mass accretion rate is near the Eddington limit, which is probably the case for several sources, the spin parameter can be easily overestimated if we employ a model that assumes a thin disk~\cite{Riaz:2019kat}. More details on the possible interpretation of current spin measurements of supermassive black holes can be found in~\cite{r-bh-i1}.

\subsection{Quasi periodic oscillations}\label{ss-qpos}

\begin{figure}[b]
\begin{center}
\includegraphics[width=8.7cm]{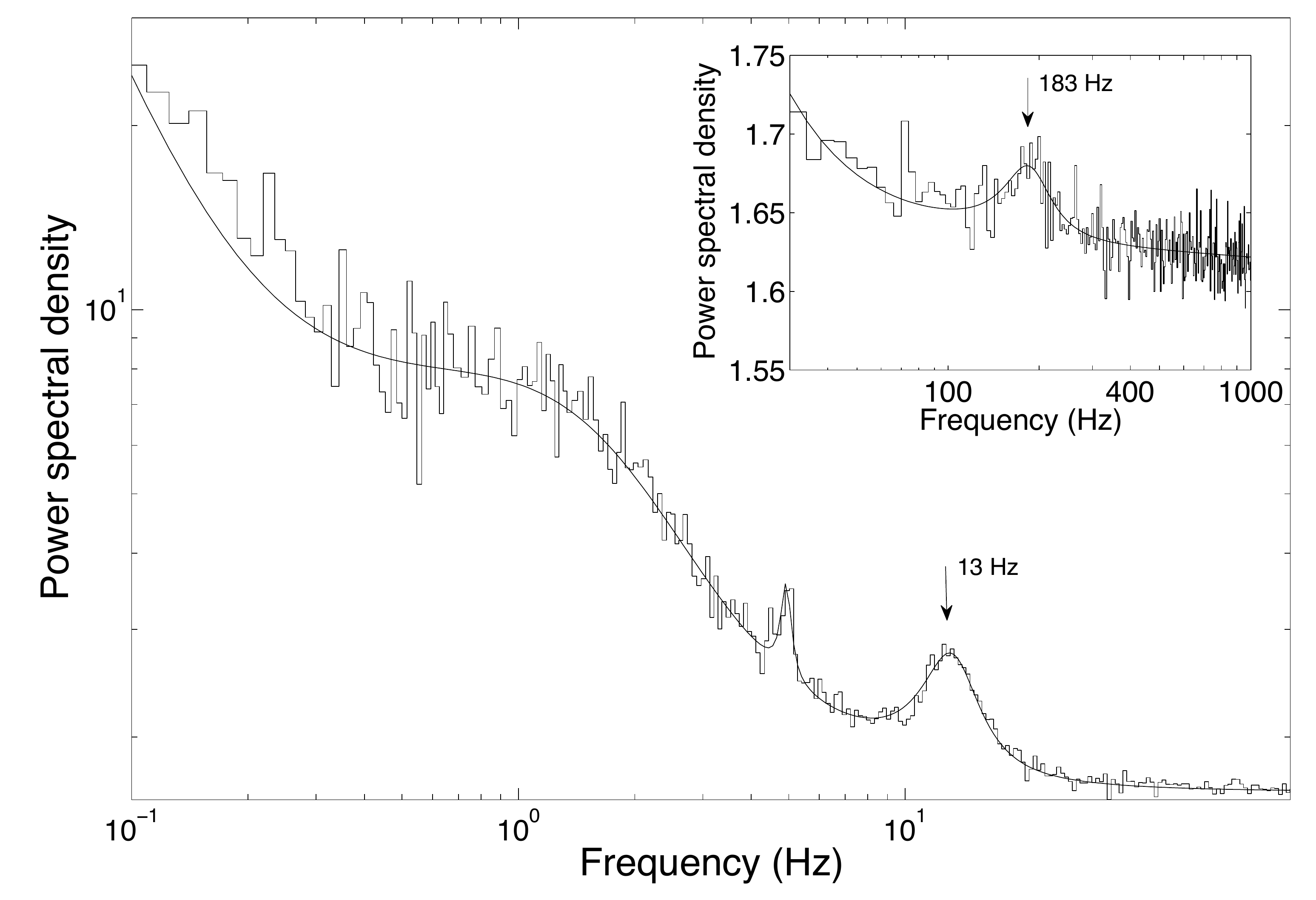}
\end{center}
\vspace{-0.2cm}
\caption{Power density spectrum from an observation of XTE~J1550-564. We see a QPO around 5~Hz, a QPO at 13~Hz (marked by an arrow), and a QPO at 183~Hz in the inset (marked by an arrow). Fig.~1 from~\cite{r-bh-motta14}, reproduced by permission of Oxford University Press. \label{f-qpos}}
\end{figure}

Quasi-periodic oscillations (QPOs) are a common feature in the X-ray power density spectrum of neutron stars and stellar-mass black holes~\cite{r-bh-vdk}. The power density spectrum $P(\nu)$ is the square of the Fourier transform of the photon count rate $C(t)$. If we use the Leahy normalization, we have
\be
P ( \nu ) = \frac{2}{N} 
\left| \int_0^T C(t) e^{-2 \pi i \nu t} dt \right|^2 \, ,
\ee
where $N$ is the total number of counts and $T$ is the duration of the observation. QPOs are narrow features in the X-ray power density spectrum of a source. Fig.~\ref{f-qpos} shows the power density spectrum obtained from an observation of the stellar-mass black hole XTE~J11550-564, where we can see a QPO around 5~Hz, one at 13~Hz, and one at 183~Hz in the inset.

In the case of black hole binaries, QPOs can be grouped into two classes: low-frequency QPOs (0.1-30~Hz) and high-frequency QPOs (40-450~Hz). The exact nature of these QPOs is currently unknown, but there are several proposals in the literature. In most scenarios, the frequencies of the QPOs are related to the fundamental frequencies of a particle orbiting the black hole~\cite{r-bh-qqq1,r-bh-qqq2,r-bh-qqq3}:
\begin{enumerate}
\item {\it Orbital frequency} $\nu_\phi$, which is the inverse of the orbital period.
\item {\it Radial epicyclic frequency} $\nu_r$, which is the frequency of radial oscillations around the mean orbit.
\item {\it Vertical epicyclic frequency} $\nu_\theta$, which is the frequency of vertical oscillations around the mean orbit.
\end{enumerate}
In the Kerr metric, we have a compact analytic form for the expression of these frequencies
\be
\hspace{-0.4cm}
\nu_\phi &=& \frac{c}{2\pi} \sqrt{\frac{r_{\rm g}}{r^3}}
\left[1 \pm a_* \left(\frac{r_{\rm g}}{r}\right)^{3/2} \right]^{-1} \, , \\
\hspace{-0.4cm}
\nu_r &=& \nu_\phi \sqrt{1 - 6 \, \frac{r_{\rm g}}{r} 
\pm 8 a_* \left(\frac{r_{\rm g}}{r}\right)^{3/2} 
- 3 a^2_* \left(\frac{r_{\rm g}}{r}\right)^2} \, , \\
\hspace{-0.4cm}
\nu_\theta &=& \nu_\phi \sqrt{1 \mp 4 a_* \left(\frac{r_{\rm g}}{r}\right)^{3/2} 
+ 3 a^2_* \left(\frac{r_{\rm g}}{r}\right)^2} \, ,
\ee
where $r$ is the orbital radius in Boyer-Lindquist coordinates. To have an idea of the order of magnitude of these frequencies, we can write the orbital frequency for a Schwarzschild black hole
\be
\nu_\phi (a_* = 0) = 220 \left(\frac{10 \, M_\odot}{M}\right) 
\left(\frac{6 \, r_{\rm g}}{r}\right)^{3/2} \text{Hz} \, .
\ee
High-frequency QPOs at 40-450~Hz are thus of the right magnitude to be associated to the orbital frequencies near the ISCO radius of stellar-mass black holes. Interestingly, we also have evidence of high-frequency QPOs in supermassive black holes ($< 1$~mHz)~\cite{r-bh-qpo-smbh} and intermediate-mass black holes ($\sim 1$~Hz)~\cite{r-bh-i-qpo}.

Since it is often possible to measure the frequencies of QPOs with quite a good precision, if we knew the exact relation between QPOs and fundamental frequencies, it could be possible to measure black hole spins with high precision. For instance, in~\cite{r-bh-gro-motta} the authors interpret the observed QPOs of the black hole binary GRO~J1655-40 within the relativistic precession model and obtain the mass measurement $M/M_\odot = 5.31 \pm 0.07$ and the spin measurement $a_* = 0.290 \pm 0.003$.


\subsection{Direct imaging}

\begin{figure}[b]
\begin{center}
\includegraphics[width=10.0cm]{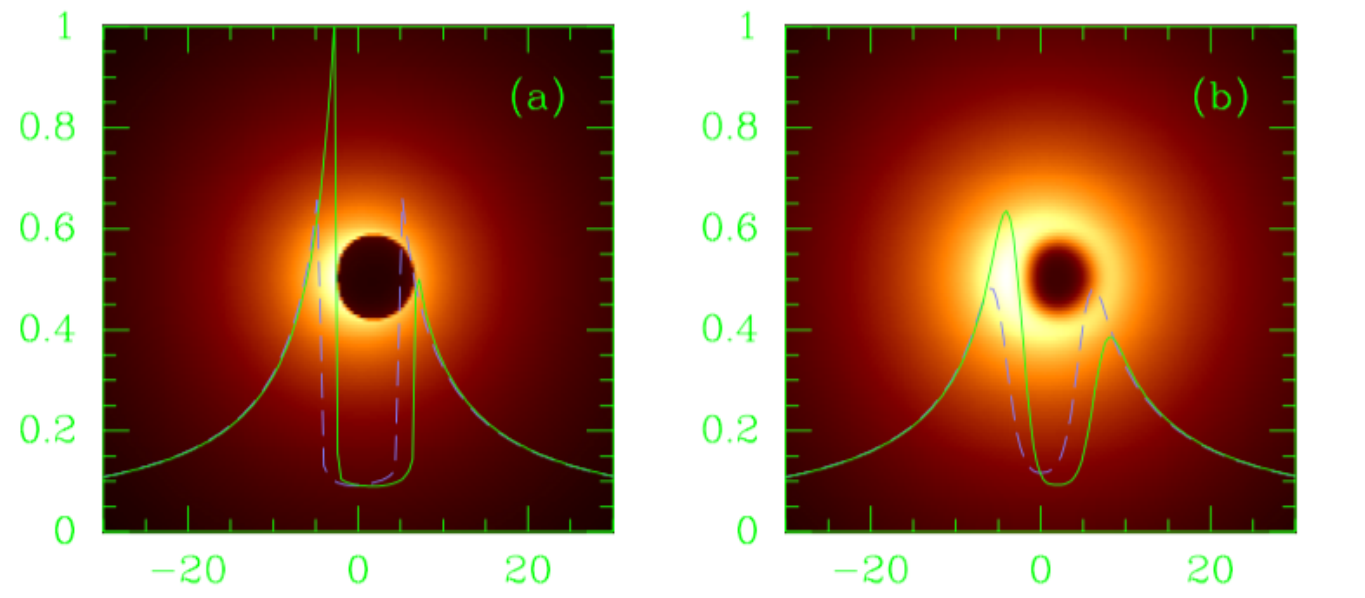}
\end{center}
\vspace{-0.2cm}
\caption{Direct image of a black hole surrounded by an optically thin emitting medium with the characteristics of that of Sgr~A$^*$. The black hole spin parameter is $a_* = 0.998$ and the viewing angle is $i=45^\circ$. Panel~$a$: image from ray-tracing calculations. Panel~$b$: image from a simulated observation of an idealized VLBI experiment at 0.6~mm wavelength taking interstellar scattering into account. The solid green curve and the dashed purple curve show, respectively, the intensity variations of the image along the $x$-axis and the $y$-axis. From~\cite{r-bh-shadow}. \copyright AAS. Reproduced with permission. \label{f-shadow}}
\end{figure}

Since the black hole does not allow any light to come out from inside the event horizon, and the disk outside this region is radiating, an interesting possibility is to observe the \emph{black hole shadow}. Depending on the geometry of the accretion disk and on its optical properties (thin/thick), if we could image the accretion flow around a black hole with a resolution of at least some gravitational radii, we would observe a dark area in the middle of a brighter surrounding. The dark area is usually referred to as the black hole shadow (see Fig.~\ref{f-shadow}). The shape of the shadow is determined by the bending of light in the strong gravity region~\cite{r-bh-bardeen73}.

Very long baseline interferometry (VLBI) uses several radio telescopes located in various continents and combines the data to mimic a single telescope of the size of the Earth. This helps achieving much smaller angular resolutions than a single telescope. The Event Horizon Telescope (EHT)\footnote{http://www.eventhorizontelescope.org/} is an international collaboration that uses mm and sub-mm VLBI techniques to image supermassive black holes. They released the image of the supermassive black hole at the center of the galaxy M87 in April 2019~\cite{Akiyama:2019cqa} and work is underway to get a similar image of SgrA*, the supermassive black hole at the center of the Galaxy.

The mass of Sgr~A* is about $4 \cdot 10^6$~$M_\odot$ and it is at $d \approx 8$~kpc from us, so its angular size in the sky is roughly
\be
\theta \sim \frac{r_{\rm g}}{d} \sim 0.05 \text{ milliarcseconds} \, .
\ee 
There are three particular conditions that make the observation of the shadow of Sgr~A* achievable. $i)$ The angular resolution of VLBI facilities scales as $\lambda/D$, where $\lambda$ is the electromagnetic radiation wavelength and $D$ is the distance among different stations. For $\lambda < 1$~mm and stations located in different continents ($D > 10^3$~km), it is possible to reach an angular resolution of 0.1~milliarcseconds. $ii)$ The emitting medium around the black hole at the center of the Galaxy is optically thick at wavelengths $\lambda > 1$~mm, but becomes optically thin for $\lambda < 1$~mm. $iii)$ The interstellar scattering at the center of our Galaxy dominates over intrinsic source structures at wavelengths $\lambda > 1$~mm, but becomes subdominant for $\lambda < 1$~mm.

In the case of stellar-mass black holes in our Galaxy, the angular size is 4-5 orders of magnitude smaller. Similar angular resolutions are impossible today, but they may be possible in the future with X-ray interferometric techniques~\cite{r-bh-maxim,Uttley:2019ngm}.


\section{Astrophysical jets \label{app-2}}

A very exciting phenomenon observed in nature are astrophysical jets. Jets are collimated streams of matter emerging from a extraterrestrial object. They are usually highly ionized and relativistic, and beamed in the direction of the rotation axis. They are a common feature of several astrophysical objects, including protostars, stars, neutron stars, and black holes. Jets are observed both from stellar-mass black holes in X-ray binaries and supermassive black holes in galactic nuclei, see e.g.~\cite{r-bh-jets-mirabel,r-bh-jets-fbg04,r-bh-jets-agn}.

\subsection{Theory of jets}

The two most popular mechanisms for the formation of black hole jets are the Blandford-Znajek model~\cite{r-bh-jets-bz} and the Blandford-Payne model~\cite{r-bh-jets-bp}, both with a number of variants and extensions. There are also proposals of hybrid models, in which the two mechanisms can coexist~\cite{r-bh-jets-meier}.

In the {\it Blandford-Znajek scenario}, magnetic fields thread the black hole horizon and can extract the rotational energy of the compact object via some version of the Penrose process~\cite{r-bh-jets-bz,r-bh-jets-penrose}. This mechanism exploits the existence of the ergoregion. However, strictly speaking, the extraction of the rotational energy of a compact object may be possible even in the case of neutron stars in the presence of magnetic fields anchored on the surface of the body. The paper by Blandford and Znajek derived the jet power $P_{\rm BZ}$ perturbatively, for slowly rotating black holes. In that case, one finds $P_{\rm BZ} \propto a^2_*$. A more detailed analysis provides the following formula~\cite{r-bh-jets-formula}
\be
P_{\rm BZ} = \frac{\kappa}{16 \pi} \Phi_{\rm B}^2 \Omega_{\rm H}^2 f(\Omega_{\rm H}) \, ,
\ee
where $\kappa$ is a constant that depends on the magnetic field configuration, $\Omega_{\rm H}$ is the angular frequency at the black hole horizon and reads
\be
\Omega_{\rm H} = \frac{c a_*}{2 r_{\rm H}} 
= \frac{c}{2 r_{\rm g}} \frac{a_*}{1 + \sqrt{1 - a_*^2}} \, ,
\ee
$\Phi_{\rm B}$ is the magnetic flux threading the black hole horizon, and $f(\Omega_{\rm H})$ is a dimensionless function that takes into account higher order terms in $\Omega_{\rm H}$
\be
f(\Omega_{\rm H}) \approx 1 + c_1 \Omega_{\rm H}^2 + c_2 \Omega_{\rm H}^4 + ... \, ,
\ee
where $\{ c_i \}$ are numerical coefficients and this last formula assumes units in which $M = c = 1$. For example, for a black hole with a thin accretion disk, $c_1 = 1.38$ and $c_2 = -9.2$~\cite{r-bh-jets-formula}.

In the {\it Blandford-Payne model}, magnetic fields thread the accretion disk, corrotating with it~\cite{r-bh-jets-bp}. Now the energy is provided by the gravitational potential energy of the accretion flow. The power of the jet can be written as
\be
P_{\rm BZ} \sim \varepsilon \, L \, 
\ln \left(\frac{r_{\rm out}}{r_{\rm in}}\right) \, ,
\ee
where $\varepsilon$ is the efficiency of the transformation of the binding energy of the accreting matter into jet power at the inner radius of the disk $r_{\rm in}$, $r_{\rm out}$ is the outer radius of the disk, and $L$ is the accretion luminosity.

\subsection{Observations of jets}

We will discuss the observational aspects of astrophysical jets in the cases of black hole binaries and active galactic nuclei separately.   

In the case of black hole binaries, we observe two kinds of jets~\cite{r-bh-jets-fbg04}. {\it Steady jets} manifest when a source is in the hard state. The jet is steady, typically not very relativistic, and may extend up to a few tens of AU. {\it Transient jets} are instead observed when a source switches from the hard to the soft state and crosses the ``jet line'' (see Fig.~\ref{f-hid} and Sec.~\ref{s2-spectra}). These pc-scale jets appears as blobs of plasma emitting mainly in the radio band, and are relativistic. They have features similar to the kpc-scale jets observed in AGNs and for this reason the black hole binaries producing transient jets are also called microquasars~\cite{r-bh-jets-mirabel}.

If the mechanism responsible for the formation of jets were the Blandford-Znajek model, one may expect a correlation between black hole spin measurements and estimates of the jet power. Such a correlation has been found in some studies~\cite{r-bh-jets-nar-mcc1}, while other studies did not find any correlation~\cite{r-bh-jets-f1}. Presently, this is a controversial issue~\cite{r-bh-jets-f2,r-bh-cfm4}. Both studies are based on a small number of data with large uncertainty. Future observations are expected to provide a conclusive answer to this issue~\cite{r-bh-cfm4}.

\begin{figure}[b]
\vspace{0.5cm}
\begin{center}
\includegraphics[width=10.0cm]{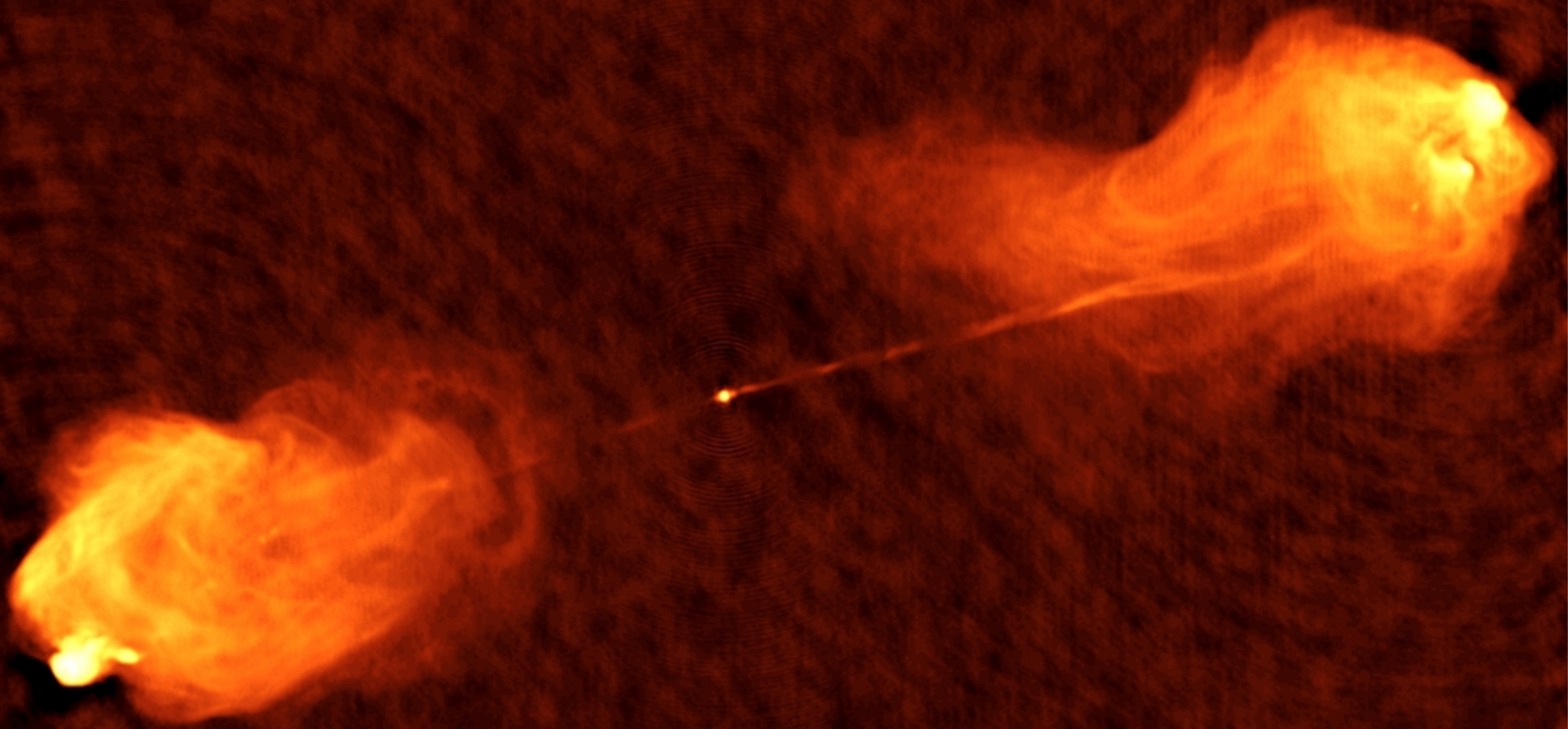}
\end{center}
\vspace{0.3cm}
\caption{Radio image of Cygnus~A. The bright dot at the center is the location of the supermassive black hole, where the two relativistic jets are generated. The jets are stopped by the intergalactic medium, forming two giant lobes. Image courtesy of NRAO/AUI.}
\label{f-cyga}
\end{figure}

In the case of AGNs, only a small fraction of them, around 10\%, exhibit relativistic, kpc-scale jets. One of the most spectacular examples is Cygnus~A (see Fig.~\ref{f-cyga}). Radio images of this object show two highly collimated jets from the very center of the galaxy, where its supermassive black hole is supposed to be located. The two jets extend well outside the galaxy, for hundreds of kpc.  
For AGNs with an accretion luminosity above 1\% of the Eddington limit, the most natural interpretation is that their jets are the counterpart of the transient jets in black hole binaries. This conclusion may be supported by the consideration that microquasars show intermittent jets for a few percent of the time, which is similar to the fraction of radio-loud AGNs~\cite{r-bh-jets-nipoti}. The time scale of these systems is proportional to their mass, so intermittent jets in black hole binaries look like persistent jets in AGNs. However, in the case of AGNs with a luminosity below 1\% of the Eddington limit, this explanation cannot work: black hole binaries with a low accretion luminosity are all radio-loud.  

Jets dominate the spectrum of AGNs at radio frequencies. There are apparently two distinct populations of AGNs: radio-loud AGNs and radio-quiet AGNs (see Figs.~\ref{f-agn1} and~\ref{f-agn2}). This classification is particularly evident when the optical luminosity and radio luminosity are plotted. For the same optical luminosity, radio-loud AGNs have a radio luminosity 3-4 orders of magnitude higher than that of radio-quiet AGNs. These two populations seem to follow two different tracks with a gap between them. The origin of this radio-quiet/radio-loud dichotomy is not understood~\cite{r-bh-jets-dichotomy}.
One popular interpretation is that the dichotomy is determined by the black hole spin. When the accretion luminosity is low, it turns out that radio-loud AGNs are in elliptical galaxies, while radio-quiet AGNs are mainly in spiral galaxies. Galaxies with different morphology have likely a different merger and accretion history. This, in turn, may have produced two populations of black holes, with high and low values of spin~\cite{r-bh-jets-marta07}. A difference in radio luminosity of 3-4 orders of magnitude between the two populations is impossible to explain if the jet power is proportional to $\Omega_{\rm H}^2$, but in the case of thick disks the jet power may scale as $\Omega_{\rm H}^6$~\cite{r-bh-jets-formula}. Another interpretation is to doubt the existence of this dichotomy, attributing it to observational bias~\cite{r-bh-dicho}.

If jets are powered by the rotational energy of the accreting compact object, it is possible to extract energy and have an accretion efficiency $\eta > 1$. Some observations indicate that some AGNs may have $\eta > 1$~\cite{r-bh-jets-eta1a,r-bh-jets-eta1b}. If these measurements are correct, the jet is extracting energy from the system, and it is likely that this is the rotational energy of the black hole; some version of the Blandford-Znajek mechanism is working. While in the past general-relativistic magnetohydrodynamic simulations have not been able to find high accretion efficiency from jets, more recent simulations have achieved $\eta > 1$~\cite{r-bh-jets-eta-sim}.



\begin{thebibliography}{99}

 \bibitem{r-bh-gw150914} 
  B.~P.~Abbott {\it et al.} [LIGO Scientific and Virgo Collaborations],
  Phys.\ Rev.\ Lett.\  {\bf 116}, 061102 (2016)
  [arXiv:1602.03837 [gr-qc]].    

\bibitem{r-bh-thick-disk2} 
  M.~A.~Abramowicz, M.~Calvani and L.~Nobili,
  Astrophys.\ J.\  {\bf 242}, 772 (1980).

\bibitem{r-bh-slim-disk} 
  M.~A.~Abramowicz, B.~Czerny, J.~P.~Lasota and E.~Szuszkiewicz,
  Astrophys.\ J.\  {\bf 332}, 646 (1988).

\bibitem{r-bh-rev-abram} 
  M.~A.~Abramowicz and P.~C.~Fragile,
  Living Rev.\ Rel.\  {\bf 16}, 1 (2013)
  [arXiv:1104.5499 [astro-ph.HE]].

\bibitem{r-bh-qqq2} 
  M.~A.~Abramowicz and W.~Kluzniak,
  Astron.\ Astrophys.\  {\bf 374}, L19 (2001)
  [astro-ph/0105077].

\bibitem{r-bh-qqq3} 
  M.~A.~Abramowicz, W.~Kluzniak, Z.~Stuchlik and G.~Torok,
  Astron.\ Astrophys.\  {\bf 436}, 1 (2005)
  [astro-ph/0401464].
  
\bibitem{r-bh-rin-thick} 
  M.~A.~Abramowicz and J.~P.~Lasota,
  Acta Astron.\  {\bf 30}, 35 (1980).   

 \bibitem{r-bh-lensing} 
  E.~Agol, M.~Kamionkowski, L.~V.~E.~Koopmans and R.~D.~Blandford,
  Astrophys.\ J.\  {\bf 576}, L131 (2002)
  [astro-ph/0203257].  
  
\bibitem{Akiyama:2019cqa} 
  K.~Akiyama {\it et al.} [Event Horizon Telescope Collaboration],
  Astrophys.\ J.\  {\bf 875}, L1 (2019)
  [arXiv:1906.11238 [astro-ph.GA]].  
  
\bibitem{r-bh-super-e2} 
  M.~Bachetti {\it et al.},
  Nature {\bf 514}, 202
  [arXiv:1410.3590 [astro-ph.HE]].  
  
\bibitem{r-bh-book} 
  C.~Bambi,
  {\it Black Holes: A Laboratory for Testing Strong Gravity}
  (Springer Singapore, 2017),  doi:10.1007/978-981-10-4524-0  
 
\bibitem{r-bh-bardeen73}  
  J.~M.~Bardeen, 
  {\it Timelike and null geodesics in the Kerr metric}
  in ``Black Holes'', ed. C.~DeWitt \& B.~S.~DeWitt,
  (Gordon \& Breach, 1973), pp. 215-239. 
 
\bibitem{r-bh-thin-bp} 
  J.~M.~Bardeen and J.~A.~Petterson,
  Astrophys.\ J.\  {\bf 195}, L65 (1975).
  
\bibitem{r-bh-fuel2} 
  J.~E.~Barnes and L.~Hernquist,
  Astrophys.\ J.\  {\bf 471}, 115 (1996).  
  
\bibitem{r-bh-mass} 
  K.~Belczynski, T.~Bulik, C.~L.~Fryer, A.~Ruiter, J.~S.~Vink and J.~R.~Hurley,
  Astrophys.\ J.\  {\bf 714}, 1217 (2010)
  [arXiv:0904.2784 [astro-ph.SR]].  
  
\bibitem{r-bh-spst} 
  T.~M.~Belloni,
  Lect.\ Notes Phys.\  {\bf 794}, 53 (2010)
  [arXiv:0909.2474 [astro-ph.HE]].  
  
\bibitem{r-bh-thin-berti} 
  E.~Berti and M.~Volonteri,
  Astrophys.\ J.\  {\bf 684}, 822 (2008)
  [arXiv:0802.0025 [astro-ph]].  
  
\bibitem{r-bh-jets-bp} 
  R.~D.~Blandford and D.~G.~Payne,
  Mon.\ Not.\ Roy.\ Astron.\ Soc.\  {\bf 199}, 883 (1982).

\bibitem{r-bh-jets-bz}
  R.~D.~Blandford and R.~L.~Znajek,
  Mon.\ Not.\ Roy.\ Astron.\ Soc.\  {\bf 179}, 433  (1977).  
  
\bibitem{r-bh-bondi1} 
  H.~Bondi,
  Mon.\ Not.\ Roy.\ Astron.\ Soc.\  {\bf 112}, 195 (1952).    
  
\bibitem{r-bh-i2} 
  L.~Brenneman,
  {\it Measuring Supermassive Black Hole Spins in Active Galactic Nuclei}
  (Springer New York, 2013)
  [arXiv:1309.6334 [astro-ph.HE]].  
  
\bibitem{r-bh-mcg63015a} 
  L.~W.~Brenneman and C.~S.~Reynolds,
  Astrophys.\ J.\  {\bf 652}, 1028 (2006)
  [astro-ph/0608502].

\bibitem{r-bh-ngc1365b} 
  L.~W.~Brenneman, G.~Risaliti, M.~Elvis and E.~Nardini,
  Mon.\ Not.\ Roy.\ Astron.\ Soc.\  {\bf 429}, 2662 (2013)
  [arXiv:1212.0772 [astro-ph.HE]].  
  
\bibitem{r-bh-ngc3783} 
  L.~W.~Brenneman {\it et al.},
  Astrophys.\ J.\  {\bf 736}, 103 (2011)
  [arXiv:1104.1172 [astro-ph.HE]].   
  
\bibitem{r-bh-mass2} 
  J.~Casares and P.~G.~Jonker,
  Space Sci.\ Rev.\  {\bf 183}, 223 (2014)
  [arXiv:1311.5118 [astro-ph.HE]].  
  
\bibitem{r-bh-cfm-gou_novamus}
  Z.~Chen, L.~Gou, J.~E.~McClintock, J.~F.~Steiner, J.~Wu, W.~Xu, J.~Orosz and Y.~Xiang,
  arXiv:1601.00615 [astro-ph.HE]. 
  
\bibitem{r-bh-cfm-1652} 
  C.~Y.~Chiang, R.~C.~Reis, D.~J.~Walton and A.~C.~Fabian,
  Mon.\ Not.\ Roy.\ Astron.\ Soc.\  {\bf 425}, 2436 (2012)
  [arXiv:1207.0682 [astro-ph.HE]].    
  
\bibitem{r-bh-ulxs} 
  E.~J.~M.~Colbert and R.~F.~Mushotzky,
  Astrophys.\ J.\  {\bf 519}, 89 (1999)
  [astro-ph/9901023].  
  
\bibitem{r-bh-inter}
  M.~Coleman Miller and E.~J.~M.~Colbert,
  Int.\ J.\ Mod.\ Phys.\  D {\bf 13}, 1 (2004)
  [arXiv:astro-ph/0308402].   
  
\bibitem{r-bh-lamppost} 
  T.~Dauser, J.~Garcia, J.~Wilms, M.~Bock, L.~W.~Brenneman, M.~Falanga, K.~Fukumura and C.~S.~Reynolds,
  Mon.\ Not.\ Roy.\ Astron.\ Soc.\  {\bf 430}, 1694 (2013)
  [arXiv:1301.4922 [astro-ph.HE]].    
  
\bibitem{r-bh-cfm-gs1354} 
  A.~M.~El-Batal {\it et al.},
  Astrophys.\ J.\  {\bf 826}, L12 (2016)
  [arXiv:1607.00343 [astro-ph.HE]].   
  
\bibitem{r-bh-i2003} 
  A.~C.~Fabian, K.~Iwasawa, C.~S.~Reynolds and A.~J.~Young,
  Publ.\ Astron.\ Soc.\ Pac.\  {\bf 112}, 1145 (2000)
  [astro-ph/0004366]. 
  
\bibitem{r-bh-corona--3} 
  A.~C.~Fabian, A.~Lohfink, E.~Kara, M.~L.~Parker, R.~Vasudevan and C.~S.~Reynolds,
  Mon.\ Not.\ Roy.\ Astron.\ Soc.\  {\bf 451}, 4375 (2015)
  [arXiv:1505.07603 [astro-ph.HE]].  
  
\bibitem{r-bh-cfm-cyg3}  
  A.~C.~Fabian {\it et al.},
  Mon.\ Not.\ Roy.\ Astron.\ Soc.\  {\bf 424}, 217 (2012)
  [arXiv:1204.5854 [astro-ph.HE]].   
  
\bibitem{r-bh-shadow} 
  H.~Falcke, F.~Melia and E.~Agol,
  Astrophys.\ J.\  {\bf 528}, L13 (2000)
  [astro-ph/9912263].    
  
\bibitem{r-bh-mass-gap} 
  W.~M.~Farr, N.~Sravan, A.~Cantrell, L.~Kreidberg, C.~D.~Bailyn, I.~Mandel and V.~Kalogera,
  Astrophys.\ J.\  {\bf 741}, 103 (2011)
  [arXiv:1011.1459 [astro-ph.GA]].   
  
\bibitem{r-bh-jets-fbg04} 
  R.~P.~Fender, T.~M.~Belloni and E.~Gallo,
  Mon.\ Not.\ Roy.\ Astron.\ Soc.\  {\bf 355}, 1105 (2004)
  [astro-ph/0409360]. 
 
\bibitem{r-bh-jets-f1} 
  R.~Fender, E.~Gallo and D.~Russell,
  Mon.\ Not.\ Roy.\ Astron.\ Soc.\  {\bf 406}, 1425 (2010)
  [arXiv:1003.5516 [astro-ph.HE]].   
  
\bibitem{r-bh-ferrarese} 
  L.~Ferrarese {\it et al.},
  Astrophys.\ J.\  {\bf 644}, L21 (2006)
  [astro-ph/0603840].  
  
\bibitem{r-bh-thin-fragile1} 
  P.~C.~Fragile, O.~M.~Blaes, P.~Anninois and J.~D.~Salmonson,
  Astrophys.\ J.\  {\bf 668}, 417 (2007)
  [arXiv:0706.4303 [astro-ph]].  
  
\bibitem{r-bh-thin-fragos2010}
  T.~Fragos, M.~Tremmel, E.~Rantsiou and K.~Belczynski,
  Astrophys.\ J.\  {\bf 719}, L79 (2010)
  [arXiv:1001.1107 [astro-ph.HE]].  
  
\bibitem{r-bh-thin-acc-fragos} 
  T.~Fragos and J.~E.~McClintock,
  Astrophys.\ J.\  {\bf 800}, 17 (2015)
  [arXiv:1408.2661 [astro-ph.HE]].    

\bibitem{r-bh-mrk79} 
  L.~C.~Gallo, G.~Miniutti, J.~M.~Miller, L.~W.~Brenneman, A.~C.~Fabian, M.~Guainazzi and C.~S.~Reynolds,
  Mon.\ Not.\ Roy.\ Astron.\ Soc.\  {\bf 411}, 607 (2011)
  [arXiv:1009.2987 [astro-ph.HE]].      

\bibitem{r-bh-amuse} 
  E.~Gallo, T.~Treu, J.~Jacob, J.~H.~Woo, P.~Marshall and R.~Antonucci,
  Astrophys.\ J.\  {\bf 680}, 154 (2008)
  [arXiv:0711.2073 [astro-ph]].  
  
 \bibitem{r-bh-ref} 
  J.~Garcia, T.~Dauser, C.~S.~Reynolds, T.~R.~Kallman, J.~E.~McClintock, J.~Wilms and W.~Eikmann,
  Astrophys.\ J.\  {\bf 768}, 146 (2013)
  [arXiv:1303.2112 [astro-ph.HE]].    
  
\bibitem{r-bh-cfm-gx339c} 
  J.~Garcia {\it et al.},
  Astrophys.\ J.\ {\bf 813}, 84 (2015)
  [arXiv:1505.03607[astro-ph.HE]].   
  
 \bibitem{r-bh-i-1752} 
  J.~A.~Garcia {\it et al.},
  Astrophys.\ J.\  {\bf 864}, 25 (2018)
  [arXiv:1807.01949 [astro-ph.HE]].   
  
\bibitem{r-bh-clu1} 
  K.~Gebhardt, R.~M.~Rich and L.~C.~Ho,
  Astrophys.\ J.\  {\bf 578}, L41 (2002)
  [astro-ph/0209313].  
  
\bibitem{r-bh-clu2} 
  K.~Gebhardt, R.~M.~Rich and L.~C.~Ho,
  Astrophys.\ J.\  {\bf 634}, 1093 (2005)
  [astro-ph/0508251].   
  
\bibitem{r-bh-c-ghez} 
  A.~M.~Ghez, S.~Salim, S.~D.~Hornstein, A.~Tanner, M.~Morris, E.~E.~Becklin and G.~Duchene,
  Astrophys.\ J.\  {\bf 620}, 744 (2005)
  [astro-ph/0306130].
  
\bibitem{r-bh-jets-eta1a} 
  G.~Ghisellini, F.~Tavecchio, L.~Foschini, G.~Ghirlanda, L.~Maraschi and A.~Celotti,
  Mon.\ Not.\ Roy.\ Astron.\ Soc.\  {\bf 402}, 497 (2010)
  [arXiv:0909.0932 [astro-ph.CO]].   
  
\bibitem{r-bh-qpo-smbh} 
  M.~Gierlinski, M.~Middleton, M.~Ward and C.~Done,
  Nature {\bf 455}, 369 (2008)
  [arXiv:0807.1899 [astro-ph]].    
  
\bibitem{r-bh-cfm-lmcx1} 
  L.~Gou, J.~E.~McClintock, J.~Liu, R.~Narayan, J.~F.~Steiner, R.~A.~Remillard, J.~A.~Orosz and S.~W.~Davis,
  Astrophys.\ J.\  {\bf 701}, 1076 (2009)
  [arXiv:0901.0920 [astro-ph.HE]].    
  
\bibitem{r-bh-cfm-62} 
  L.~Gou, J.~E.~McClintock, J.~F.~Steiner, R.~Narayan, A.~G.~Cantrell, C.~D.~Bailyn and J.~A.~Orosz,
  Astrophys.\ J.\  {\bf 718}, L122 (2010)
  [arXiv:1002.2211 [astro-ph.HE]].    
  
\bibitem{r-bh-cfm-cyg1} 
  L.~Gou {\it et al.},
  Astrophys.\ J.\  {\bf 742}, 85 (2011)
  [arXiv:1106.3690 [astro-ph.HE]].  
  
\bibitem{r-bh-cfm-cyg2}
  L.~Gou {\it et al.},
  Astrophys.\ J.\  {\bf 790}, 29 (2014)
  [arXiv:1308.4760 [astro-ph.HE]].    
  
\bibitem{r-bh-corona--1} 
  F.~Haardt and L.~Maraschi,
  Astrophys.\ J.\  {\bf 380}, L51 (1991).       
  
\bibitem{r-bh-sandwich} 
  F.~Haardt and L.~Maraschi,
  Astrophys.\ J.\  {\bf 413}, 507 (1993).   
  
\bibitem{r-bh-gg2} 
  A.~Heger, C.~L.~Fryer, S.~E.~Woosley, N.~Langer and D.~H.~Hartmann,
  Astrophys.\ J.\  {\bf 591}, 288 (2003)
  [astro-ph/0212469].

\bibitem{r-bh-gg1} 
  A.~Heger and S.~E.~Woosley,
  Astrophys.\ J.\  {\bf 567}, 532 (2002)
  [astro-ph/0107037].  
  
\bibitem{r-bh-spst2} 
  J.~Homan and T.~Belloni,
  Astrophys.\ Space Sci.\  {\bf 300}, 107 (2005)
  [astro-ph/0412597].  
  
\bibitem{r-bh-thin-scott} 
  S.~A.~Hughes and R.~D.~Blandford,
  Astrophys.\ J.\  {\bf 585}, L101 (2003)
  [astro-ph/0208484].  
  
\bibitem{r-bh-thin-ingram}
  A.~Ingram, C.~Done and P.~C.~Fragile,
  Mon.\ Not.\ Roy.\ Astron.\ Soc.\  {\bf 397}, L101 (2009).     
  
\bibitem{r-bh-thick-disk1} 
  M.~Jaroszynski, M.~A.~Abramowicz and B.~Paczynski,
  Acta Astron.\  {\bf 30}, 1 (1980).  
  
\bibitem{r-bh-bh2}
  V.~Kalogera and G.~Baym,
  Astrophys.\ J.\  {\bf 470}, L61 (1996).         
  
\bibitem{r-bh-lm2} 
  P.~D.~Kiel and J.~R.~Hurley,
  Mon.\ Not.\ Roy.\ Astron.\ Soc.\  {\bf 369}, 1152 (2006)
  [astro-ph/0605080].  
  
\bibitem{r-bh-thin-acc-king} 
  A.~R.~King and U.~Kolb,
  Mon.\ Not.\ Roy.\ Astron.\ Soc.\  {\bf 305}, 654 (1999)
  [astro-ph/9901296].  
  
\bibitem{r-bh-thin-r_w1} 
  A.~R.~King, S.~H.~Lubow, G.~I.~Ogilvie and J.~E.~Pringle,
  Mon.\ Not.\ Roy.\ Astron.\ Soc.\  {\bf 363}, 49 (2005)
  [astro-ph/0507098].  
  
\bibitem{r-bh-thin-chaotic} 
  A.~R.~King and J.~E.~Pringle,
  Mon.\ Not.\ Roy.\ Astron.\ Soc.\  {\bf 373}, L93 (2006)
  [astro-ph/0609598].  
  
\bibitem{r-bh-King:2014sja} 
  A.~L.~King {\it et al.},
  Astrophys.\ J.\  {\bf 784}, L2 (2014)
  [arXiv:1401.3646 [astro-ph.HE]].  
  
\bibitem{r-bh-cfm-gx339} 
  M.~Kolehmainen and C.~Done,
  Mon.\ Not.\ Roy.\ Astron.\ Soc.\  {\bf 406}, 2206 (2010)
  [arXiv:0911.3281 [astro-ph.HE]].    
  
\bibitem{r-bh-k-r} 
  J.~Kormendy and D.~Richstone,
  Ann.\ Rev.\ Astron.\ Astrophys.\  {\bf 33}, 581 (1995).  
  
\bibitem{r-bh-edd1b} 
  A.~K.~Kulkarni {\it et al.},
  Mon.\ Not.\ Roy.\ Astron.\ Soc.\  {\bf 414}, 1183 (2011)
  [arXiv:1102.0010 [astro-ph.HE]].  
  
\bibitem{r-bh-thin-bp2}
  S.~Kumar and J.~E.~Pringle,
  Mon.\ Not.\ Roy.\ Astron.\ Soc.\  {\bf 213}, 435 (1985).  
  
\bibitem{r-bh-lasota4} 
  J.~P.~Lasota,
  {\it Black hole accretion discs},
  doi:10.1007/978-3-662-52859-4\_1
  arXiv:1505.02172 [astro-ph.HE].   

\bibitem{r-bh-bh3} 
  J.~M.~Lattimer,
  Ann.\ Rev.\ Nucl.\ Part.\ Sci.\  {\bf 62}, 485 (2012)
  [arXiv:1305.3510 [nucl-th]].

\bibitem{r-bh-cfm2} 
  L.~X.~Li, E.~R.~Zimmerman, R.~Narayan and J.~E.~McClintock,
  Astrophys.\ J.\ Suppl.\  {\bf 157}, 335 (2005)
  [astro-ph/0411583]. 
  
\bibitem{r-bh-cfm-liu08} 
  J.~Liu, J.~McClintock, R.~Narayan, S.~Davis and J.~Orosz,
  Astrophys.\ J.\  {\bf 679}, L37 (2008) [Erratum: Astrophys.\ J.\  {\bf 719}, L109 (2010)]
  [arXiv:0803.1834 [astro-ph]].   

\bibitem{r-bh-thin-r_w2} 
  G.~Lodato and J.~E.~Pringle,
  Mon.\ Not.\ Roy.\ Astron.\ Soc.\  {\bf 368}, 1196 (2006)
  [astro-ph/0602306]. 
  
\bibitem{r-bh-3c120} 
  A.~M.~Lohfink {\it et al.},
  Astrophys.\ J.\  {\bf 772}, 83 (2013)
  [arXiv:1305.4937 [astro-ph.HE]].  
  
\bibitem{r-bh-fairall9} 
  A.~Lohfink {\it et al.},
  Astrophys.\ J.\  {\bf 821}, 11 (2016)
  [arXiv:1602.05589 [astro-ph.GA]].  
  
\bibitem{r-bh-super-e1} 
  P.~Madau, F.~Haardt and M.~Dotti,
  Astrophys.\ J.\  {\bf 784}, L38 (2014)
  [arXiv:1402.6995 [astro-ph.CO]].

\bibitem{r-bh-maoz} 
  E.~Maoz,
  Astrophys.\ J.\  {\bf 494}, L181 (1998)
  [astro-ph/9710309].  
  
\bibitem{r-bh-mcg63015b} 
  A.~Marinucci {\it et al.},
  Astrophys.\ J.\  {\bf 787}, 83 (2014)
  [arXiv:1404.3561 [astro-ph.HE]].     
  
\bibitem{r-bh-thin-rm07m} 
  R.~G.~Martin, J.~E.~Pringle and C.~A.~Tout,
  Mon.\ Not.\ Roy.\ Astron.\ Soc.\  {\bf 381}, 1617 (2007)
  [arXiv:0708.2034 [astro-ph]].

\bibitem{r-bh-thin-rm08m} 
  R.~G.~Martin, C.~A.~Tout and J.~E.~Pringle,
  Mon.\ Not.\ Roy.\ Astron.\ Soc.\  {\bf 387}, 188 (2008)
  [arXiv:0802.3912 [astro-ph]].  
  
\bibitem{r-bh-fuel3} 
  L.~Mayer, S.~Kazantzidis, P.~Madau, M.~Colpi, T.~R.~Quinn and J.~Wadsley,
  Science {\bf 316}, 1874 (2007)
  [arXiv:0706.1562 [astro-ph]]. 

\bibitem{r-bh-cfm4} 
  J.~E.~McClintock, R.~Narayan and J.~F.~Steiner,
  Space Sci.\ Rev.\  {\bf 183}, 295 (2014)
  [arXiv:1303.1583 [astro-ph.HE]].  
  
\bibitem{r-bh-cfm-1915} 
  J.~E.~McClintock, R.~Shafee, R.~Narayan, R.~A.~Remillard, S.~W.~Davis and L.~X.~Li,
  Astrophys.\ J.\  {\bf 652}, 518 (2006)
  [astro-ph/0606076].    
  
\bibitem{r-bh-cfm3} 
  J.~E.~McClintock {\it et al.},
  Class.\ Quant.\ Grav.\  {\bf 28}, 114009 (2011)
  [arXiv:1101.0811 [astro-ph.HE]].  
  
\bibitem{r-bh-jets-eta1b} 
  B.~R.~McNamara, M.~Rohanizadegan and P.~E.~J.~Nulsen,
  Astrophys.\ J.\  {\bf 727}, 39 (2011)
  [arXiv:1007.1227 [astro-ph.CO]].    
  
\bibitem{r-bh-jets-meier} 
  D.~L.~Meier,
  Astrophys.\ J.\  {\bf 548}, L9 (2001)
  [astro-ph/0010231].    
  
\bibitem{r-bh-cfm-m31} 
  M.~Middleton, J.~Miller-Jones and R.~Fender,
  Mon.\ Not.\ Roy.\ Astron.\ Soc.\  {\bf 439}, 1740 (2014)
  [arXiv:1401.1829 [astro-ph.HE]].
  
\bibitem{r-bh-cfm-1915b} 
  J.~M.~Miller {\it et al.},
  Astrophys.\ J.\  {\bf 775}, L45 (2013)
  [arXiv:1308.4669 [astro-ph.HE]]. 
  
\bibitem{r-bh-Miller:2014sla} 
  J.~M.~Miller {\it et al.},
  Astrophys.\ J.\  {\bf 799}, L6 (2015)
  [arXiv:1411.1921 [astro-ph.HE]].  
  
\bibitem{r-bh-iron-maxi15b} 
  J.~M.~Miller {\it et al.},
  Astrophys.\ J.\  {\bf 860}, L28 (2018)
  [arXiv:1806.04115 [astro-ph.HE]].   
  
\bibitem{r-bh-swift2127} 
  G.~Miniutti, F.~Panessa, A.~De Rosa, A.~C.~Fabian, A.~Malizia, M.~Molina, J.~M.~Miller and S.~Vaughan,
  Mon.\ Not.\ Roy.\ Astron.\ Soc.\  {\bf 398}, 255 (2009)
  [arXiv:0905.2891 [astro-ph.HE]].  
  
\bibitem{r-bh-jets-mirabel} 
  I.~F.~Mirabel and L.~F.~Rodriguez,
  Ann.\ Rev.\ Astron.\ Astrophys.\  {\bf 37}, 409 (1999)
  [astro-ph/9902062].  
  
\bibitem{r-bh-multicolor} 
  K.~Mitsuda {\it et al.},
  Publ.\ Astron.\ Soc.\ Jap.\  {\bf 36}, 741 (1984).  
\bibitem{r-bh-Mori:2019iwz} 
  K.~Mori {\it et al.},
  arXiv:1910.03459 [astro-ph.HE].

\bibitem{r-bh-gro-motta} 
  S.~E.~Motta, T.~M.~Belloni, L.~Stella, T.~Muñoz-Darias and R.~Fender,
  Mon.\ Not.\ Roy.\ Astron.\ Soc.\  {\bf 437}, 2554 (2014)
  [arXiv:1309.3652 [astro-ph.HE]].  
  
\bibitem{r-bh-motta14} 
  S.~E.~Motta, T.~Munoz-Darias, A.~Sanna, R.~Fender, T.~Belloni and L.~Stella,
  Mon.\ Not.\ Roy.\ Astron.\ Soc.\  {\bf 439}, 65 (2014)
  [arXiv:1312.3114 [astro-ph.HE]].  
  
\bibitem{r-bh-jets-nar-mcc1} 
  R.~Narayan and J.~E.~McClintock,
  Mon.\ Not.\ Roy.\ Astron.\ Soc.\  {\bf 419}, L69 (2012)
  [arXiv:1112.0569 [astro-ph.HE]].  
  
\bibitem{r-bh-adaf-disk-1} 
  R.~Narayan and I.~Yi,
  Astrophys.\ J.\  {\bf 428}, L13 (1994)
  [astro-ph/9403052].

\bibitem{r-bh-adaf-disk-2} 
  R.~Narayan and I.~Yi,
  Astrophys.\ J.\  {\bf 452}, 710 (1995)
  [astro-ph/9411059].  
  
\bibitem{r-bh-ark120} 
  E.~Nardini, A.~C.~Fabian, R.~C.~Reis and D.~J.~Walton,
  Mon.\ Not.\ Roy.\ Astron.\ Soc.\  {\bf 410}, 1251 (2011)
  [arXiv:1008.2157 [astro-ph.HE]].

\bibitem{r-bh-jets-nipoti} 
  C.~Nipoti, K.~M.~Blundell and J.~Binney,
  Mon.\ Not.\ Roy.\ Astron.\ Soc.\  {\bf 361}, 633 (2005)
  [astro-ph/0505280].     
  
\bibitem{r-bh-ntm}
  I.~D.~Novikov and K.~S.~Thorne,
  {\it Astrophysics and black holes}, in {\it Black Holes}, edited by C.~De~Witt and B.~De~Witt
  (Gordon and Breach, New York, New York, 1973). 
  
\bibitem{r-bh-mine} 
  K.~Ohsuga and S.~Mineshige,
  Astrophys.\ J.\  {\bf 736}, 2 (2011)
  [arXiv:1105.5474 [astro-ph.HE]].    
  
\bibitem{r-bh-ntm2}
  D.~N.~Page and K.~S.~Thorne,
  Astrophys.\ J.\  {\bf 191}, 499 (1974).  
  
\bibitem{r-bh-mrk335} 
  M.~L.~Parker {\it et al.},
  Mon.\ Not.\ Roy.\ Astron.\ Soc.\  {\bf 443}, no. 2, 1723 (2014)
  [arXiv:1407.8223 [astro-ph.HE]].  

\bibitem{r-bh-cfm-cyg5} 
  M.~L.~Parker {\it et al.},
  Astrophys.\ J.\  {\bf 808}, 9 (2015)
  [arXiv:1506.00007 [astro-ph.HE]].
  
\bibitem{r-bh-cfm-gx339d} 
  M.~L.~Parker {\it et al.},
  Astrophys.\ J.\  {\bf 821}, L6 (2016)
  [arXiv:1603.03777 [astro-ph.HE]].    

\bibitem{r-bh-i-qpo} 
  D.~R.~Pasham, T.~E.~Strohmayer and R.~F.~Mushotzky,
  Nature {\bf 513}, 74 (2014)
  [arXiv:1501.03180 [astro-ph.HE]].   
  
\bibitem{r-bh-ngc4051} 
  A.~R.~Patrick, J.~N.~Reeves, D.~Porquet, A.~G.~Markowitz, V.~Braito and A.~P.~Lobban,
  Mon.\ Not.\ Roy.\ Astron.\ Soc.\  {\bf 426}, 2522 (2012)
  [arXiv:1208.1150 [astro-ph.HE]].   
  
\bibitem{r-bh-edd1a} 
  R.~F.~Penna, J.~C.~McKinney, R.~Narayan, A.~Tchekhovskoy, R.~Shafee and J.~E.~McClintock,
  Mon.\ Not.\ Roy.\ Astron.\ Soc.\  {\bf 408}, 752 (2010)
  [arXiv:1003.0966 [astro-ph.HE]].  

\bibitem{r-bh-jets-penrose} 
  R.~Penrose,
  Riv.\ Nuovo Cim.\  {\bf 1}, 252 (1969)
  [Gen.\ Rel.\ Grav.\  {\bf 34}, 1141 (2002)].   

\bibitem{r-bh-postman} 
  M.~Postman {\it et al.},
  Astrophys.\ J.\  {\bf 756}, 159 (2012)
  [arXiv:1205.3839 [astro-ph.CO]].
  
\bibitem{r-bh-Qin:2018sxk} 
  Y.~Qin, P.~Marchant, T.~Fragos, G.~Meynet and V.~Kalogera,
  arXiv:1810.13016 [astro-ph.SR].  
  
\bibitem{r-bh-thin-acc-reid} 
  M.~J.~Reid, J.~E.~McClintock, J.~F.~Steiner, D.~Steeghs, R.~A.~Remillard, V.~Dhawan and R.~Narayan,
  Astrophys.\ J.\  {\bf 796}, 2 (2014)
  [arXiv:1409.2453 [astro-ph.GA]].  
  
\bibitem{r-bh-cfm-swift}   
  R.~C.~Reis, A.~C.~Fabian, R.~R.~Ross and J.~M.~Miller,
  Mon.\ Not.\ Roy.\ Astron.\ Soc.\  {\bf 395}, 1257 (2009).
  
\bibitem{r-bh-cfm-gx339b} 
  R.~C.~Reis, A.~C.~Fabian, R.~Ross, G.~Miniutti, J.~M.~Miller and C.~Reynolds,
  Mon.\ Not.\ Roy.\ Astron.\ Soc.\  {\bf 387}, 1489 (2008)
  [arXiv:0804.0238 [astro-ph]].   
  
\bibitem{r-bh-corona--2} 
  R.~C.~Reis and J.~M.~Miller,
  Astrophys.\ J.\  {\bf 769}, L7 (2013)
  [arXiv:1304.4947 [astro-ph.HE]].   
  
\bibitem{r-bh-cfm-maxi} 
  R.~C.~Reis, J.~M.~Miller, M.~T.~Reynolds, A.~C.~Fabian and D.~J.~Walton,
  Astrophys.\ J.\  {\bf 751}, 34 (2012)
  [arXiv:1111.6665 [astro-ph.HE]].   
  
\bibitem{r-bh-cfm-1752} 
  R.~C.~Reis {\it et al.},
  Mon.\ Not.\ Roy.\ Astron.\ Soc.\  {\bf 410}, 2497 (2011)
  [arXiv:1009.1154 [astro-ph.HE]].     

\bibitem{r-bh-re-mc} 
  R.~A.~Remillard and J.~E.~McClintock,
  Ann.\ Rev.\ Astron.\ Astrophys.\  {\bf 44}, 49 (2006)
  [astro-ph/0606352]. 
  
\bibitem{r-bh-i1} 
  C.~S.~Reynolds,
  Space Sci.\ Rev.\  {\bf 183}, 277 (2014)
  [arXiv:1302.3260 [astro-ph.HE]].   
  
\bibitem{r-bh-bh1}
  C.~E.~Rhoades and R.~Ruffini,
  Phys.\ Rev.\ Lett.\  {\bf 32}, 324 (1974).  

\bibitem{Riaz:2019kat} 
  S.~Riaz, D.~Ayzenberg, C.~Bambi and S.~Nampalliwar,
  arXiv:1911.06605 [astro-ph.HE].

\bibitem{r-bh-ngc1365a} 
  G.~Risaliti {\it et al.},
  Nature {\bf 494}, 449 (2013)
  [arXiv:1302.7002 [astro-ph.HE]].  

\bibitem{r-bh-jets-f2} 
  D.~M.~Russell, E.~Gallo and R.~P.~Fender,
  Mon.\ Not.\ Roy.\ Astron.\ Soc.\  {\bf 431}, 405 (2013)
  [arXiv:1301.6771 [astro-ph.HE]].       
  
\bibitem{r-bh-cfm-sh06} 
  R.~Shafee, J.~E.~McClintock, R.~Narayan, S.~W.~Davis, L.~X.~Li and R.~A.~Remillard,
  Astrophys.\ J.\  {\bf 636}, L113 (2006)
  [astro-ph/0508302].      

\bibitem{r-bh-thin-disk} 
  N.~I.~Shakura and R.~A.~Sunyaev,
  Astron.\ Astrophys.\  {\bf 24}, 337 (1973).

\bibitem{r-bh-bondi2} 
  S.~L.~Shapiro and S.~A.~Teukolsky,
  {\it Black holes, white dwarfs, and neutron stars: The physics of compact objects}
  (Wiley-VCH, 1983).    
  
\bibitem{r-bh-fuel1} 
  I.~Shlosman, M.~C.~Begelman and J.~Frank,
  Nature {\bf 345}, 679 (1990).
  
\bibitem{r-bh-jets-dichotomy} 
  M.~Sikora, L.~Stawarz and J.~P.~Lasota,
  Astrophys.\ J.\  {\bf 658}, 815 (2007)
  [astro-ph/0604095].       
  
\bibitem{r-bh-dicho} 
  J.~Singal, V.~Petrosian, A.~Lawrence and L.~Stawarz,
  Astrophys.\ J.\  {\bf 743}, 104 (2011)
  [arXiv:1101.2930 [astro-ph.CO]].  
  
\bibitem{r-bh-gg3} 
  M.~Spera, M.~Mapelli and A.~Bressan,
  Mon.\ Not.\ Roy.\ Astron.\ Soc.\  {\bf 451}, no. 4, 4086 (2015)
  [arXiv:1505.05201 [astro-ph.SR]].

\bibitem{r-bh-thin-6849} 
  J.~F.~Steiner and J.~E.~McClintock,
  Astrophys.\ J.\  {\bf 745}, 136 (2012)
  [arXiv:1110.6849 [astro-ph.HE]].   
  
\bibitem{r-bh-cfm-lmcx3} 
  J.~F.~Steiner, J.~E.~McClintock, J.~A.~Orosz, R.~A.~Remillard, C.~D.~Bailyn, M.~Kolehmainen and O.~Straub,
  Astrophys.\ J.\  {\bf 793}, L29 (2014)
  [arXiv:1402.0148 [astro-ph.HE]].    
  
\bibitem{r-bh-thin-constant} 
  J.~F.~Steiner, J.~E.~McClintock, R.~A.~Remillard, L.~Gou, S.~Yamada and R.~Narayan,
  Astrophys.\ J.\  {\bf 718}, L117 (2010)
  [arXiv:1006.5729 [astro-ph.HE]].  
  
\bibitem{r-bh-cfm-h1743} 
  J.~F.~Steiner, J.~E.~McClintock and M.~J.~Reid,
  Astrophys.\ J.\  {\bf 745}, L7 (2012)
  [arXiv:1111.2388 [astro-ph.HE]].    
  
 \bibitem{r-bh-cfm-st16}
  J.~F.~Steiner, D.~J.~Walton, J.~A.~Garcia, J.~E.~McClintock, S.~G.~T.~Laycock, M.~J.~Middleton, R.~Barnard and K.~K.~Madsen,
  arXiv:1512.03414 [astro-ph.HE].    
  
\bibitem{r-bh-cfm-xte} 
  J.~F.~Steiner {\it et al.},
  Mon.\ Not.\ Roy.\ Astron.\ Soc.\  {\bf 416}, 941 (2011)
  [arXiv:1010.1013 [astro-ph.HE]].  
  
\bibitem{r-bh-cfm-lmcx1b}   
  J.~F.~Steiner {\it et al.},
  Mon.\ Not.\ Roy.\ Astron.\ Soc.\  {\bf 427}, 2552 (2012)
  [arXiv:1209.3269 [astro-ph.HE]].   
  
\bibitem{r-bh-qqq1} 
  L.~Stella, M.~Vietri and S.~Morsink,
  Astrophys.\ J.\  {\bf 524}, L63 (1999)
  [astro-ph/9907346].  

\bibitem{r-bh-shangyu} 
  S.~Sun, M.~Guainazzi, Q.~Ni, J.~Wang, C.~Qian, F.~Shi, Y.~Wang and C.~Bambi,
  Mon.\ Not.\ Roy.\ Astron.\ Soc.\  {\bf 478}, 1900 (2018)
  [arXiv:1704.03716 [astro-ph.HE]].  

\bibitem{r-bh-compton1} 
  R.~A.~Sunyaev and J.~Truemper,
  Nature {\bf 279}, 506 (1979).

\bibitem{r-bh-compton2} 
  R.~A.~Sunyaev and L.~G.~Titarchuk,
  Astron.\ Astrophys.\  {\bf 86}, 121 (1980).

\bibitem{r-bh-iras521} 
  Y.~Tan, J.~Wang, X.~Shu and Y.~Zhou,
  Astrophys.\ J.\  {\bf 747}, L11 (2012)
  [arXiv:1202.0400 [astro-ph.HE]].  
  
\bibitem{r-bh-Tao:2019yhu} 
  L.~Tao, J.~A.~Tomsick, J.~Qu, S.~Zhang, S.~Zhang and Q.~Bu,
  arXiv:1910.11979 [astro-ph.HE].  

\bibitem{r-bh-jets-formula} 
  A.~Tchekhovskoy, R.~Narayan and J.~C.~McKinney,
  Astrophys.\ J.\  {\bf 711}, 50 (2010)
  [arXiv:0911.2228 [astro-ph.HE]].
  
\bibitem{r-bh-jets-eta-sim} 
  A.~Tchekhovskoy, R.~Narayan and J.~C.~McKinney,
  Mon.\ Not.\ Roy.\ Astron.\ Soc.\  {\bf 418}, L79 (2011)
  [arXiv:1108.0412 [astro-ph.HE]].  
  
\bibitem{r-bh-thorne}
  K.~S.~Thorne,
  Astrophys.\ J.\  {\bf 191}, 507 (1974).  

\bibitem{r-bh-bhnum} 
  F.~X.~Timmes, S.~E.~Woosley and T.~A.~Weaver,
  Astrophys.\ J.\  {\bf 457}, 834 (1996)
  [astro-ph/9510136].
     
\bibitem{r-bh-cfm-cyg4} 
  J.~A.~Tomsick {\it et al.},
  Astrophys.\ J.\  {\bf 780}, 78 (2014)
  [arXiv:1310.3830 [astro-ph.HE]].    

\bibitem{r-bh-Urry:1995mg} 
  C.~M.~Urry and P.~Padovani,
  Publ.\ Astron.\ Soc.\ Pac.\  {\bf 107}, 803 (1995)
  [astro-ph/9506063]. 

\bibitem{Uttley:2019ngm} 
  P.~Uttley {\it et al.},
  arXiv:1908.03144 [astro-ph.HE].

\bibitem{r-bh-bhnum0} 
  E.~P.~J.~van den Heuvel,
  {\it Endpoints of stellar evolution: The incidence of stellar mass black holes in the galaxy},
  in ``Environment Observation and Climate Modelling Through International Space Projects'', 29 (1992).

\bibitem{r-bh-vdk} 
  M.~van der Klis,
  astro-ph/0410551.   
  
\bibitem{r-bh-vasudevan} 
  R.~V.~Vasudevan, A.~C.~Fabian, C.~S.~Reynolds, J.~Aird, T.~Dauser and L.~C.~Gallo,
  Mon.\ Not.\ Roy.\ Astron.\ Soc.\  {\bf 458}, 2012 (2016)
  [arXiv:1506.01027 [astro-ph.HE]].  

\bibitem{r-bh-mv} 
  M.~Volonteri,
  Astron.\ Astrophys.\ Rev.\  {\bf 18}, 279 (2010)
  [arXiv:1003.4404 [astro-ph.CO]]. 

\bibitem{r-bh-mv07b} 
  M.~Volonteri, F.~Haardt and K.~Gultekin,
  Mon.\ Not.\ Roy.\ Astron.\ Soc.\  {\bf 384}, 1387 (2008)
  [arXiv:0710.5770 [astro-ph]].

\bibitem{r-bh-mv07a} 
  M.~Volonteri, G.~Lodato and P.~Natarajan,
  Mon.\ Not.\ Roy.\ Astron.\ Soc.\  {\bf 383}, 1079 (2008)
  [arXiv:0709.0529 [astro-ph]].
  
\bibitem{r-bh-jets-marta07} 
  M.~Volonteri, M.~Sikora and J.~P.~Lasota,
  Astrophys.\ J.\  {\bf 667}, 704 (2007)
  [arXiv:0706.3900 [astro-ph]].    
  
\bibitem{r-bh-cfm-1650} 
  D.~J.~Walton, R.~C.~Reis, E.~M.~Cackett, A.~C.~Fabian and J.~M.~Miller,
  Mon.\ Not.\ Roy.\ Astron.\ Soc.\  {\bf 422}, 2510 (2012)
  [arXiv:1202.5193 [astro-ph.HE]].  

\bibitem{r-bh-suzaku} 
  D.~J.~Walton, E.~Nardini, A.~C.~Fabian, L.~C.~Gallo and R.~C.~Reis,
  Mon.\ Not.\ Roy.\ Astron.\ Soc.\  {\bf 428}, 2901 (2013)
  [arXiv:1210.4593 [astro-ph.HE]].     

\bibitem{r-bh-cfm-cyg6} 
  D.~J.~Walton {\it et al.},
  Astrophys.\ J.\  {\bf 826}, 87 (2016)
  [arXiv:1605.03966 [astro-ph.HE]].   

\bibitem{r-bh-Walton:2016fso} 
  D.~J.~Walton {\it et al.},
  Astrophys.\ J.\  {\bf 839}, 110 (2017)
  [arXiv:1609.01293 [astro-ph.HE]].

\bibitem{r-bh-thin-agn-soltan} 
  J.~M.~Wang, Y.~M.~Chen, L.~C.~Ho and R.~J.~McLure,
  Astrophys.\ J.\  {\bf 642}, L111 (2006)
  [astro-ph/0603813].

\bibitem{r-bh-maxim} 
  N.~White,
  Nature {\bf 407}, 146 (2000).

\bibitem{r-bh-thin-acc-rem1} 
  S.~E.~Woosley and J.~S.~Bloom,
  Ann.\ Rev.\ Astron.\ Astrophys.\  {\bf 44}, 507 (2006)
  [astro-ph/0609142].  
  
\bibitem{r-bh-iron-maxi15a} 
  Y.~Xu {\it et al.},
  Astrophys.\ J.\  {\bf 852}, L34 (2018)
  [arXiv:1711.01346 [astro-ph.HE]].
  
\bibitem{r-bh-iron-swift-xu} 
  Y.~Xu {\it et al.},
  arXiv:1805.07705 [astro-ph.HE].    
  
\bibitem{r-bh-thin-acc-rem2} 
  S.~C.~Yoon, N.~Langer and C.~Norman,
  Astron.\ Astrophys.\  {\bf 460}, 199 (2006)
  [astro-ph/0606637].
  
\bibitem{r-bh-lm1} 
  L.~R.~Yungelson, J.-P.~Lasota, G.~Nelemans, G.~Dubus, E.~P.~J.~van den Heuvel, J.~Dewi and S.~Portegies Zwart,
  Astron.\ Astrophys.\  {\bf 454}, 559 (2006)
  [astro-ph/0604434].  

\bibitem{r-bh-jets-agn} 
  J.~A.~Zensus,
  Ann.\ Rev.\ Astron.\ Astrophys.\  {\bf 35}, 607 (1997).

\bibitem{r-bh-cfm1} 
  S.~N.~Zhang, W.~Cui and W.~Chen,
  Astrophys.\ J.\  {\bf 482}, L155 (1997)
  [astro-ph/9704072]. 

\bibitem{r-bh-thin-fragile2} 
  V.~V.~Zhuravlev, P.~B.~Ivanov, P.~C.~Fragile and D.~M.~Teixeira,
  Astrophys.\ J.\  {\bf 796}, 104 (2014)
  [arXiv:1406.5515 [astro-ph.HE]].  
  
\bibitem{r-bh-1h0707} 
  A.~Zoghbi, A.~Fabian, P.~Uttley, G.~Miniutti, L.~Gallo, C.~Reynolds, J.~Miller and G.~Ponti,
  Mon.\ Not.\ Roy.\ Astron.\ Soc.\  {\bf 401}, 2419 (2010)
  [arXiv:0910.0367 [astro-ph.HE]].    

\end{thebibliography}
\end{document}